\documentclass[12pt,prd,superscriptaddress,preprintnumbers,amsmath,amssymb,nofootinbib]{revtex4-2}
\usepackage{xcolor}
\usepackage{graphicx}
\usepackage{tabularx}
\usepackage{latexsym}
\usepackage{float}
\usepackage{epsfig}
\usepackage{subfigure}
\usepackage{longtable}
\usepackage[colorlinks=true,
            linkcolor=blue,
            citecolor=blue,
            urlcolor=blue,
            filecolor=blue,
            anchorcolor=blue]{hyperref}
\usepackage[outdir=./]{epstopdf}
\usepackage{verbatim}
\usepackage{braket}
\usepackage{ragged2e}
\usepackage{tikz}
\usetikzlibrary{positioning, arrows.meta}
\allowdisplaybreaks

\begin{abstract}

\end{abstract}
\begin{document}
\title{
 A comprehensive study of $\Lambda_c^- \to \Lambda (\to p \pi) \mu^- \bar \nu_{\mu}$ incorporating SMEFT implications and right-handed neutrino\\}
%%%%%%%%%%%%%%%%%%%%%%%%%%%%%%%%%%%%%%%%%%%%%%%%%%%%%%%%%%%%%%%%%%%%%%%%%%%%%%%%%%%%%%%%%%%%%%%%%%%%%%%%%%%%%%%%%%%%%%%%%%%%%
\author{Priyanka Boora}
\email{2020rpy9601@mnit.ac.in}
\affiliation{Department of Physics, Malaviya National Institute of Technology Jaipur, India}
%%%%%%%%%%%%%%%%%%%%%%%%%%%%%%%%%%%%%%%%%%%%%%%%%%%%%%%%%%%%%%%%%%
\author{Siddhartha Karmakar}
\email{siddharthak@iitk.ac.in (Presently at IIT Kanpur)}
\affiliation{Tata Institute of Fundamental Research, Mumbai, India}
%%%%%%%%%%%%%%%%%%%%%%%%%%%%%%%%%%%%%%%%%%%%%%%%%%%%%%%%%%%%%%%
\author{Dinesh Kumar}
\email{dinesh@uniraj.ac.in}
\affiliation{Department of Physics, University of Rajasthan, Jaipur 302004, India}
%%%%%%%%%%%%%%%%%%%%%%%%%%%%%%%%%%%%%%%%%%%%%%%%%%%%%%%%%%%%%%%
\author{Kavita Lalwani}
\email{kavita.phy@mnit.ac.in}
\affiliation{Department of Physics, Malaviya National Institute of Technology Jaipur, India}
%%%%%%%%%%%%%%%%%%%%%%%%%%%%%%%%%%%%%%%%%%%%%%%%%%%%%%%%%%%%%%
\maketitle
%%%%%%%%%%%%%%%%%%%%%%%%%%%%%%%%%%%%%%%%%%%%%%%%%%%%%%%%%%%%%%%%%%
\section*{Abstract} \label{abs}
%%%%%%%%%%%%%%%%%%%%%%%%%%%%%%%%%%%%%%%%%%%%%%%%%%%%%%%%%%%%%
We present a comprehensive analysis of the decay $\Lambda_c^- \to \Lambda(\to p\pi)\,\mu^- \bar\nu_\mu$ within a model-independent effective field theory framework. Previous studies have been restricted to the three-body decay $\Lambda_c^+ \to \Lambda \mu^+ \nu_\mu$ and considered only left-handed neutrinos within the Low-Energy Effective Theory (LEFT). In this work, we extend the analysis to the complete four-body angular distribution for the first time, incorporating the indirect constraints implied by the Standard Model Effective Field Theory (SMEFT) on the LEFT Wilson coefficients. We also include the effects of right-handed neutrino (RHN) operators, enabling a unified treatment of both left- and right-handed neutrino interactions in the $c \to s \mu \nu_\mu$ transition.
Using the global bounds derived from LEFT observables and their SMEFT correlations, we study the impact of allowed new-physics scenarios on a variety of observables, including differential decay rates, forward–backward asymmetries, polarization asymmetries of the final-state hadron and lepton, and the angular coefficients (${\cal M}_0$ to ${\cal M}_9$) in the 4-body angular distribution. Our analysis reveals significant deviations from the Standard Model in the observables $\mathcal{P}^\Lambda_L$ and ${\cal M}_1$ for $C^V_{RL}$ and $C^V_{RR}$, and striking $T$-odd effects in ${\cal M}_7$ for complex $C^V_{RL}$. These features provide sensitive probes of vector-type new physics, such as leptoquark or right-handed current models. The predicted patterns can be tested in forthcoming measurements at BESIII, LHCb, and Belle~II, where polarization-sensitive observables in $\Lambda_c$ decays are becoming experimentally accessible.
\clearpage
\tableofcontents

\newpage
%%%%%%%%%%%%%%%%%%%%%%%%%%%%%%%%%%%%%%%%%%%%%%%%%%%%%%%%%%%%%%%%
\section{Introduction}  \label{introduction} 
%%%%%%%%%%%%%%%%%%%%%%%%%%%%%%%%%%%%%%%%%%%%%%%%%%%%%%%%%
The Standard Model (SM) of particle physics is remarkably successful in describing the interactions of fundamental particles. However, it is accompanied by persistent tensions between theoretical predictions and experimental results. This motivates to study the consistencies of experimental results with the SM and search for possible new physics (NP) effects. 

Among the various probes of new physics, flavor physics offers a particularly sensitive testing ground. Since flavor-changing processes are highly suppressed in the Standard Model (SM), even small deviations from theoretical expectations can signal the presence of new interactions. Recent measurements have revealed intriguing anomalies in both charged- and neutral-current transitions, such as $R_{D^{(*)}}$~\cite{BaBar:2012obs,BaBar:2013mob,Belle:2015qfa,LHCb:2015gmp}, $R_{J/\psi}$~\cite{LHCb:2017vlu}, $P'_5$~\cite{LHCb:2013ghj,Descotes-Genon:2012isb,Descotes-Genon:2013wba}, and the branching ratios of $B\to K^{(*)}\ell^+\ell^-$~\cite{LHCb:2014cxe,LHCb:2022qnv,Belle-II:2023esi}. These persistent tensions highlight the potential of precision flavor observables to reveal the flavor structure of NP and motivate the exploration of complementary processes beyond the mesonic sector.

Baryonic modes provide such a complementary direction, offering rich kinematic structures and additional polarization observables that can help to disentangle the Lorentz nature of possible NP interactions. In recent years, several theoretical and phenomenological studies have investigated semileptonic baryon decays~\cite{Dutta:2015ueb, DiSalvo:2016rfi, Ray:2018hrx, Bernlochner:2018bfn, Hu:2018veh, Ferrillo:2019owd, Boer:2019zmp, Mu:2019bin, Becirevic:2022bev, Karmakar:2023rdt, Nandi:2024aia}, demonstrating their sensitivity to both vector and scalar operators and their potential as complementary probes of flavor dynamics.

While most efforts have focused on the bottom sector, the charm sector offers a complementary and comparatively clean environment to test the flavor structure of possible NP scenarios. The weak decays of charm mesons and baryons serve as sensitive probes of the SM, providing an opportunity to search for deviations that may originate from non-standard interactions \cite{Wang:2014uiz, Fajfer:2015ixa, Fleischer:2019wlx, Alvarado:2024lpq, Boora:2024nsx, Zhang:2025tki, Faustov:2019ddj, BESIII:2022qaf, Gutsche:2015rrt, Huang:2021ots, Li:2016qai, Zhao:2018zcb, Gutsche:2014zna, Geng:2020fng}.
With the increasing precision of experimental measurements in the charm sector, particularly from BESIII, Belle~II, and LHCb, it has become possible to perform detailed studies analogous to those in the bottom sector. In particular, the semileptonic decays of charm baryons such as $\Lambda_c$ provide a unique laboratory to test the SM consistency and to search for possible signatures of NP \cite{Richman:1995wm, Bianco:2003vb, Li:2021iwf, Cheng:2021qpd, BESIII:2023jxv, Alvarado:2024lpq, Zhang:2025tki, BESIII:2015ysy, BESIII:2016ffj, BESIII:2022ysa}.

A systematic interpretation of such potential deviations requires a framework that connects measurable low-energy observables to possible high-scale dynamics in a model-independent way. Since new heavy degrees of freedom may not be directly accessible at current colliders, their effects can be captured through Effective Field Theories (EFTs), where higher-dimensional operators built from SM fields encode the influence of heavy states. The Standard Model Effective Field Theory (SMEFT)~\cite{Grzadkowski:2010es, Isidori:2023pyp} parametrizes these interactions above the electroweak scale, while the Low-Energy Effective Theory (LEFT)~\cite{Buchalla:1995vs, Jenkins:2017jig} describes the corresponding dynamics below it. Matching SMEFT onto LEFT links high-scale new physics to precision flavor observables, enabling unified analyses across different sectors and decay modes~\cite{Alonso:2014csa, Cata:2015lta, Azatov:2018knx, Fuentes-Martin:2020lea, Bause:2020auq, Bause:2020xzj, Bissmann:2020mfi, Bause:2021cna, Bause:2021ihn, Bruggisser:2021duo, Bause:2022rrs, Sun:2023cuf, Grunwald:2023nli, Greljo:2023bab, Fajfer:2012vx, Bause:2023mfe, Bhattacharya:2023beo, Chen:2024jlj, Fernandez-Martinez:2024bxg, Karmakar:2024dml}.

The EFT framework can be further extended by including right-handed neutrino fields, which appear naturally in many extensions of the Standard Model aiming to explain neutrino masses and mixings. The presence of such sterile or right-handed neutrinos in semileptonic transitions has been widely investigated in both mesonic and baryonic decays, with significant implications for new physics in the bottom and charm sectors \cite{Alvarado:2024lpq, Penalva:2021wye, Boora:2024nsx, Mandal:2020htr, Datta:2022czw}. Their inclusion enlarges the operator basis of SMEFT and LEFT, allowing for additional Lorentz structures and novel interference patterns with the SM amplitudes. This makes semileptonic charm decays 
particularly suitable for testing scenarios involving right-handed neutrino couplings.

In this work, we focus on the semileptonic decay $\Lambda_c^- \to \Lambda(\to p\pi)\,\mu^- \bar\nu_{\mu}$, which involves second-generation quarks and leptons and provides a clean environment to probe new physics in the $c \to s \mu \nu_{\mu}$ transition. This channel is of particular interest, as it allows us to test possible flavor-dependent structures of new interactions that may differ from those observed in the bottom sector.  This decay mode has received relatively less attention in the literature. Previous studies~\cite{Richman:1995wm, Bianco:2003vb, Li:2021iwf, Cheng:2021qpd, BESIII:2023jxv, Alvarado:2024lpq, Zhang:2025tki, BESIII:2015ysy, BESIII:2016ffj, BESIII:2022ysa} have analyzed it within the LEFT framework, focusing primarily on the two-fold angular distribution. In the present work, we extend the analysis to the complete four-fold angular distribution for the first time, incorporating SMEFT-implied constraints on the LEFT Wilson coefficients. Furthermore, we include the effects of RHN operators within the general EFT framework.

We begin with the most general low-energy effective Hamiltonian for the $c \to s \mu \nu_{\mu}$ process, including scalar and vector operators. The operators and the corresponding Wilson coefficients (WCs) are considered in the LEFT framework. For both the left-handed neutrino (LHN) and right-handed neutrino (RHN) scenarios, we first determine the direct bounds from observables that depend explicitly on the $c\to s \mu \nu_\mu$ transition. For LHN operators, we further do the matching between the LEFT operators the corresponding SMEFT operators. Based on the SMEFT implied correlations, we consider several observables apart from the $c\to s \mu \nu_\mu$ transition and do a global-fit analysis on the SMEFT and thus on the LEFT operators. The resulting bounds, which we call `indirect bounds', are then compared to the direct bounds obtained for the LEFT WCs earlier. We find that for several occasions, these indirect bounds give relatively tighter allowed regions compared to the direct bounds. 

Using the allowed regions of the WCs, we compute a set of key observables for both the three-body decay $\Lambda_c^- \to \Lambda \mu^- \bar\nu_{\mu}$ and the four-body decay $\Lambda_c^- \to \Lambda(\to p\pi),\mu^- \bar\nu_{\mu}$. The observables considered include the differential branching fraction, forward–backward asymmetry, and the polarization asymmetries of the $\Lambda$ baryon and the lepton, along with angular coefficients that characterize the full four-body kinematics of $\Lambda_c^-\to \Lambda(\to p \pi) \mu^- \bar \nu_{\mu}$. We discuss the observables that show sensitivity to different Lorentz structures of the NP operators and serve as powerful probes for testing SM consistency and possible NP searches in future measurements at BESIII, Belle~II, and LHCb.

The paper is structured as follows: In section \ref{TF}, we present the low-energy effective Hamiltonian with matching between EFT and SMEFT operators. In section \ref{npc}, we present the constraints on the NP operators. In section \ref{observables}, we list the observables considered in our analysis. Results of the study are presented in the section \ref{result}. Finally, in section \ref{conclusion} we present the concluding remarks. Other required technical information are provided in Appendices \ref{app-A0}, \ref{app-A}, \ref{app-B}, \ref{app-kin}, and \ref{app-C}.

%%%%%%%%%%%%%%%%%%%%%%%%%%%%%%%%%%%%%%%%%%%%%%%%%%%%%%%%%%%%%
\section{Theoretical Framework} \label{TF}
%%%%%%%%%%%%%%%%%%%%%%%%%%%%%%%%%%%%%%%%%%%%%
In this section, we set up the effective description used throughout the analysis. We begin with the low-energy operator basis (LEFT) for the $\bar c \to \bar s \mu^- \bar\nu_\mu$ transition, specify the subset of SMEFT operators that match onto these interactions, and summarize the tree-level matching relevant for our study. We then collect the hadronic input in terms of $\Lambda_c\to\Lambda$ form factors and, finally, define the helicity amplitudes for both hadronic and leptonic currents.

The low energy effective Hamiltonian for the $ \bar c \to \bar s \mu^- \bar\nu_{\mu}$ transition is written as \cite{Mandal:2020htr, Boora:2024nsx, Zhang:2025tki}
\begin{equation}
\label{eqn-1}
\begin{aligned}
H_{\text{eff}} &= \frac{4 G_F V_{cs}}{\sqrt{2}} \bigg[
    (1 + C_{LL}^V) \mathcal{O}_{LL}^V + C_{RL}^V \mathcal{O}_{RL}^V + C_{LL}^S \mathcal{O}_{LL}^S + C_{RL}^S \mathcal{O}_{RL}^S \\
&\quad  + C_{LR}^V \mathcal{O}_{LR}^V + C_{RR}^V \mathcal{O}_{RR}^V 
+ C_{LR}^S \mathcal{O}_{LR}^S + C_{RR}^S \mathcal{O}_{RR}^S \bigg] + \text{h.c.}
\end{aligned}
\end{equation}

where, $G_F$ is the Fermi coupling constant and $V_{cs}$ is the CKM matrix element. The dimension-six four-fermion operators for the left-handed and right-handed neutrinos are given as follows:
\begin{eqnarray}
\label{equ-2}
\begin{aligned}
&\mathcal{O}_{LL}^V = (\bar{c} \gamma^{\alpha}P_{L} s) (\bar{\mu} \gamma_{\alpha}P_{L} \nu_{\mu}) \quad \quad 
&\mathcal{O}_{LR}^V = (\bar{c} \gamma^{\alpha}P_{L} s) (\bar{\mu} \gamma_{\alpha}P_{R} \nu_{\mu}) \\
&\mathcal{O}_{RL}^V = (\bar{c} \gamma^{\alpha}P_{R} s) (\bar{\mu} \gamma_{\alpha}P_{L} \nu_{\mu})  \quad \quad 
&\mathcal{O}_{RR}^V = (\bar{c} \gamma^{\alpha}P_{R} s) (\bar{\mu} \gamma_{\alpha}P_{R} \nu_{\mu}) \\
&\mathcal{O}_{LL}^S = (\bar{c} P_{L} s) (\bar{\mu} P_L \nu_{\mu}) \quad \quad  
&\mathcal{O}_{LR}^S = (\bar{c} P_{L} s) (\bar{\mu} P_R \nu_{\mu})\\
 &\mathcal{O}_{RL}^S = (\bar{c} P_{R} s) (\bar{\mu} P_L \nu_{\mu})  \quad \quad 
&\mathcal{O}_{RR}^S = (\bar{c} P_{R} s) (\bar{\mu} P_R \nu_{\mu}) 
\end{aligned}
\end{eqnarray}

The SMEFT operators relevant to $\bar c \to \bar s \mu^- \bar\nu_{\mu}$ transitions are as follows:

\begin{eqnarray}
\label{equ-3}
\begin{aligned}
    \mathcal{O}_{lq}^{(3)} &=  \left(\bar{\ell}_i \gamma_\mu \tau^I \ell_j\right)\left(\bar{q}_k \gamma^\mu \tau^I q_l\right), \quad \quad
    \mathcal{O}_{ledq} = \left(\bar{\ell}_i^a e_j\right)\left(\bar{d}_k q_l^a\right)  \\
    \mathcal{O}_{lequ}^{(1)} &= \left(\bar{\ell}_i^a e_j\right) \epsilon_{a b}\left(\bar{q}_k^b u_l\right) , \quad \quad 
    \mathcal{O}_{\phi ud} = i\left(\tilde{\phi}^{\dagger} D_\mu \phi\right)\left(\bar{u}_i \gamma^\mu d_j\right)
\end{aligned}
\end{eqnarray}
where $q$, $\ell$, and $\phi$ represent the quark, lepton, and Higgs doublets, respectively. Whereas $u$, $d$, and $e$ denote the right-handed quark and lepton singlets. In this work, we exclude the tensor operator. One practical reason is the lack of reliable Lattice QCD determinations for the corresponding tensor form factors in $\Lambda_c \to \Lambda$ transitions~\cite{Detmold:2015aaa, Meinel:2016dqj}. In addition, tensor interactions are already tightly constrained by precision studies of nuclear and kaon decays~\cite{Gonzalez-Alonso:2013uqa, Bhattacharya:2011qm}, leaving little room for sizable effects in the charm sector. From a theoretical perspective, such operators are not generated at tree level in most ultraviolet completions that generate semileptonic interactions, and receive only suppressed contributions through renormalization-group mixing~\cite{Aebischer:2015fzz, Jenkins:2017jig}. For these reasons, our analysis focuses on vector and scalar operators, which capture the dominant phenomenological effects.

Gauge-invariant SMEFT operators defined at the high scale $\Lambda$ are matched onto LEFT at $\mu_W$ by integrating out the heavy SM fields ($W^\pm$, $Z$, $t$, $h$)~\cite{Aebischer:2015fzz}. The subsequent RG running to the hadronic scale relates the SMEFT coefficients to the LEFT Wilson coefficients relevant for observables. At the tree level, the matching for left-handed neutrino operators is given as~\cite{Aebischer:2015fzz}

\begin{equation}
\label{eqn-4}
\begin{aligned}
    C_{LL}^V &= \frac{v^2}{\Lambda^2} \tilde{C}_{lq}^{(3)\, ll22}, &\quad 
    C_{RL}^V = - \frac{v^2}{2 \Lambda^2 V_{cs}} \tilde{C}_{\phi ud}^{ 22},  \\
    C_{LL}^S &=  \frac{v^2}{2 \Lambda^2} \tilde{C}_{lequ}^{(1)* \, ll22}, &\quad
    C_{RL}^S = \frac{v^2}{2 \Lambda^2 } \tilde{C}_{ledq}^{* l l 22}
\end{aligned}
\end{equation}

where $v$ denotes the Higgs vacuum expectation value, $\Lambda$ represents the characteristic new-physics scale, and $\tilde{C}$ are the Wilson coefficients in the fermion-mass basis. The relations in eq.~\ref{eqn-4} correspond to the $c \to s \mu^- \bar{\nu}_{\mu}$ ($2222$) transition, for which the CKM matrix elements cancel in all cases except in $C_{RL}^V$.

%%%%%%%%%%%%%%%%%%%%%%%%%%%%%%%%%%%%%%%%%%%%%%%%%%%%%%%%%%%%%%%%%%%%%%
\section{Constraining the NP Wilson coefficients} \label{npc}
%%%%%%%%%%%%%%%%%%%%%%%%%%%%%%%%%%%%%%%%%%%%%%%%%%%%%%%%%%%%%%%%%%%%

In this section, we determine the allowed regions for the LEFT Wilson coefficients: 
$C_{LL}^V$, $C_{RL}^V$, $C_{LL}^S$, and $C_{RL}^S$ in the LHN scenario, and 
$C_{LR}^V$, $C_{RR}^V$, $C_{LR}^S$, and $C_{RR}^S$ in the RHN scenario. 
These coefficients contribute directly to the $c \to s \mu \nu_{\mu}$ mediated observables listed as 1-6 in Table~\ref{tab:observables}. 
All of these observables are mutually independent.  

To extract the constraints, we perform a $\chi^2$ minimization and obtain the best-fit values along with their $1\sigma$ allowed regions. 
The generic definition of $\chi^2$ is given by  
\begin{equation}
    \label{eqn-chi2}
    \chi^2\!\left(C^i\right) \;=\; 
    \frac{\big(\mathcal{O}_{th}(C^i) - \mathcal{O}_{exp}\big)^2}{\sigma_{th}^2 + \sigma_{exp}^2},
\end{equation}
where $\mathcal{O}_{th}$ denotes the theoretical prediction as a function of the Wilson coefficient $C^i$, 
and $\mathcal{O}_{exp}$ denotes the corresponding experimental measurement. 
The quantities $\sigma_{th}$ and $\sigma_{exp}$ represent the theoretical and experimental uncertainties, respectively. 
The uncertainties in the best-fit values are obtained from the likelihood estimates using the \texttt{MINUIT} package~\cite{James:1975dr, James:1994vla}.

%\begin{table}[H]
%	\begin{center}
%		\resizebox{0.9\textwidth}{!}{%
			%\setlength{\tabcolsep}{12pt} 
			\begin{longtable}{|c|c|c|}
				%\hline\noalign{\smallskip}
				\hline
				\textbf{S. No.} & \textbf{Mode} & \textbf{Experimental Measurement}   \\
				\hline
				\multicolumn{3}{|c|}{\textbf{$c \to s $ Observables}} \\
				\hline
				1  &$\mathcal{B}(D_{s}^{+} \to \mu^{+} \nu_{\mu})$ & (5.35 $\pm$ 0.12) $\times 10^{-3}$ \cite{PhysRevD.110.030001}\\
				
				2  &$\mathcal{B}(D^{0} \to K^{*-} \mu^{+} \nu_{\mu})$ & (1.89 $\pm$ 0.24) $\times 10^{-2}$ \cite{PhysRevD.110.030001} \\ 
				3  &	$\mathcal{B}(D^{0} \to K^{-} \mu^{+} \nu_{\mu})$ & (3.41 $\pm$ 0.04) $\times 10^{-2}$ \cite{PhysRevD.110.030001} \\
				4  &	$\mathcal{B}(D^{+} \to \Bar{K^{*0}} \mu^{+} \nu_{\mu})$ & (5.27 $\pm$ 0.15) $\times 10^{-2}$ \cite{PhysRevD.110.030001} \\
				5  &	$\mathcal{B}(D^{+} \to \Bar{K^{0}} \mu^{+} \nu_{\mu})$ &(8.76 $\pm$ 0.19) $\times 10^{-2}$ \cite{PhysRevD.110.030001} \\
				6  &	$\mathcal{B}(D_{s}^{+} \to \phi \mu^{+} \nu_{\mu})$ & (2.24 $\pm$ 0.11) $\times 10^{-2}$ \cite{PhysRevD.110.030001} \\
				\hline
				\multicolumn{3}{|c|}{\textbf{$c \to d $ Observables}} \\
				\hline
				7 & $\mathcal{B}(D^+ \to \mu^+ \nu_{\mu})$& $(3.74 \pm 0.17) \times 10^{-4}$ \cite{PhysRevD.110.030001}   \\ 
				8 & $\mathcal{B}(D^+ \to \pi^0 \mu^+ \nu_{\mu})$& $(3.50 \pm 0.15) \times 10^{-3}$ \cite{PhysRevD.110.030001}   \\ 
				9 & $\mathcal{B}(D^0 \to \pi^- \mu^+ \nu_{\mu})$ & $(2.67 \pm 0.12) \times 10^{-3}$ \cite{PhysRevD.110.030001}   \\ 
				\hline
				\multicolumn{3}{|c|}{\textbf{$b \to c $ Observables}} \\
				\hline
				10 &   $R_D^{\mu e}$ & (0.993 $\pm$ 0.0089 $\pm$ 0.0187) \cite{PhysRevD.110.030001} \\ 
				11 & $R_{D^*}^{\mu e}$ & (1.002 $\pm$ 0.009 $\pm$ 0.02) \cite{PhysRevD.110.030001} \\ 
				\hline
				\multicolumn{3}{|c|}{\textbf{$b \to s $ Observables}} \\
				\hline
				12 &  $\mathcal{B}(B_s \to \mu^+ \mu^-)$ & ( 3.34 $\pm$ 0.27) $\times 10^{-9} $ \cite{PhysRevD.110.030001}  \\ 
				13 &  $\mathcal{B}(B^+ \to K^+ \nu \nu)$ & (2.3 $\pm$ 0.7) $\times 10^{-5}$ \cite{Belle-II:2023esi}  \\ 
				14 &  $P_5^{\prime}$ \text{[4 - 6]} GeV$^2$ & $-0.439 \pm 0.111 \pm 0.036$ \cite{LHCb:2020lmf}  \\ 
				15 &  $R_{\phi}^{-1}$ [0.1,1.1] GeV$^2$ & $1.57^{+0.28}_{-0.25} \pm 0.05$  \cite{LHCb:2024rto}  \\ 
				16 &   $R_K \, [0.1, 1.1] $ GeV$^2$ & $0.994_{-0.082}^{+0.090}(\text { stat })_{-0.027}^{+0.029}($ syst $)$ \cite{LHCb:2022qnv}  \\ 
				17 &  $R_{K^*} \, [0.1, 1.1]$ GeV$^2$ & $0.927_{-0.087}^{+0.093}$ (stat) ${ }_{-0.035}^{+0.036}$ (syst) \cite{LHCb:2022qnv}  \\ 
				18 &  $R_K \, [1.1, 6.0]$ GeV$^2$ & $0.949_{-0.041}^{+0.042}(\mathrm{stat})_{-0.022}^{+0.022}(\mathrm{syst})$ \cite{LHCb:2022qnv}   \\ 
				19 &  $R_{K^*} \, [1.1, 6.0]$ GeV$^2$ & $1.027_{-0.068}^{+0.072}(\mathrm{stat})_{-0.026}^{+0.027}(\mathrm{syst})$  \cite{LHCb:2022qnv}  \\
				20 & $\frac{d\mathcal{B}}{dq^2}$ ($B_s \to \phi \mu^+ \mu^-$) [1.1, 6.0] GeV$^2$ & ($ 2.88 \pm 0.15 \pm 0.05 \pm 0.14 $) $\times 10^{-8}$ \cite{LHCb:2021zwz}  \\ 
				21  & $\frac{d\mathcal{B}}{dq^2}$ ($B_s \to \phi \mu^+ \mu^-$) [15.0, 19.0] GeV$^2$ & ($ 4.63 \pm 0.20 \pm 0.11 \pm 0.22$) $\times 10^{-8}$ \cite{LHCb:2021zwz} \\ 
				\hline
				\multicolumn{3}{|c|}{\textbf{$b \to u $ Observables}} \\
				\hline
				22 & $\mathcal{B}(B^0 \to \pi^- \ell^+ \nu_{\ell})$ & $(1.50 \pm 0.06) \times 10^{-4}$ \cite{PhysRevD.110.030001}  \\
				\hline
				\multicolumn{3}{|c|}{\textbf{$s \to u $ Observables}} \\
				\hline
				23 & $\mathcal{B}(K^+ \to \mu^+ \nu_{\mu})$& (63.56 $\pm$ 0.11) $\times 10^{-2}$ \cite{PhysRevD.110.030001}   \\ 
				\hline
				\multicolumn{3}{|c|}{\textbf{$s \to d $ Observables}} \\
				\hline
				24 &  $\mathcal{B}(K^+ \to \pi^0 \mu^+ \nu_{\mu})$ & (3.352 $\pm$ 0.034) $\times 10^{-2}$ \cite{PhysRevD.110.030001} \\
				\hline   
%\captionsetup{width=\linewidth}
\caption{List of the observables for calculating bounds of NP operators.}
\label{tab:observables}   
			\end{longtable}
		
%	\end{center}
%\end{table}

\begin{figure} [h!]
    \centering
    \includegraphics[width=0.49\linewidth]{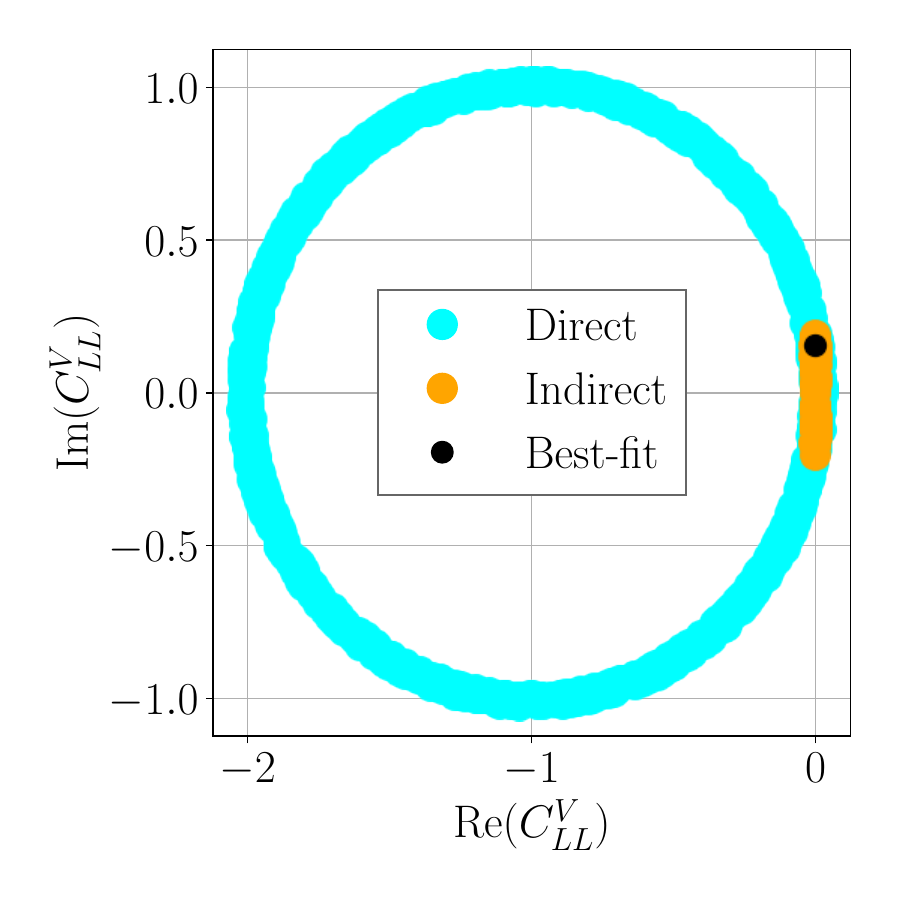}
    \includegraphics[width = 0.49\linewidth]{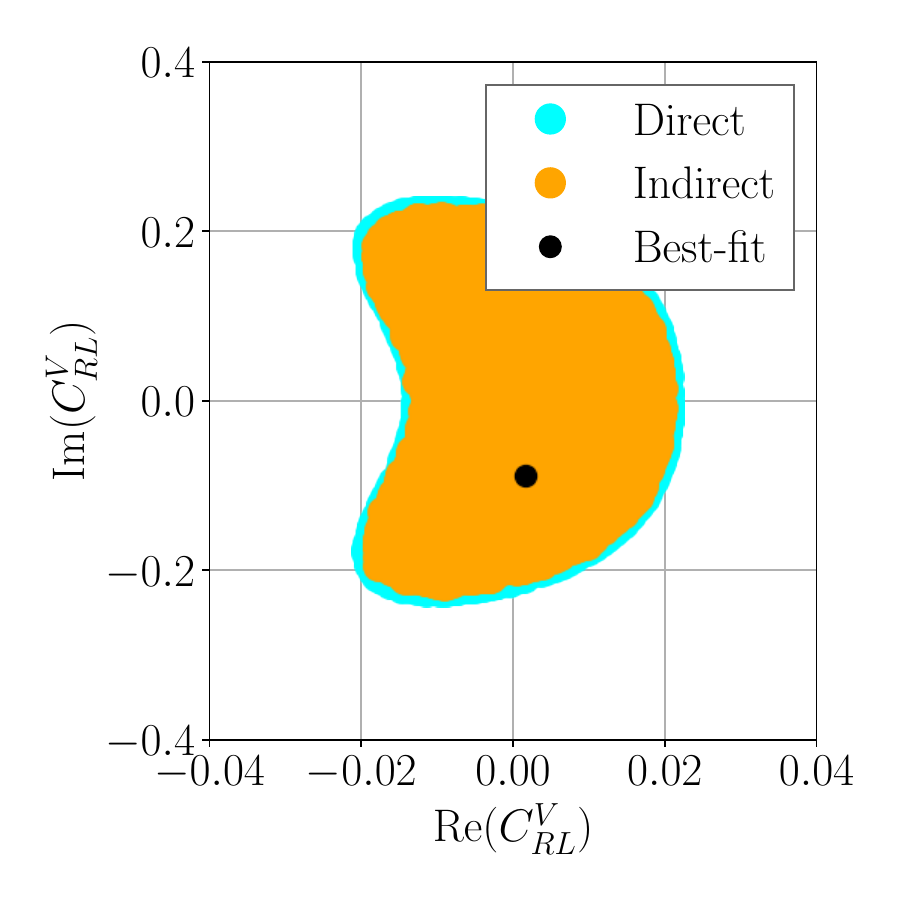}\\
\includegraphics[width=0.49\linewidth]{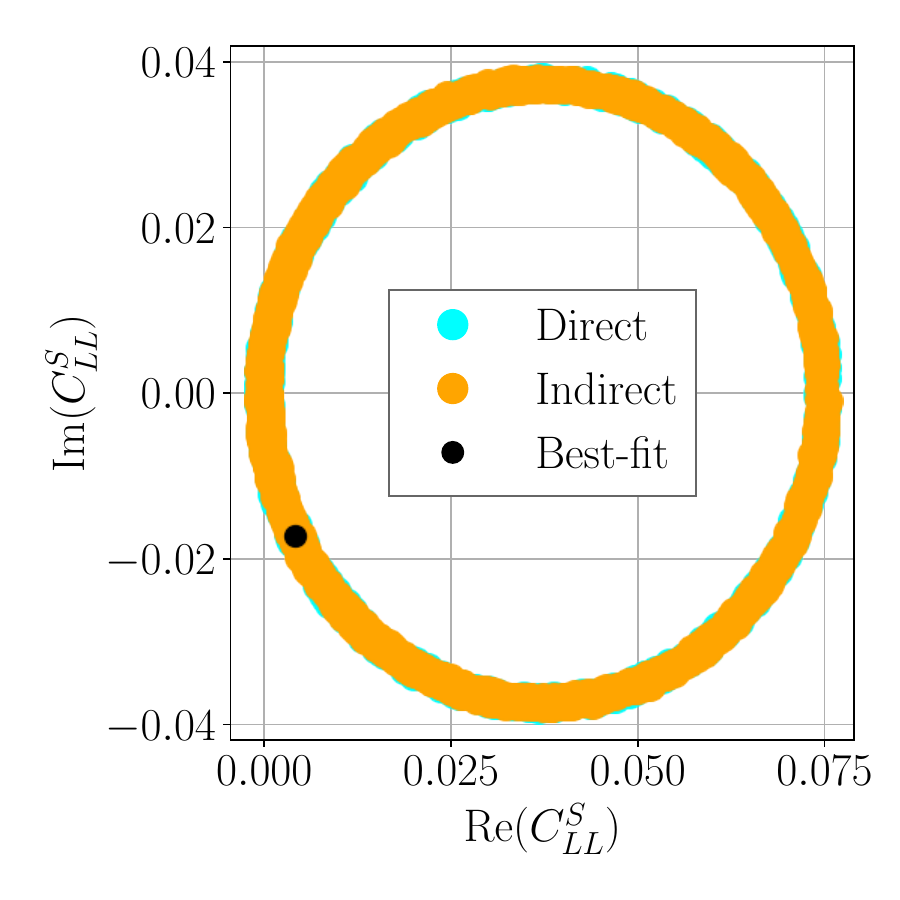}
    \includegraphics[width = 0.49\linewidth]{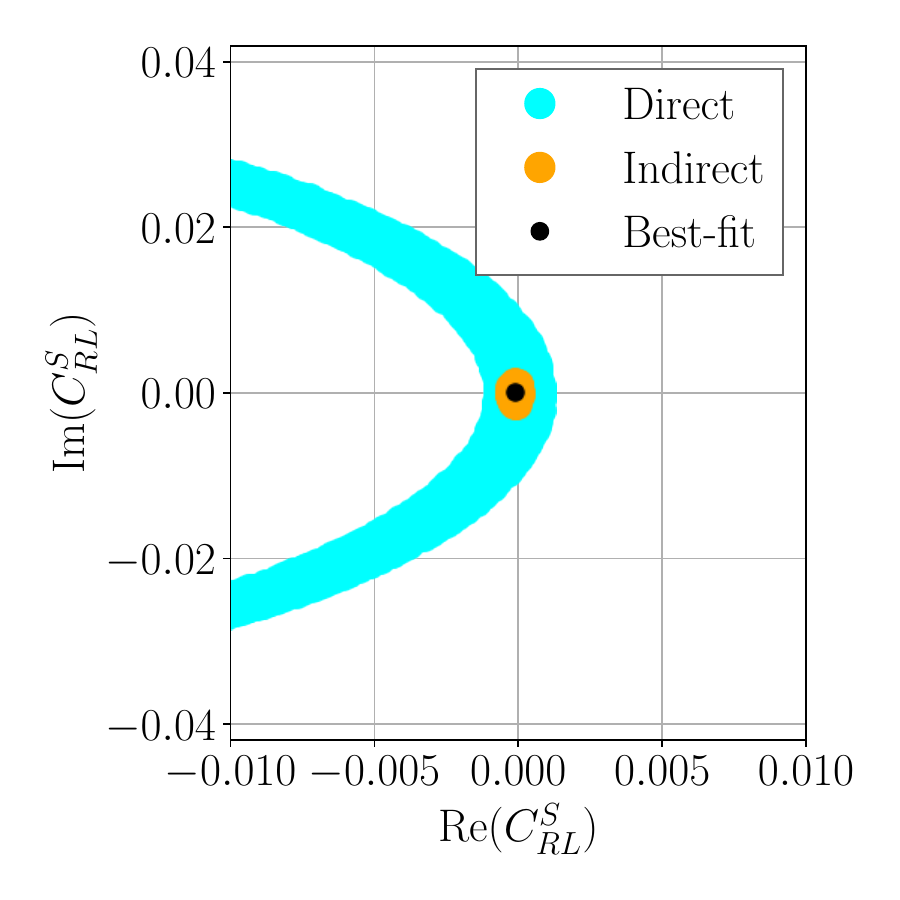}\\
    \caption{\small Constraints on the LEFT WCs involving LHNs. The cyan regions denote that constraining observables are mediated directly via $c\to s\mu \nu_\nu$ transition. Orange regions denote SMEFT implied constraints.}
    \label{fig:LHNbounds}
\end{figure}

\begin{figure}[h!]
    \centering
    \includegraphics[width=0.49\linewidth]{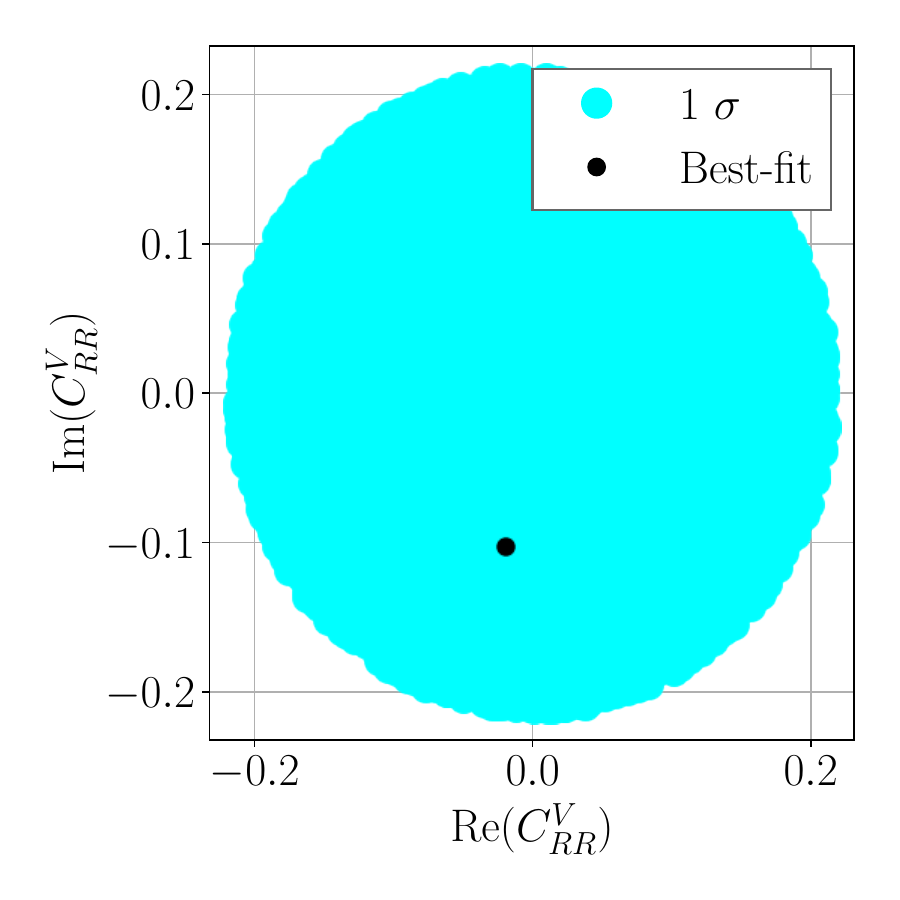}
    \includegraphics[width = 0.49\linewidth]{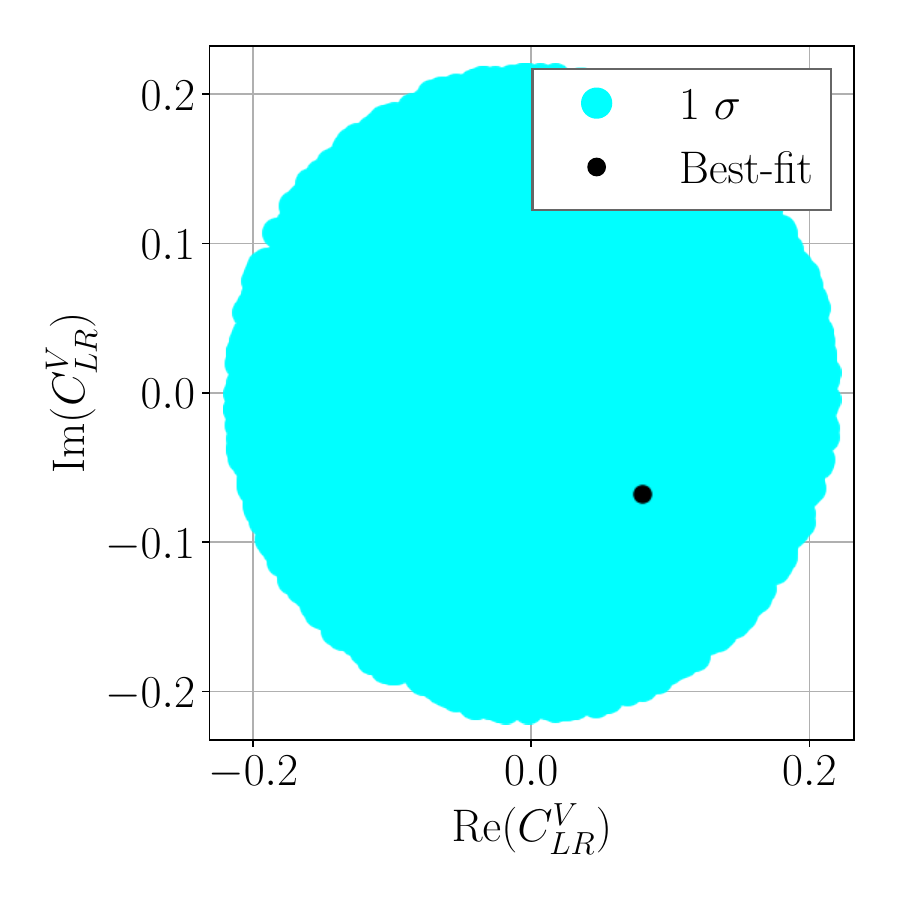}\\
\includegraphics[width=0.49\linewidth]{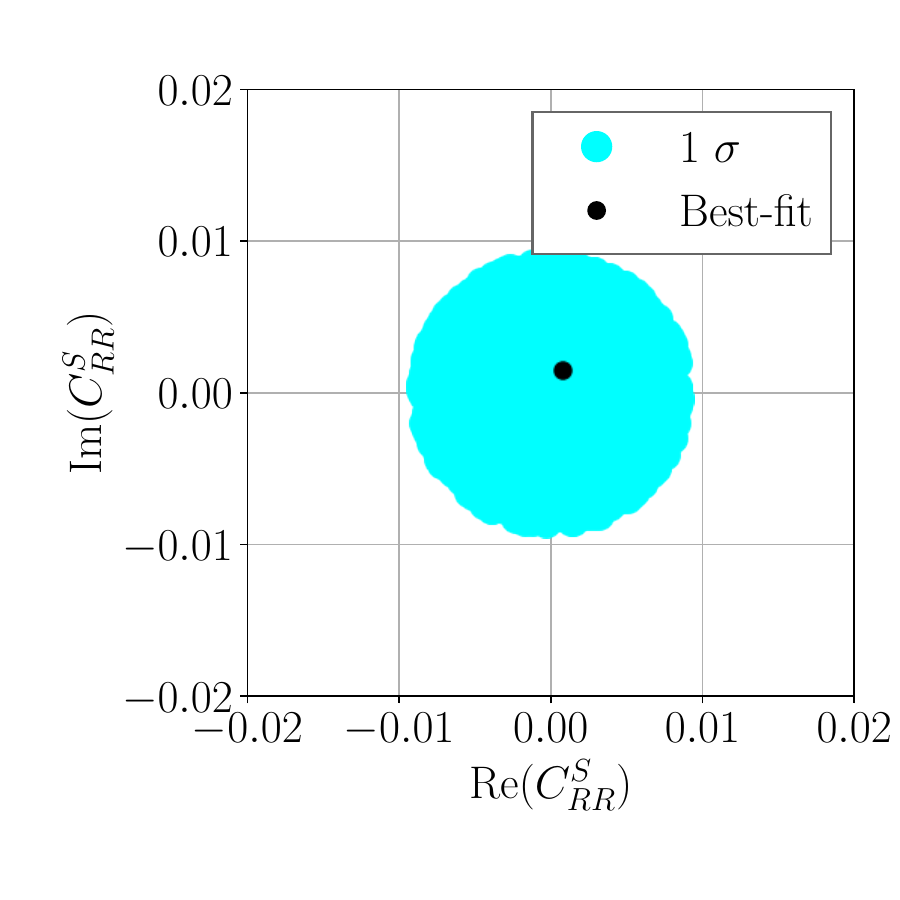}
    \includegraphics[width = 0.49\linewidth]{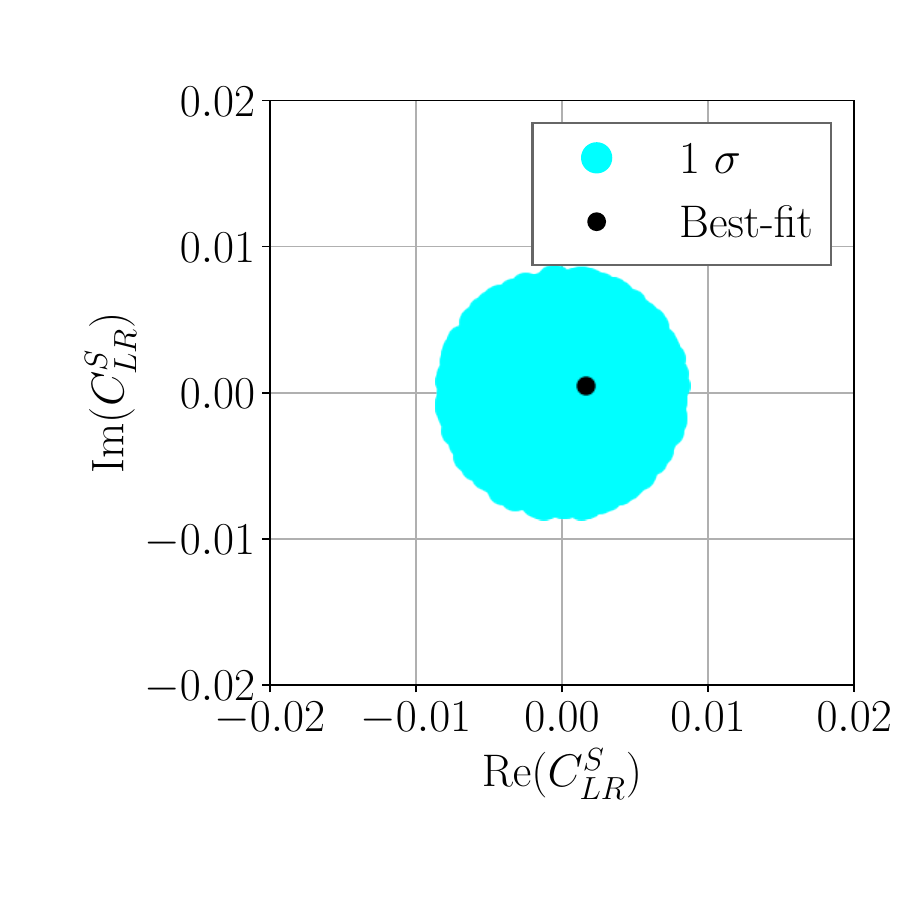}\\
    \caption{\small Constraints on the LEFT WCs involving RHNs. The constraining observables are mediated directly via $c\to s\mu \nu_\nu$ transition.}
    \label{fig:RHNbounds}
\end{figure}

When we use only the observables 1--6 from Table \ref{tab:observables} and apply the $\chi^2$ minimization of eq.~(\ref{eqn-chi2}), to constrain the LEFT Wilson coefficients, we refer to the resulting limits as the \textbf{direct bounds}. 
These direct bounds are displayed as the cyan regions in Figs.~\ref{fig:LHNbounds} and \ref{fig:RHNbounds}. 

Within the SMEFT framework, the LEFT operators are generated by certain SMEFT operators defined at a higher energy scale, as listed in eq.(\ref{eqn-4}). However, these SMEFT operators, in addition to generating the LEFT operators relevant for $c \to s \mu \nu_{\mu}$, also induce several other operators through renormalization group (RG) running, matching, and the transition from the flavor to the mass basis  \cite{Alonso:2014csa, Henning:2014wua, Cata:2015lta, Azatov:2018knx, Fuentes-Martin:2020lea, Bause:2020auq, Bause:2020xzj, Bissmann:2020mfi, Bause:2021cna, Bause:2021ihn, Bruggisser:2021duo, Bause:2022rrs, Sun:2023cuf, Grunwald:2023nli, Greljo:2023bab, Fajfer:2012vx, Bause:2023mfe, Bhattacharya:2023beo, Chen:2024jlj, Fernandez-Martinez:2024bxg, Karmakar:2024gla}. As a result, they contribute to a wide range of observables beyond those directly mediated by $c \to s \mu \nu_{\mu}$.

In our SMEFT analysis, we begin with the SMEFT Wilson coefficients that directly match onto the $c \to s \mu \nu_{\mu}$ LEFT operators. We then include, along with the $c \to s \mu \nu_{\mu}$ observables, all additional observables that receive significant indirect contributions from the considered SMEFT operators via operator mixing. All of these observables are treated as independent.
We perform a global $\chi^2$ fit for the SMEFT Wilson coefficients using the set of observables listed in Table \ref{tab:observables}, applying the $\chi^2$ definition of eq.(\ref{eqn-chi2}). The $c \to s \mu \nu_{\mu}$ observables are computed using the available analytic expressions \cite{Fleischer:2019wlx, Leng:2020fei, Boora:2024nsx}, while the remaining observables are evaluated with the \texttt{flavio} package~\cite{Straub:2018kue}. From this fit, we extract constraints on the SMEFT Wilson coefficients at the scale $\Lambda = 1$~TeV. The corresponding bounds on the LEFT coefficients are then obtained by evolving and matching the SMEFT results using the \texttt{wilson} package~\cite{Aebischer:2018bkb}.
We refer to these SMEFT-implied constraints on the LEFT coefficients as \textbf{indirect bounds}. The indirect bounds are shown in Fig.~\ref{fig:LHNbounds} in the orange region. In this work, we present the indirect bounds only for the LHN scenario. For the RHN case, we restrict ourselves to the direct bounds and leave the calculation of the corresponding indirect bounds to future work.

In Fig.~\ref{fig:LHNbounds}, we compare the indirect bounds with the direct bounds. We find that for all Wilson coefficients, the indirect bounds are consistently tighter than the direct ones. This improvement arises because many of the observables entering the indirect analysis are measured with much higher precision than those used in the direct bound analysis. 

The observables mediated via $c \to s \mu \nu_\mu$ (i.e., direct-bound observables) receive a large SM contribution from a tree-level diagram (with no CKM suppression). In such modes, NP effects are extracted as small deviations on top of a large signal, which naturally leads to weaker direct bounds. The same SMEFT operator, under RG running, induces operators that contribute to $b\to s \mu^+\mu^-$ transitions. The induced Wilson coefficient scales as $\propto y_t^2\,V_{tb}V_{ts}^\ast \ln(\Lambda/\mu_{\rm EW})$,\cite{Jenkins:2013zja} and can be sizable enough to affect $b\to s \mu^+\mu^-$ observables. In particular, $P_5^\prime$ and $\frac{d \cal B}{dq^2}(B_{s}\to \phi \mu^+\mu^-)$ are loop-suppressed and CKM-suppressed in the SM, and are therefore sensitive even to small NP contributions, especially through the vector operators $\mathcal{O}_{LL}^V$ and $\mathcal{O}_{RL}^V$. Consequently, the indirect bounds on $C_{LL}^V$ and $C_{RL}^V$ driven by these observables are significantly stronger than the corresponding direct bounds.

Among the scalar operators, $\mathcal{O}_{LL}^S$ is already strongly constrained by the direct observable $\mathcal{B}(D^-\to \mu^-\nu_\mu)$, and the indirectly constraining modes do not lead to a notable improvement. This is because the RG running of the corresponding SMEFT operator $\mathcal{O}_{lequ}^{(1),2222}$ does not generate sizable mixing into operators involving different quark generations. As a result, the indirectly constrained region almost overlaps with the direct bound, as shown in Fig.~\ref{fig:LHNbounds}.

On the other hand, the SMEFT operator $\mathcal{O}_{ledq}^{2222}$ induces operators relevant for the $s\to u$ transition. This leads to a strong indirect constraint on $\mathcal{O}_{RL}^S$ from the precisely measured $\mathcal{B}(K^+ \to \mu^+ \nu_\mu)$. An analogous improvement does not arise for the vector operators $\mathcal{O}^V_{LL}$ and $\mathcal{O}^V_{RL}$, even though operator mixing is present. The corresponding $s \to u \mu \nu$ modes do not impose comparably strong constraints because any NP contribution in the vector channel is suppressed relative to the dominant SM amplitude (with $C^{V}_{LL,\text{SM}} = 1$).

In Fig.~\ref{fig:RHNbounds}, we present the direct bounds for the WCs corresponding to RHN. We note that for both the LHN and the RHN scenarios, the allowed regions for the vector operators ($C^V_{LL}$, $C^V_{RL}$, $C^V_{LR}$ and $C^V_{RR}$ ) are larger compared to the allowed regions for the scalar operators ($C^S_{LL}$, $C^S_{RL}$, $C^S_{LR}$ and $C^S_{RR}$).

%%%%%%%%%%%%%%%%%%%%%%%%%%%%%%%%%%%%%%%%%%%%%%%%%%%%%%%%%%%%%%%%%%%%%%%%%
\section{Angular Distributions of the Decay} \label{observables}
%%%%%%%%%%%%%%%%%%%%%%%%%%%%%%%%%%%%%%%%%%%%%%%%%%%%%%%%%%%%%%%%%%%
In this work, we analyze the full decay distribution for the three-body $\Lambda_c^- \to \Lambda \mu^- \nu_{\mu}$ and four-body $\Lambda_c^- \to \Lambda (\to p \pi) \mu^- \nu_{\mu}$ decay processes. In order to analyze the decay systematically, we first describe the observables for the three-body decay and then, in line, formulate the four-body decay. 

%%%%%%%%%%%%%%%%%%%%%%%%%%%%%%%%%%%%%%%%%%%%%%%%%%%%%%%%%%%%%%%%%%%%%%%%%%%%%
\subsection{$\Lambda_c^- \to \Lambda \mu^- \nu_{\mu}$ Decay} \label{3_body_obs}
%%%%%%%%%%%%%%%%%%%%%%%%%%%%%%%%%%%%%%%%%%%%%%%%%%%%%%%%%%%%%%%%%%%%%%%%%%%%
The angular distribution of the decay $\Lambda_c^-\to \Lambda \mu^- \bar \nu_\mu$ is written in terms of the helicity amplitudes \cite{kutsckhe1996angular, Richman:1984gh} as
\begin{align}
	I(\Omega) &= \sum_{\lambda_i, \lambda_j^\prime, s_3, s_3^\prime} \left\{
	 H^*_{\lambda_2 \lambda_3} H_{\lambda_2^\prime \lambda_3^\prime} \,
	D^{s_3}_{\lambda_3, \lambda_4 - \lambda_5}(\Omega) \, D^{s_3*}_{\lambda_3^\prime, \lambda_4 - \lambda_5}(\Omega) L^*_{\lambda_4 \lambda_5} L_{\lambda_4 \lambda_5} \right\}\,
\end{align}
The Wigner $D$-matrix is defined as
\begin{align}
	D^j_{m, m^\prime}(\Omega_i) \equiv e^{-i m \phi_i} d^j_{m, m^\prime}(\theta_i) e^{i m^\prime \phi_i}~
\end{align}
where $\Omega_i = (\phi_i, \theta_i, -\phi_i)$ are the Euler angles chosen following the Jacob-Wick convention~\cite{Jacob:1959at}. Here $\theta$ is the angle of $\mu^-$ momentum with respect to the $W^{*-}$ direction, whereas the angle $\phi$ is not physical.
The polarization index $s_3$ associated with the virtual $W^*$ can take values $0$ and $1$, corresponding to longitudinal and transverse polarizations, respectively.
The quantities $H_{\lambda_i \lambda_j}$ represent the hadronic matrix elements in the helicity basis for the decay $\Lambda_c \rightarrow \Lambda(\lambda_2)\, W^*(\lambda_3)$. The leptonic matrix element for the decay $W^* \rightarrow \mu(\lambda_4)\, \nu_\mu(\lambda_5)$ is denoted by $L_{\lambda_4 \lambda_5}$. Note that we use both the notations $L_{\lambda_\mu \lambda_\nu}$ and $L_{\lambda_\mu\lambda_{W}}$ for the leptonic amplitudes interchangeably. Here $\lambda_i$ is the helicity of the corresponding particle. The details of the helicity amplitudes are given in the Appendices \ref{decay_amp} and \ref{app-A}. In this work for the analysis of $\Lambda_c \to \Lambda$ decay, we used the results of form factors from the Lattice QCD, and the expressions are given in the Appendix \ref{ffs}.

By summing over the helicities of the hadrons and leptons, the three-body decay can be described in terms of the angle $\theta$ and variable $q^2$. Therefore, 
two-fold angular distribution, including left-handed and right-handed neutrinos, is written as follows~\cite{Zhang:2025tki}

\begin{equation}
\label{eqn-2fold}
\begin{aligned} 
\frac{d^2 \varGamma\left(\varLambda_c^- \rightarrow \varLambda \mu^{-} \bar \nu_{\mu}\right)}{d q^2 d \cos \theta} = & \frac{G_F^2\left|V_{c s}\right|^2 q^2 \sqrt{Q_{+} Q_{-}}}{2^{10} \pi^3 m_{\varLambda_c}^3}\left(1-\frac{m_{\mu}^2}{q^2}\right)^2 \mathcal{A}_{total}
\end{aligned}
\end{equation}

The total amplitude function $\mathcal{A}_{total}$ is defined as
\begin{equation}
\begin{aligned}
\mathcal{A}_{total} = & \Bigg\{ \Big[ \left|1+C_{LL}^V\right|^2 \mathcal{A}_{V L}^{\nu_L}+\left|C_{RL}^V\right|^2 \mathcal{A}_{V R}^{\nu_L} + \left|C_{LL}^S\right|^2 \mathcal{A}_{S L}^{\nu_L} +\left|C_{RL}^S\right|^2 \mathcal{A}_{S R}^{\nu_L} \\ & + 2 \operatorname{\Re}\left[\left(1+C_{LL}^V\right)^* C_{RL}^V\right] \mathcal{A}_{V L, V R}^{\nu_L, \mathrm{int}}  + 2 \operatorname{\Re}\left[\left(1+C_{LL}^V\right)^* C_{LL}^S\right] \mathcal{A}_{V L, S L}^{\nu_L, \mathrm{int}} \\ & + 2 \operatorname{\Re}\left[\left(1+C_{LL}^V\right)^* C_{RL}^S\right] \mathcal{A}_{V L, S R}^{\nu_L, \mathrm{int}} + 2 \operatorname{\Re}\left[C_{RL}^{V*}\, C_{LL}^S\right] \mathcal{A}_{V R, S L}^{\nu_L, \mathrm{int}} \\ & +2 \operatorname{\Re}\left[C_{RL}^{V*}\, C_{RL}^S\right] \mathcal{A}_{V R, S R}^{\nu_L, \mathrm{int}}  + 2 \operatorname{\Re}\left[C_{L L}^{S*}\, C_{RL}^S\right] \mathcal{A}_{S L, S R}^{\nu_L, \mathrm{int}} \Big] \\ & +  \Big[ \left|C_{LR}^V\right|^2 {\mathcal{A}_{V L}^{\nu_R}}+ \left|C_{RR}^V\right|^2 \mathcal{A}_{V R}^{\nu_R}  + \left|C_{LR}^S\right|^2 \mathcal{A}_{S L}^{\nu_R} + \left|C_{RR}^S\right|^2 \mathcal{A}_{S R}^{\nu_R} \\ & + 2 \operatorname{\Re}\left[C_{LR}^{V*}\, C_{RR}^V\right] \mathcal{A}_{V L, V R}^{\nu_R, \mathrm{int}} + 2 \operatorname{\Re}\left[C_{LR}^{V*}\, C_{LR}^S\right] \mathcal{A}_{V L, S L}^{\nu_R, \mathrm{int}}  \\ & + 2 \operatorname{\Re}\left[C_{LR}^{V*} \,C_{RR}^S\right] \mathcal{A}_{V L, S R}^{\nu_L, \mathrm{int}}  + 2 \operatorname{\Re}\left[C_{RR}^{V*}\, C_{LR}^S\right] \mathcal{A}_{V R, S L}^{\nu_R, \mathrm{int}}  \\ & + 2 \operatorname{\Re}\left[C_{RR}^{V*}\, C_{RR}^S \right] \mathcal{A}_{V R, S R}^{\nu_R, \mathrm{int}} + 2 \operatorname{\Re}\left[C_{L R}^{S*}\, C_{RR}^S \right] \mathcal{A}_{S L, S R}^{\nu_R, \mathrm{int}} \Big] \Bigg\}  
\end{aligned}
\end{equation}

The functions $\mathcal{A}_i^{\nu_L, \nu_R}$ are the amplitude functions for the left-handed and right-handed neutrinos, respectively. These are calculated in terms of combinations of helicity amplitudes $H_{\lambda_2\lambda_3}$ and presented in the Appendix \ref{app-B}.

The integration of the eq. \ref{eqn-2fold} over cos$\theta$ $\in$ (-1, 1) gives the differential decay width, as follows: 

\begin{equation} \label{diff_decay_width}
\frac{d \Gamma \left(\varLambda_c^- \rightarrow \varLambda \mu^{-} \bar \nu_{\mu}\right)}{d q^2}=\frac{G_F^2\left|V_{c s}\right|^2 q^2 \sqrt{Q_{+} Q_{-}}}{2^{10} \pi^3 m_{\varLambda_c}^3}\left(1-\frac{m_{\mu}^2}{q^2}\right)^2 \int_{-1}^1 \mathcal{A}_{\mathrm{total}} d \cos \theta
\end{equation}

The differential branching fraction can be expressed as 

\begin{equation} \label{diff_br}
\frac{d \mathcal{B} \left(\varLambda_c^- \rightarrow \varLambda \mu^{-} \bar \nu_{\mu}\right)}{d q^2} = \tau_{\Lambda_c}  \frac{d \Gamma \left(\varLambda_c^- \rightarrow \varLambda \mu^{-} \bar \nu_{\mu}\right)}{d q^2}
\end{equation}

where $\tau_{\Lambda_c}$ is the $\Lambda_c$ life time as listed in the Table \ref{tab:parameters}.
An experimentally robust observable, the forward-backward asymmetry	is defined as
 
\begin{equation}
\mathcal{A}_{\mathrm{FB}}\left(q^2\right)=\frac{\int_0^1 \frac{d^2 \varGamma\left(\varLambda_c^- \rightarrow \varLambda \mu^{-} \bar \nu_{\mu}\right)}{d q^2 d \cos \theta} d \cos \theta-\int_{-1}^0 \frac{d^2 \varGamma\left(\varLambda_c^- \rightarrow \varLambda \mu^{-} \bar \nu_{\mu}\right)}{d q^2 d \cos \theta} d \cos \theta}{\int_0^1 \frac{d^2 \varGamma\left(\varLambda_c^- \rightarrow \varLambda \mu^{-} \bar \nu_{\mu}\right)}{d q^2 d \cos \theta} d \cos \theta+\int_{-1}^0 \frac{d^2 \varGamma\left(\varLambda_c^- \rightarrow \varLambda \mu^{-} \bar \nu_{\mu}\right)}{d q^2 d \cos \theta} d \cos \theta}
\end{equation}	

We also consider the longitudinal polarization asymmetry for the $\Lambda$ and muon, which are defined as

\begin{equation}
\mathcal{P}_L^{\varLambda}\left(q^2\right)=\frac{\mathrm{d} \varGamma^{\lambda_{\Lambda}=\frac{1}{2}} / \mathrm{d} q^2-\mathrm{d} \varGamma^{\lambda_{\Lambda} =-\frac{1}{2}} / \mathrm{d} q^2}{\mathrm{~d} \varGamma^{\lambda_{\Lambda} =\frac{1}{2}} / \mathrm{d} q^2+\mathrm{d} \varGamma^{\lambda_{\Lambda} =-\frac{1}{2}} / \mathrm{d} q^2}
\end{equation}

\begin{equation}
\mathcal{P}_L^{\mu}\left(q^2\right)=\frac{\mathrm{d} \varGamma^{\lambda_{\mu}=\frac{1}{2}} / \mathrm{d} q^2-\mathrm{d} \varGamma^{\lambda_{\mu}=-\frac{1}{2}} / \mathrm{d} q^2}{\mathrm{~d} \varGamma^{\lambda_{\mu}=\frac{1}{2}} / \mathrm{d} q^2+\mathrm{d} \varGamma^{\lambda_{\mu}=-\frac{1}{2}} / \mathrm{d} q^2}
\end{equation}

These observables are of particular interest because experimental measurements already exist~\cite{BESIII:2023jxv}. In particular, the branching fraction, forward–backward asymmetry ($\cal{A}_{\mathrm{FB}}$), and the $\Lambda$ polarization asymmetry ($\mathcal{P}_L^{\Lambda}$) have been measured by the BESIII collaboration with precisions of approximately $7\%$, $22\%$, and $10\%$, respectively.
%%%%%%%%%%%%%%%%%%%%%%%%%%%%%%%%%%%%%%%%%%%%%%%%%%%%%%%%%%%%%%%%%%%%%%%%%%%%%
\subsection{$\Lambda_c^- \to \Lambda (p \pi) \mu^- \nu_{\mu}$ Decay} \label{4_body_obs}
%%%%%%%%%%%%%%%%%%%%%%%%%%%%%%%%%%%%%%%%%%%%%%%%%%%%%%%%%%%%%%%%%%%%%%%%%%%%%%%%%
The four-fold distribution for the four-body $\Lambda_c^- \to \Lambda (\to p \pi) \mu^- \nu_{\mu}$ decay is given as~\cite{kutsckhe1996angular, Richman:1984gh}
\begin{align}
	I(\Omega_2, \Omega_3) &= \sum_{\lambda_i, \lambda_j^\prime, s_3, s_3^\prime} \left\{
	2\pi\, \delta_{(\lambda_2 - \lambda_3), (\lambda_2^\prime - \lambda_3^\prime)}\, (-1)^{s_3 + s_3^\prime} \, H^*_{\lambda_2 \lambda_3} H_{\lambda_2^\prime \lambda_3^\prime} \,
	D^{s_2}_{\lambda_2, \lambda_4 - \lambda_5}(\Omega_2) \, D^{s_2*}_{\lambda_2^\prime, \lambda_4 - \lambda_5}(\Omega_2) \nonumber \right. \\
	&\left. \times B^*_{\lambda_4 \lambda_5} B_{\lambda_4 \lambda_5} \,
	D^{s_3}_{\lambda_3, \lambda_6 - \lambda_7}(\Omega_3) \, D^{s_3^\prime*}_{\lambda_3^\prime, \lambda_6 - \lambda_7}(\Omega_3) \,
	L^*_{\lambda_6 \lambda_7} L_{\lambda_6 \lambda_7} \right\}\,.
\end{align}
Here $B_{\lambda_4 \lambda_5}$ is the helicity amplitude for $\Lambda \to p \pi$ decay. $H_{\lambda_2\lambda_3}$ and $L_{\lambda_6\lambda_7}$ are the same hadronic and leptonic helicity amplitudes defined earlier in Sec.~\ref {3_body_obs}.

The angular variables relevant to our analysis are defined as follows. In the rest frame of the parent baryon $\Lambda_c$, the angle $\theta_1$ denotes the angle between the momentum of the outgoing $\Lambda$ and the spin quantization axis $\mathbf{z_1}$ of the $\Lambda_c$. The direction of the $\Lambda_c$ momentum defines the $\mathbf{z_2}$ axis, while the opposite direction—corresponding to the momentum of the virtual $W^*$—defines the $\mathbf{z_3}$ axis. The $\mathbf{x_1}$, $\mathbf{x_2}$, and $\mathbf{x_3}$ axes are chosen arbitrarily within the planes orthogonal to their respective $\mathbf{z}$ axes, with the orientation fixed by the condition $\mathbf{x_2} = -\mathbf{x_3}$. The $\mathbf{y_1}$, $\mathbf{y_2}$, and $\mathbf{y_3}$ axes are then determined according to the right-handed coordinate system convention.

In the rest frame of the $\Lambda$, the momentum of the daughter baryon $p$ is specified by the polar and azimuthal angles $(\theta_2, \phi_2)$ with respect to the $\mathbf{z_2}$ axis. Similarly, in the rest frame of the virtual boson $W^{*}$, the direction of the charged lepton (here, the $\mu$) is described by the angles $(\theta_3, \phi_3)$ measured with respect to the $\mathbf{z_3}$ axis. The explicit expressions of the momenta of particles in the respective rest frames are given in Appendix \ref{app-kin}.

\begin{figure}[h!]
    \centering
    \includegraphics[width=1.0\linewidth]{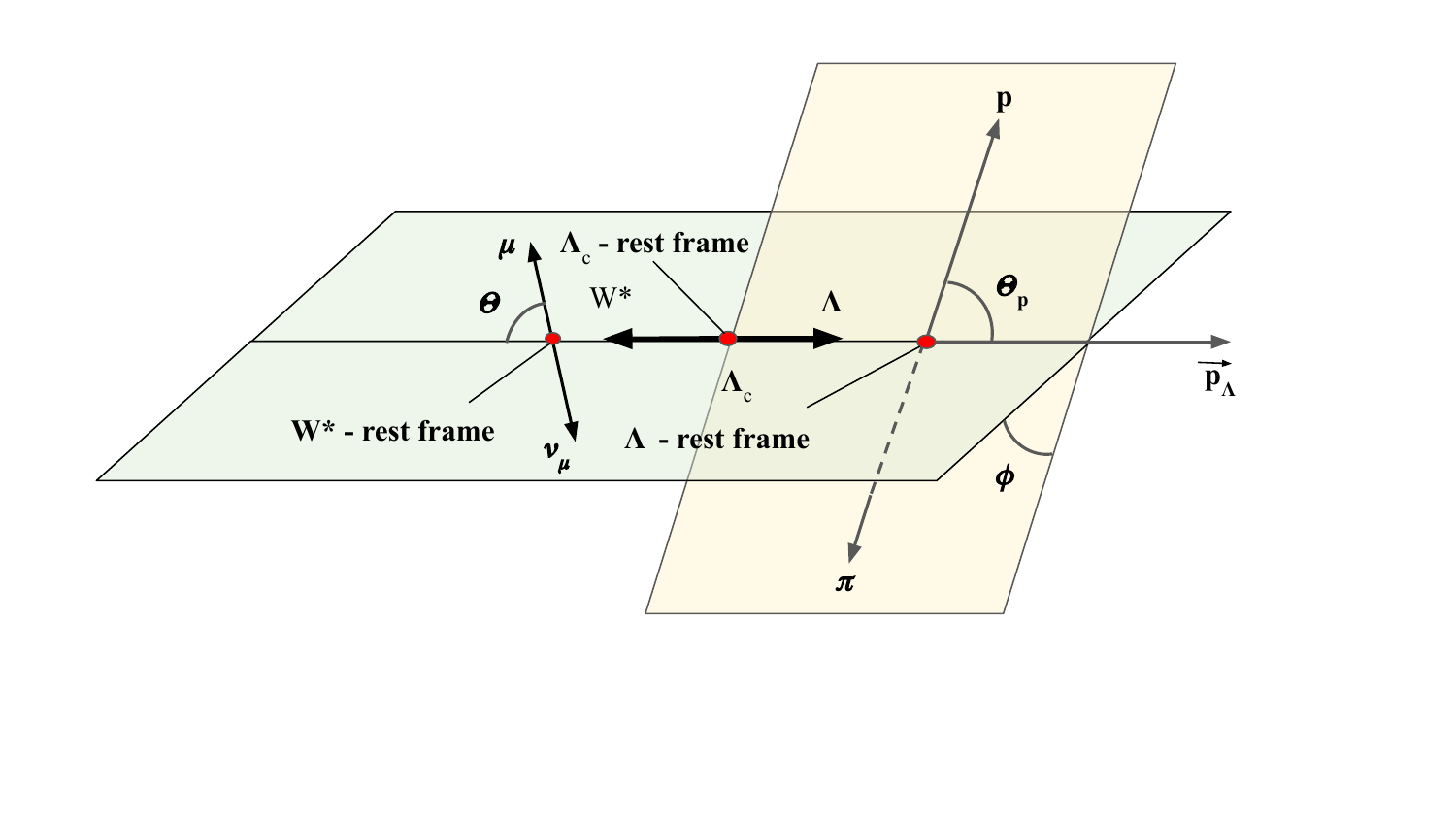}
    \caption{\small Kinematics of $\Lambda_c^- \to \Lambda \, (\to p \pi) \mu^- \bar\nu_{\mu}$ decay.}
    \label{fig:kinematics}
\end{figure}

The full decay distribution can be expressed in terms of three physical angles: $\theta_2$, $\theta_3$, and $\phi = \phi_2 + \phi_3$. From this point onward, we refer to $\theta_2$ as $\theta_p$ (the angle in the hadronic system) and $\theta_3$ as $\theta$ (the angle in the leptonic system).
After summing over the helicities of all particles and integrating over the unobserved angular variables, the normalized angular distribution can be expressed as~\cite{Karmakar:2023rdt}
\begin{eqnarray}
\label{4fold}
\begin{aligned}
& \frac{1}{\left(d \Gamma / d q^2\right)} \frac{d \Gamma}{d q^2 d \cos \theta_p d \cos \theta d \phi} =
\mathcal{M}_0^{\nu_L (\nu_R)} + \mathcal{M}_1^{\nu_L (\nu_R)} \cos \theta_p + \mathcal{M}_2^{\nu_L (\nu_R)} \cos \theta \\ &  + \mathcal{M}_3^{\nu_L (\nu_R)} \cos \theta_p \cos \theta  + \mathcal{M}_4^{\nu_L (\nu_R)} \cos ^2 \theta + \mathcal{M}_5^{\nu_L (\nu_R)} \cos \theta_p \cos ^2 \theta \\ &  + \mathcal{M}_6^{\nu_L (\nu_R)} \sin \theta_p \sin \theta \cos \phi + \mathcal{M}_7^{\nu_L (\nu_R)} \sin \theta_p \sin \theta \sin \phi \\ & + \mathcal{M}_8^{\nu_L (\nu_R)} \sin \theta_p \sin \theta \cos \theta \cos \phi + \mathcal{M}_9^{\nu_L (\nu_R)} \sin \theta_p \sin \theta \cos \theta \sin \phi
\end{aligned}
\end{eqnarray}

The coefficients $\mathcal{M}_0^{\nu_L (\nu_R)}$ to $\mathcal{M}_9^{\nu_L (\nu_R)}$ denote angular observables including the left-handed and right-handed neutrinos and their explicit expression in terms of the hadronic helicity is given in Appendix \ref{app-C}. Each of these coefficients is considered as observable and can be extracted from a fit to the observed angular distribution or by the method of angular moments~\cite{Dighe:1998vk} with proper weighting functions~\cite{Karmakar:2023rdt}.

The four-body angular distribution was also discussed in the corresponding experimental analysis, where particular attention was given to the observable $T_p$. This quantity is a time-reversal–odd observable that appears in the angular distribution as the coefficient ${\cal M}_7$.

%%%%%%%%%%%%%%%%%%%%%%%%%%%%%%%%%%%%%%%%%%%%%%%%%%%%%%%%%%%%%%%%%%%%%%%%
\section{Results} \label{result}
%%%%%%%%%%%%%%%%%%%%%%%%%%%%%%%%%%%%%%%%%%%%%%%%%%%%%%%%%%%%%%%%%%%%%%%

In this section, we study the impact of the allowed NP parameter space in the $ c \to s \mu \nu_\mu$ channel on the angular distribution of the decay $\Lambda_c^- \to \Lambda(\to p \pi)\,\mu^- \bar\nu_\mu$. The discussion of observables is organized as follows. We begin with the case where only left-handed neutrinos are present. Within this scenario, we first examine observables for the decay $\Lambda_c^- \to \Lambda \mu^- \bar\nu_\mu$, which involves three final-state particles and depends on a single angle $\theta$. We then consider the subsequent decay $\Lambda \to p \pi$, leading to a four-body final state with an angular distribution expressed in terms of three angles, as given in eq. \ref{4fold}. Finally, we extend the discussion to the case where both left- and right-handed neutrinos are present. 

\begin{table}[h!] 
    \centering
    \begin{tabular}{|c|c|}
    \hline
    Parameters & Value \\
    \hline
     $v$  & 246 GeV \\ 
     \hline
   $\Lambda$  & 1 TeV \\ 
  \hline
    $G_F$ & 1.166 $\times$ $10^{-5}$ GeV$^{-2}$ \\
     \hline
    $V_{cs}$ & 0.975 $\pm$ 0.006  \\  
     \hline
      $\alpha_P$ & 0.747 $\pm$ 0.009  \\ 
     \hline
     $m_{\mu}$ & 105.658 MeV  \\ 
     \hline
      $m_{\Lambda_c}$ & 2286.46 $\pm$ 0.14 MeV  \\ 
     \hline
      $m_{\Lambda}$ & 1115.683 $\pm$ 0.006 MeV  \\ 
     \hline
      $\tau_{\Lambda_c}$ & (202.6 $\pm$ 1.0) $\times$ $10^{-15}$ Sec  \\ 
     \hline
    \end{tabular}
    \caption{Input parameters used in our analysis \cite{PhysRevD.110.030001}.}
    \label{tab:parameters}
\end{table}

\begin{comment}
\begin{table}[h!]
\centering
\begin{tabular}{|c|cccccc|}
\hline
$q^2$ (GeV$^2$) & $f_\perp$ & $f_{+}$ & $f_{0}$ & $g_\perp$ & $g_{+}$ & $g_{0}$ \\
\hline
0.1 & 1.14 & 0.67 & 0.66 & 0.59 & 0.59 & 0.59 \\
0.4 & 1.26 & 0.75 & 0.72 & 0.64 & 0.63 & 0.66 \\
1.2 & 1.74 & 1.07 & 0.96 & 0.83 & 0.82 & 0.99 \\
\hline
\end{tabular}
\caption{Central values of the $\Lambda_c \to \Lambda$ form factors at specific $q^2$ values.}
\label{table:ff_central_val}
\end{table}

\begin{table}[H]
\centering
\textcolor{blue}{\begin{tabular}{|c|c|c|c|c|c|c|}
\hline
$q^2$ (GeV$^2$) & $f_\perp$ & $f_{+}$ & $f_{0}$ & $g_\perp$ & $g_{+}$ & $g_{0}$ \\
\hline
0.1 & 1.14 $\pm$ 0.05 & 0.67 $\pm$ 0.02 & 0.66 $\pm$ 0.02 & 0.59 $\pm$ 0.02 & 0.59 $\pm$ 0.01 & 0.59 $\pm$ 0.01 \\
0.4 & 1.26 $\pm$ 0.04 & 0.75 $\pm$ 0.02 & 0.73 $\pm$ 0.02 & 0.64 $\pm$ 0.01 & 0.63 $\pm$ 0.01 & 0.67 $\pm$ 0.01 \\
1.2 & 1.73 $\pm$ 0.07 & 1.07 $\pm$ 0.03 & 0.97 $\pm$ 0.03 & 0.82 $\pm$ 0.03 & 0.81 $\pm$ 0.02 & 1.00 $\pm$ 0.04 \\
\hline
\end{tabular} }
\caption{Values of the $\Lambda_c \to \Lambda$ form factors at specific $q^2$ values.}
\label{table:ff_central_val}
\end{table}
\end{comment}

\begin{table}[H]
\centering
\begin{tabular}{|c|c|c|c|c|c|c|}
\hline
$q^2$ (GeV$^2$) & $f_\perp$ & $f_{+}$ & $f_{0}$ & $g_\perp$ & $g_{+}$ & $g_{0}$ \\
\hline
0.1 & 1.14 $\pm$ 0.05 & 0.67 $\pm$ 0.02 & 0.66 $\pm$ 0.02 & 0.59 $\pm$ 0.02 & 0.59 $\pm$ 0.01 & 0.59 $\pm$ 0.01 \\
0.4 & 1.26 $\pm$ 0.04 & 0.75 $\pm$ 0.02 & 0.73 $\pm$ 0.02 & 0.64 $\pm$ 0.01 & 0.63 $\pm$ 0.01 & 0.67 $\pm$ 0.01 \\
1.2 & 1.73 $\pm$ 0.07 & 1.07 $\pm$ 0.03 & 0.97 $\pm$ 0.03 & 0.82 $\pm$ 0.03 & 0.81 $\pm$ 0.02 & 1.00 $\pm$ 0.04 \\
\hline
\end{tabular} 
\caption{Values of the $\Lambda_c \to \Lambda$ form factors at specific $q^2$ values.}
\label{table:ff_central_val}
\end{table}

The calculation of the observables is done following the expressions provided in Sec. \ref{observables}.  The input parameters used in this analysis are presented in Table \ref{tab:parameters}. We also provide in Table~\ref{table:ff_central_val} the form factor numerical values at three $q^2$ values $0.1$, $0.4$, and $1.2$ GeV$^2$, which are later used to present the NP WC dependency of different observables.

%%%%%%%%%%%%%%%%%%%%%%%%%%%%%%%%%%%%%%%%%%%%%%%%%%%%%%%%%%%%%%%%%%
\subsection{Predictions of observables for Left-Handed Neutrinos} \label{pre_lhn}
%%%%%%%%%%%%%%%%%%%%%%%%%%%%%%%%%%%%%%%%%%%%%%%%%%%%%%%%%%%%%%%%%%%

In this section, we focus on the scenario with only left-handed neutrinos. The NP effects are studied by switching on one operator at a time. In particular, we consider the operators $O^V_{LL}$, $O^V_{RL}$, $O^S_{LL}$, and $O^S_{RL}$. Our objective is to examine the sensitivity of the angular distributions in $\Lambda_{c}^- \to \Lambda(\to p \pi)\,\mu^- \bar \nu_{\mu}$ to each of these operators. For this purpose, we select benchmark points for the corresponding Wilson coefficients within their $1\sigma$ allowed ranges. These benchmark points are chosen to highlight the possible deviations from SM expectations that remain consistent with current constraints in the $c \to s \mu \nu_{\mu}$ channel. The chosen benchmark values are summarized in Table~\ref{tab:BPs}.

\begin{table}[h!] 
    \centering
    \begin{tabular}{|c|c|}
    \hline
 Wilson Coefficients & Benchmark Points\\
    \hline
     $C^V_{LL}$  &  -0.01 - $i$ 0.20 \\ 
     \hline
  $C^V_{RL}$  & -0.02 - $i$ 0.20  \\ 
  \hline
    $C^S_{LL}$ & 0.07 - $i$ 0.03  \\
     \hline
   $C^S_{RL}$ &  -0.0003 - $i$ 0.001   \\  
     \hline
     $C^V_{LR}$ & 0.16 - $i$ 0.16 \\ 
     \hline
     $C^V_{RR}$ &  0.16 - $i$ 0.16  \\ 
     \hline
     $C^S_{LR}$ &  0.01 - $i$ 0.01  \\ 
     \hline
     $C^S_{RR}$ &  0.01 + $i$ 0.01  \\ 
     \hline
    \end{tabular}
    \caption{Benchmark points for the complex WCs.}
    \label{tab:BPs}
\end{table}

%%%%%%%%%%%%%%%%%%%%%%%%%%%%%%%%%%%%%%%%%%%%%%%%%%%%%%%%%%%%%%%%%%%%%%%%%
\subsubsection{$\Lambda_c^- \to \Lambda \mu^- \nu_{\mu}$ Decay}
%%%%%%%%%%%%%%%%%%%%%%%%%%%%%%%%%%%%%%%%%%%%%%%%%%%%%%%%%%%%%%%%%%%%%%%%%

\begin{figure}[h!]
	\centering
    \includegraphics[width=0.49\textwidth]{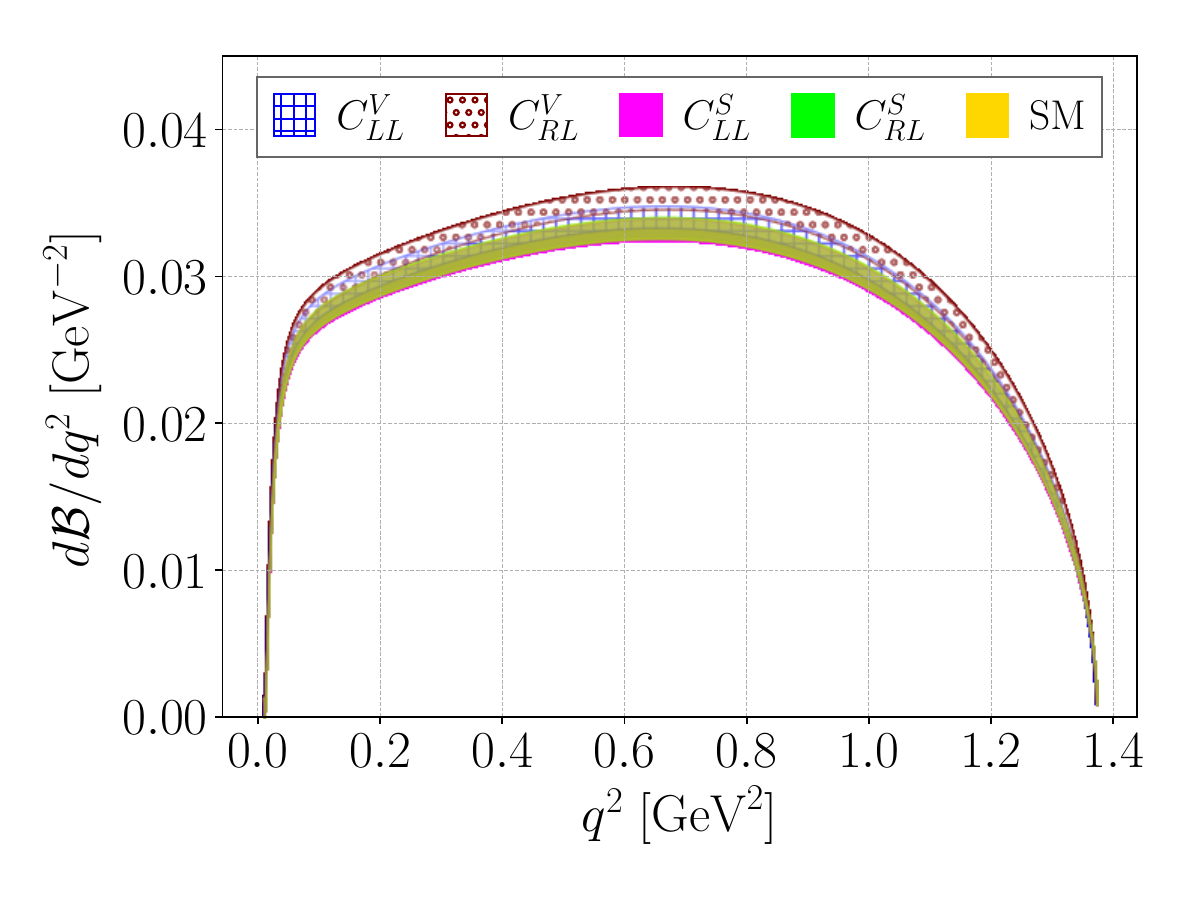}
    \includegraphics[width=0.49\textwidth]{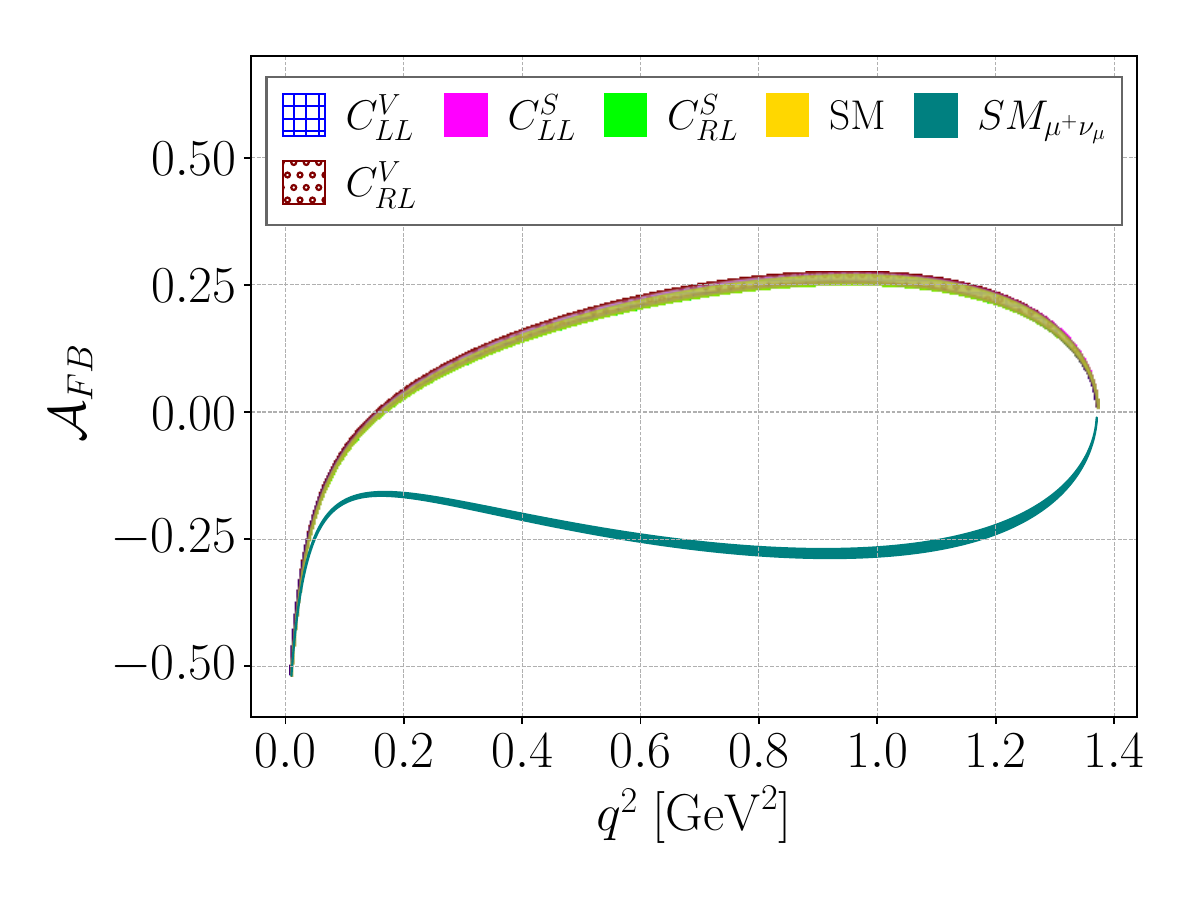}
    \includegraphics[width=0.49\textwidth]{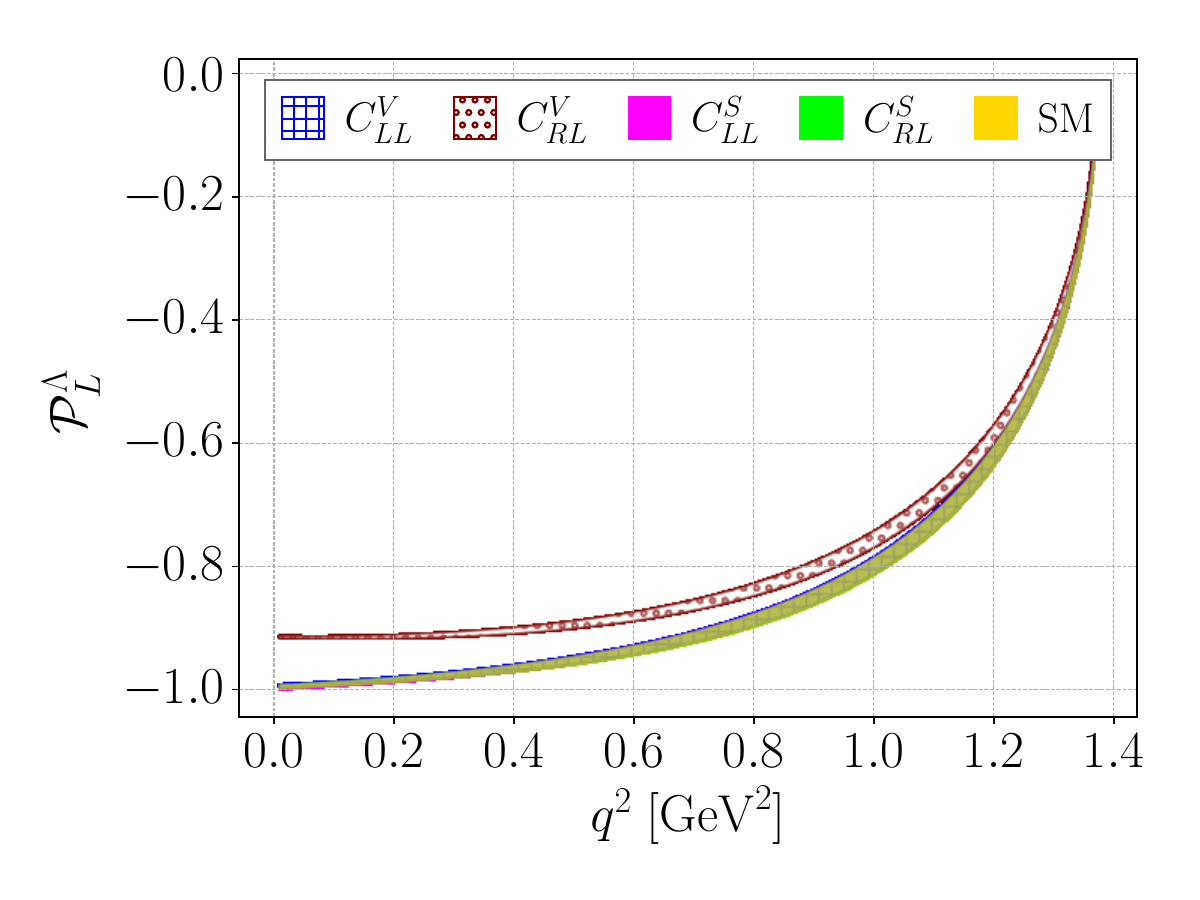}
    \includegraphics[width=0.49\textwidth]{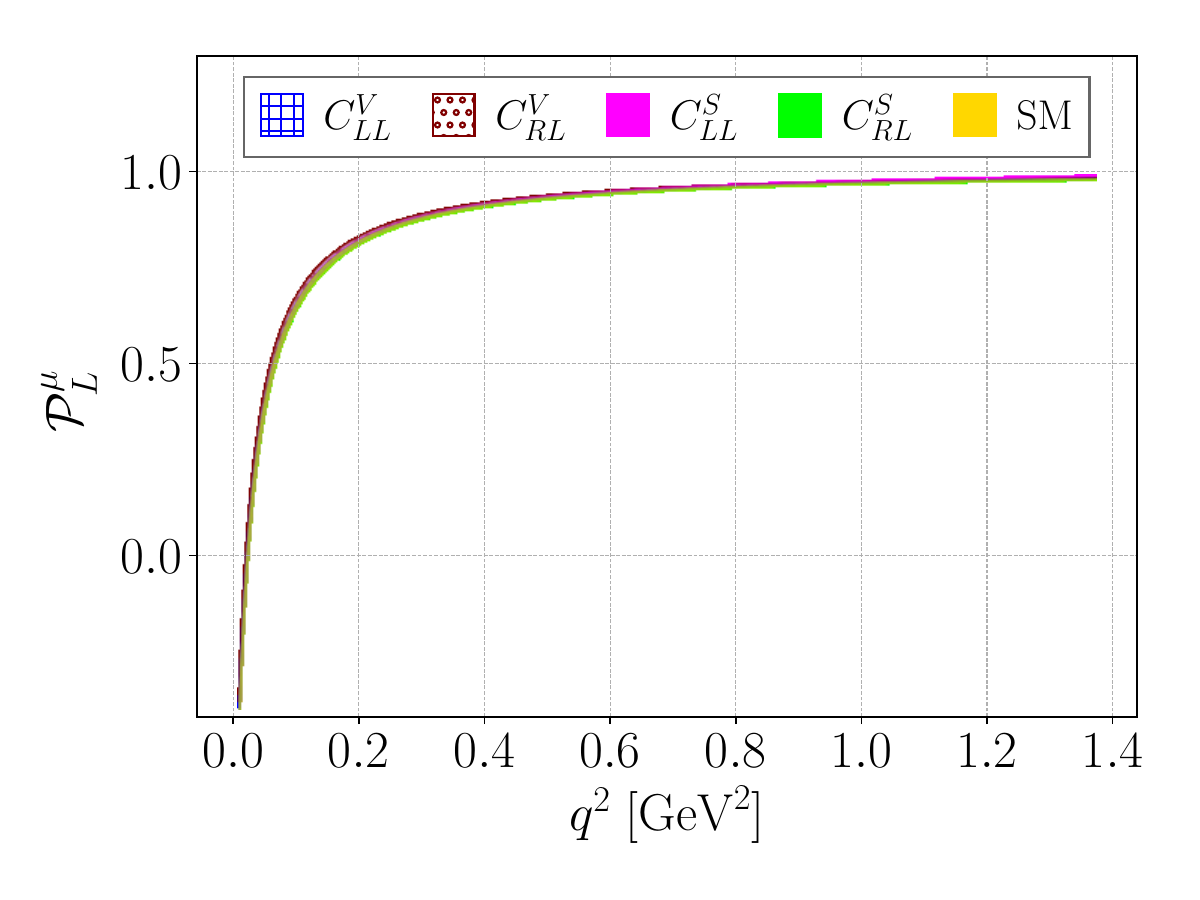}
	\caption{LHN: $q^2$ spectra for the differential branching fraction (top left), forward-backward asymmetry (top right), $\Lambda$ polarization asymmetry (bottom left), and muon polarization asymmetry (bottom right) for three-body decay.}
	\label{fig-3body_lhn}
\end{figure}

In Fig~\ref{fig-3body_lhn}, we present the differential branching fraction, forward-backward asymmetry ($\mathcal{A}_{FB}$), $\Lambda$ polarization asymmetry ($\mathcal{P}^\Lambda_L$) and muon polarization asymmetry ($\mathcal{P}^\mu_L$) in $\Lambda_c^- \to \Lambda \mu^- \bar \nu_\mu$. The bands for each scenario correspond to $1 \sigma$ uncertainties coming from the form factors and other input parameters. For the differential branching fraction, $\mathcal{A}_{FB}$ and $\mathcal{P}^\mu_L$, the NP scenarios almost overlap with the SM expectations. Note that we have presented $\mathcal{A}_{FB}$ plot for both the modes $\Lambda_c^- \to  \Lambda \mu^- \bar \nu_\mu$ and its charge conjugated mode, $\Lambda_c^+ \to \Lambda \mu^+ \nu_\mu$, both of which can be compared with results in earlier literature \cite{Mu:2019bin} and \cite{Zhang:2025tki}, respectively.

We find significant deviations of $\mathcal{P}^\Lambda_L$ from its SM expectation for the $C^V_{RL}$ scenario, particularly in the lower $q^2$ regions. This separation between $C^V_{RL}$ and the SM prediction can be understood by looking at the expression of $\mathcal{P}^\Lambda_L$ in terms of the NP parameters at $q^2 = 0.1$ GeV$^{2}$: 

\begin{equation}
\label{ex_P_lambda}
    \begin{aligned}
       \frac{\mathrm{d} \varGamma^{\lambda_{\Lambda}=\frac{1}{2}}} {\mathrm{d} q^2} - \frac{\mathrm{d} \varGamma^{\lambda_{\Lambda} =-\frac{1}{2}}} {\mathrm{d} q^2} \;=\;&
-838.15\,\bigl|1 + C^V_{LL}\bigr|^2
+ 722.03\,\bigl|C^V_{RL}\bigr|^2
+ 58.71\,\bigl|C^S_{LL}\bigr|^2 \\&
- 58.71\,\bigl|C^S_{RL}\bigr|^2
+ 116.11\,\Re\!\bigl[(1+C^V_{LL})^{\ast}\,C^V_{RL}\bigr]
\\& + 11.58\,\Re\!\bigl[(1+C^V_{LL})^{\ast}\,C^S_{LL}\bigr]
+ 158.08\,\Re\!\bigl[(1+C^V_{LL})^{\ast}\,C^S_{RL}\bigr]
\\&- 158.08\,\Re\!\bigl[C^{V \ast}_{RL}\,C^S_{LL}\bigr]
- 11.58\,\Re\!\bigl[C^{V \ast}_{RL}\,C^S_{RL}\bigr] 
    \end{aligned}
\end{equation}

\begin{equation}
\label{ex_P_lambda2}
\begin{aligned}
 \frac{\mathrm{d} \varGamma^{\lambda_{\Lambda}=\frac{1}{2}}} {\mathrm{d} q^2}  + \frac{\mathrm{d} \varGamma^{\lambda_{\Lambda} =-\frac{1}{2}}} {\mathrm{d} q^2}
 \;=\;&
848.04\,\bigl|1 + C^V_{LL}\bigr|^2
+ 848.04\,\bigl|C^V_{RL}\bigr|^2
+ 61.25\,\bigl|C^S_{LL}\bigr|^2 \\ &
+ 61.25\,\bigl|C^S_{RL}\bigr|^2
+ 30.82\,\Re\!\bigl[(1+C^V_{LL})^{\ast}\,C^V_{RL}\bigr]
 \\ & - 34.87\,\Re\!\bigl[(1+C^V_{LL})^{\ast}\,C^S_{LL}\bigr]
- 161.46\,\Re\!\bigl[(1+C^V_{LL})^{\ast}\,C^S_{RL}\bigr]
\\ & - 161.46\,\Re\!\bigl[C^{V\ast}_{RL}\,C^S_{LL}\bigr]
- 34.87\,\Re\!\bigl[C^{V\ast}_{RL}\,C^S_{RL}\bigr] \\ &
+ 34.88\,\Re\!\bigl[C^{S\ast}_{LL}\,C^S_{RL}\bigr] \,
\end{aligned}
\end{equation}

From eqs. \ref{ex_P_lambda}, \ref{ex_P_lambda2}, we note that the observable $\mathcal{P}^\Lambda_L$ is most sensitive to the NP WC $C^V_{RL}$. Moreover, the $1 \sigma$ allowed parameter range for $C^V_{RL}$ is significantly larger than $C^{S}_{LL}$ and $C^{S}_{RL}$ -- thus the observed deviation in $\mathcal{P}^\Lambda_L$ is more prominent for  $C^V_{RL}$. Note that  $\frac{\mathrm{d} \varGamma^{\lambda_{\Lambda}=\frac{1}{2}}} {\mathrm{d} q^2} - \frac{\mathrm{d} \varGamma^{\lambda_{\Lambda} =-\frac{1}{2}}} {\mathrm{d} q^2}$ is also sensitive to large allowed $C^V_{LL}$ values. However, in the definition of $\mathcal{P}^\Lambda_L$, the numerator and the denominator cancel out the $C^V_{LL}$ dependency.

The strong sensitivity of $\mathcal{P}^\Lambda_L$ to $C^V_{RL}$ can also be
understood from the helicity structure of the decay. The longitudinal
polarization of the $\Lambda$ is determined by the imbalance between the
$\lambda_\Lambda=+1/2$ and $\lambda_\Lambda=-1/2$ helicity rates, which in the SM is driven by the purely left-handed $V\!-\!A$ current. Introducing the right-handed quark current through $C^V_{RL}$ flips the relative sign between
the vector and axial-vector helicity amplitudes, thereby modifying this
imbalance. Since these interference terms enter the numerator of
$\mathcal{P}^\Lambda_L$ with opposite signs compared to the SM contribution,
even moderate values of $C^V_{RL}$ lead to a sizeable shift. This explains why
the deviation is already visible at low $q^2$, where the helicity asymmetry is
kinematically most pronounced, while the denominator remains dominated by the
large SM-like contributions. In contrast, the scalar operators contribute
primarily to timelike amplitudes, which do not significantly alter the
longitudinal helicity balance of the $\Lambda$, making their impact on
$\mathcal{P}^\Lambda_L$ much less visible.

The same deviation is not observed for $\mathcal{P}^\mu_L$, which can be seen
from its dependence on the NP Wilson coefficients in
eq.\,(\ref{ex_P_mu}) at $q^2=0.1~\text{GeV}^2$.
\begin{equation}
\label{ex_P_mu}
\begin{aligned}
 \frac{\mathrm{d} \varGamma^{\lambda_{\mu}=\frac{1}{2}}} {\mathrm{d} q^2} - \frac{\mathrm{d} \varGamma^{\lambda_{\mu} =-\frac{1}{2}}} {\mathrm{d} q^2} =\;&
557.02\,\bigl|1+C^V_{LL}\bigr|^2
+ 557.02\,\bigl|C^V_{RL}\bigr|^2
- 61.59\,\bigl|C^S_{LL}\bigr|^2 \\&
- 61.59\,\bigl|C^S_{RL}\bigr|^2
- 31.26\,\Re\!\bigl[(1+C^V_{LL})^{\ast}\,C^V_{RL}\bigr] \\&
+ 35.00\,\Re\!\bigl[(1+C^V_{LL})^{\ast}\,C^S_{LL}\bigr]
+ 162.40\,\Re\!\bigl[(1+C^V_{LL})^{\ast}\,C^S_{RL}\bigr] \\&
+ 162.40\,\Re\!\bigl[C^{V \ast}_{RL}\,C^S_{LL}\bigr]
+ 35.00\,\Re\!\bigl[C^{V \ast}_{RL}\,C^S_{RL}\bigr] \,.
\end{aligned}
\end{equation}

Experimental measurements of $\cal{A}_{FB}$ and $\mathcal{P}^{\Lambda}_L$ are consistent with the SM within the current experimental uncertainties~\cite{BESIII:2023jxv}. In particular, the deviation in $\mathcal{P}_L^{\Lambda}$ predicted in our analysis for nonzero $C^V_{RL}$ at low $q^2$ is not observed in the presently available $q^2$-differential spectra of $\mathcal{P}_L^{\Lambda}$.  This comparison indicates that the magnitude of the corresponding Wilson coefficient, $|C^V_{RL}|$, is further constrained by the angular observables of the $\Lambda_c \to \Lambda \mu \nu_{\mu}$ decay.

%%%%%%%%%%%%%%%%%%%%%%%%%%%%%%%%%%%%%%%%%%%%%%%%%%%%
\subsubsection{$\Lambda_c^- \to \Lambda \,(p \pi)\, \mu^- \nu_{\mu}$ Decay}
%%%%%%%%%%%%%%%%%%%%%%%%%%%%%%%%%%%%%%%%%%%%%%%%%%%%

\begin{figure}[h!]
	\centering
    \includegraphics[width=0.49\textwidth]{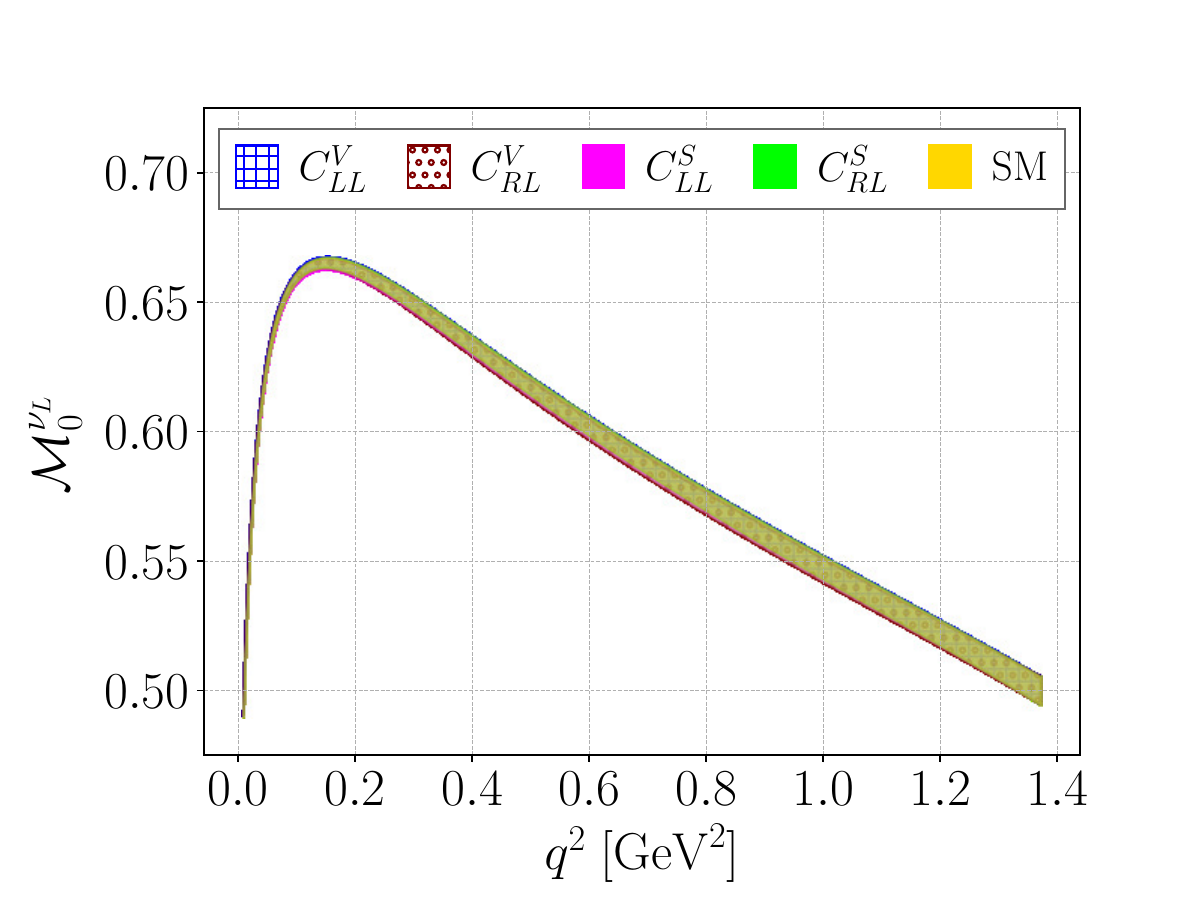}
    \includegraphics[width=0.49\textwidth]{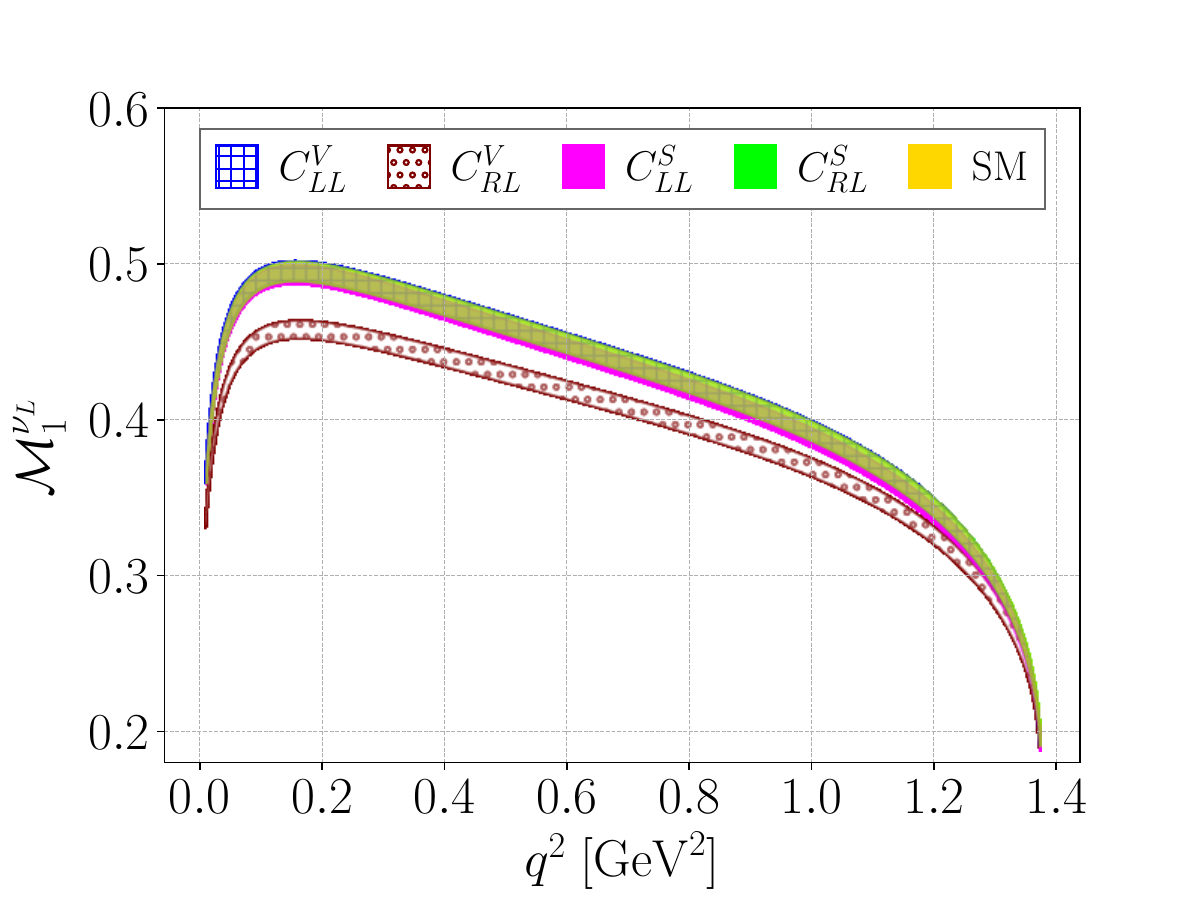}
    \includegraphics[width=0.49\textwidth]{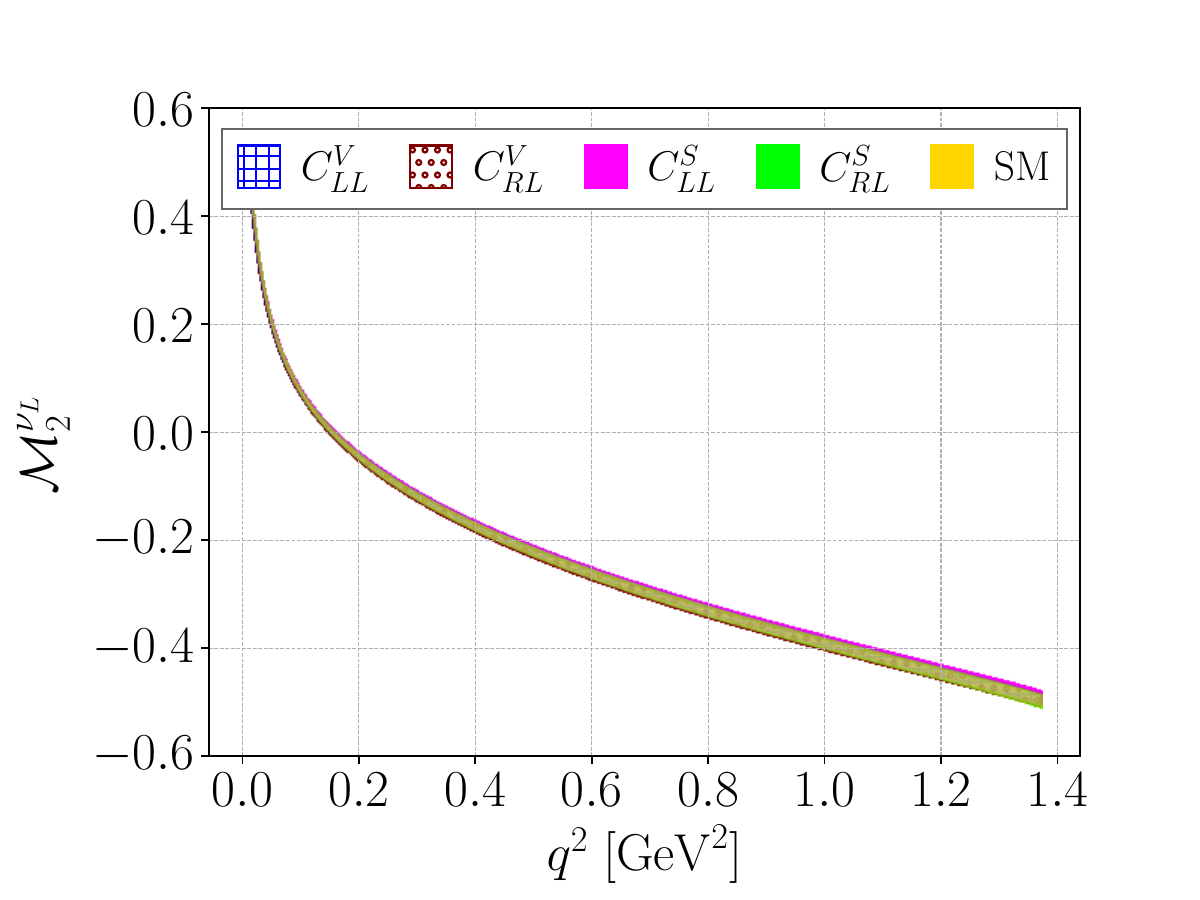}
    \includegraphics[width=0.49\textwidth]{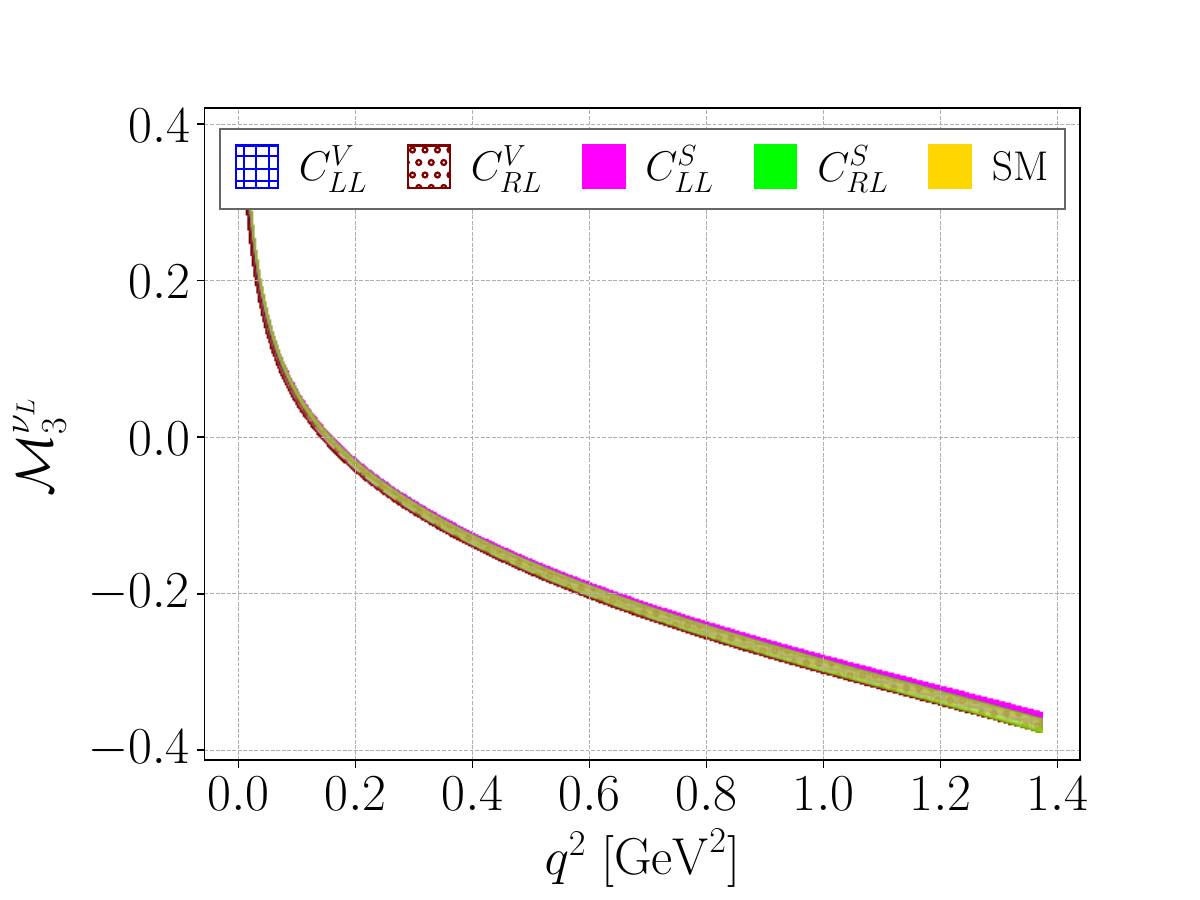}
	\caption{LHN: $q^2$ spectra for the $\mathcal{M}_0^{\nu_L}$ (top left), $\mathcal{M}_1^{\nu_L}$ (top right), $\mathcal{M}_2^{\nu_L}$ (bottom left), and $\mathcal{M}_3^{\nu_L}$ (bottom right) for four-body decay.}
	\label{fig-4body_0to3}
\end{figure}

\begin{figure}[h!]
	\centering
    \includegraphics[width=0.49\textwidth]{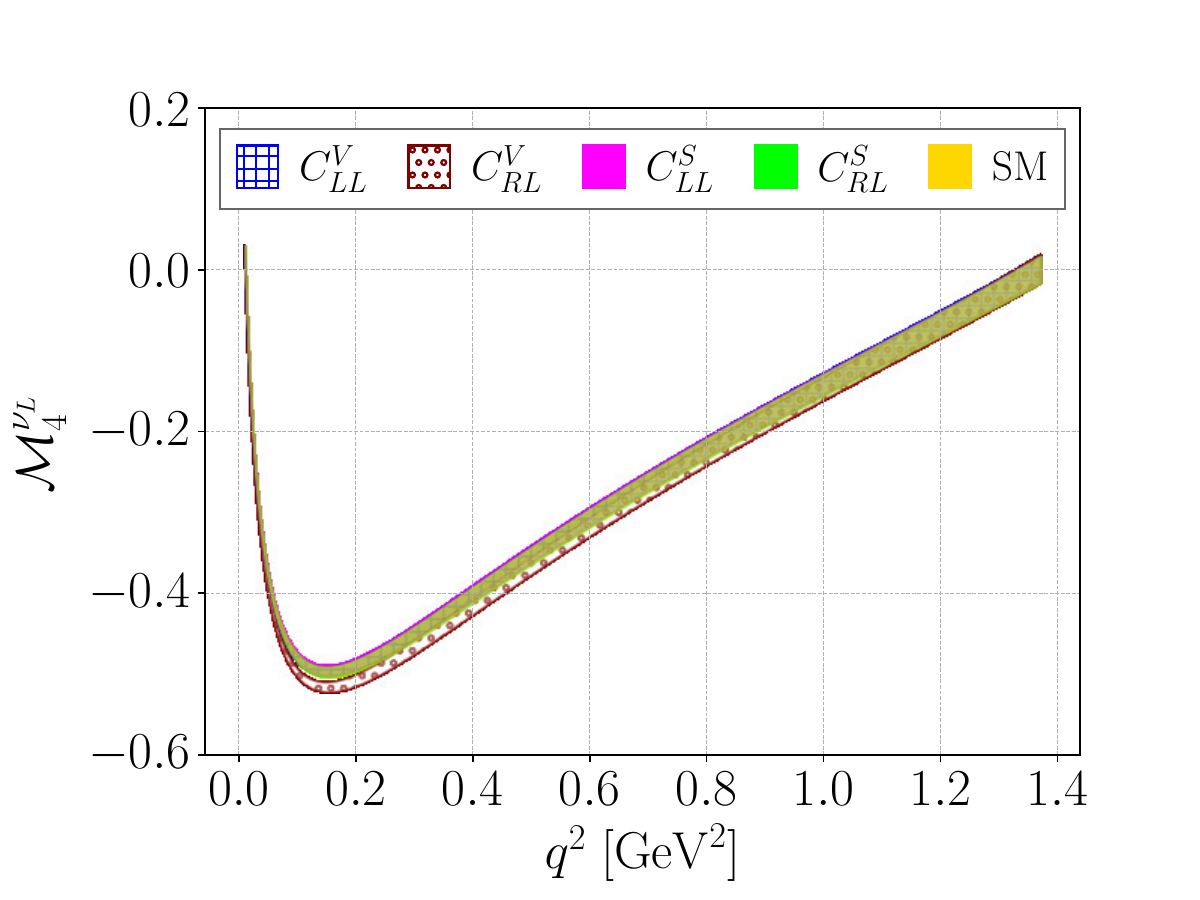}
    \includegraphics[width=0.49\textwidth]{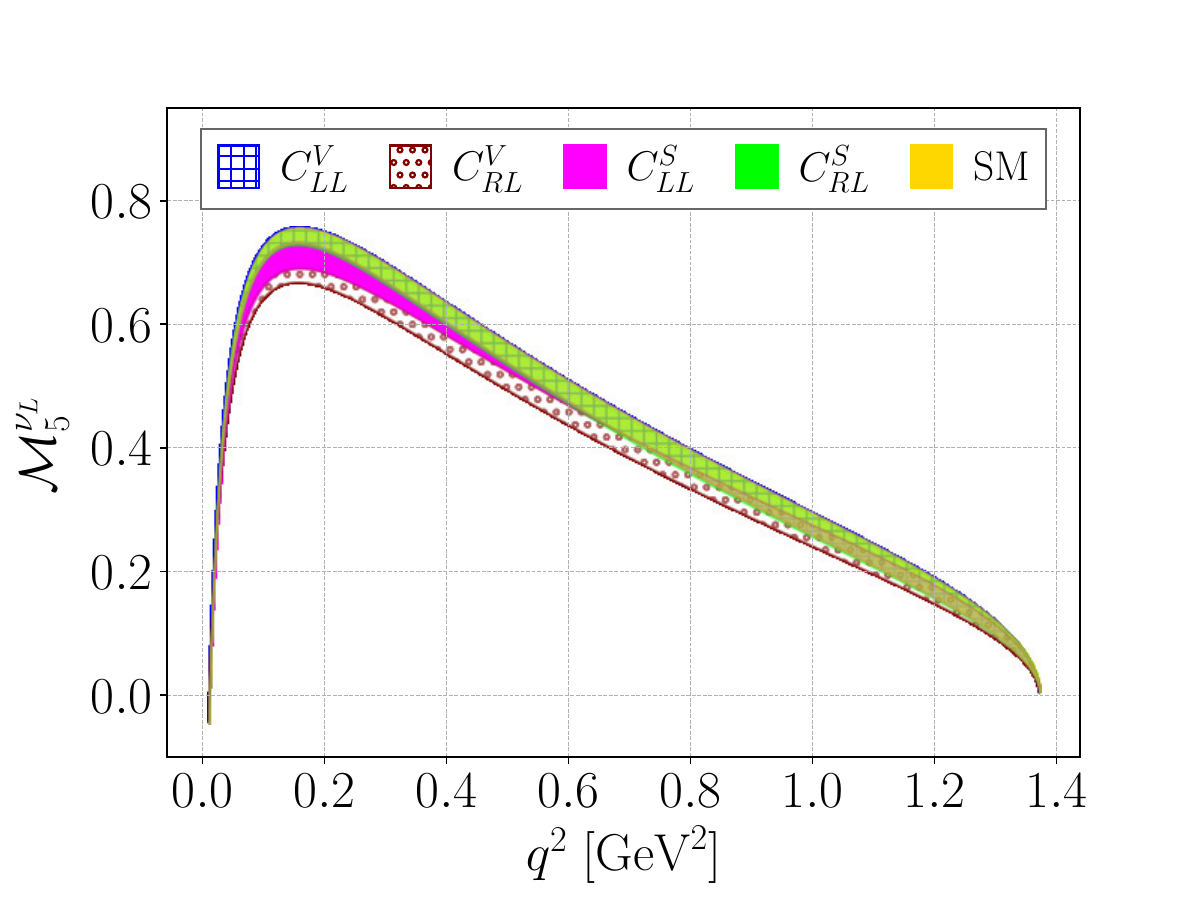}
    \includegraphics[width=0.49\textwidth]{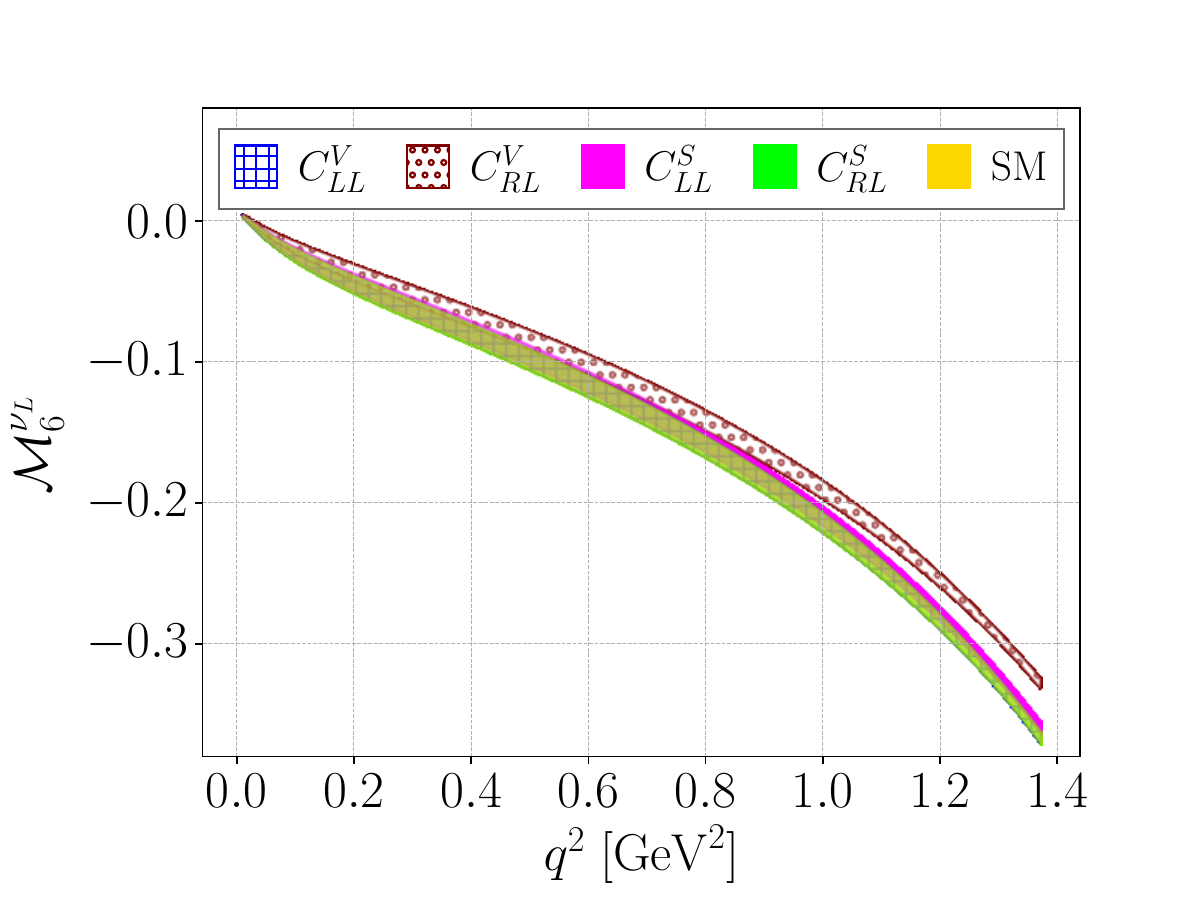}
    \includegraphics[width=0.49\textwidth]{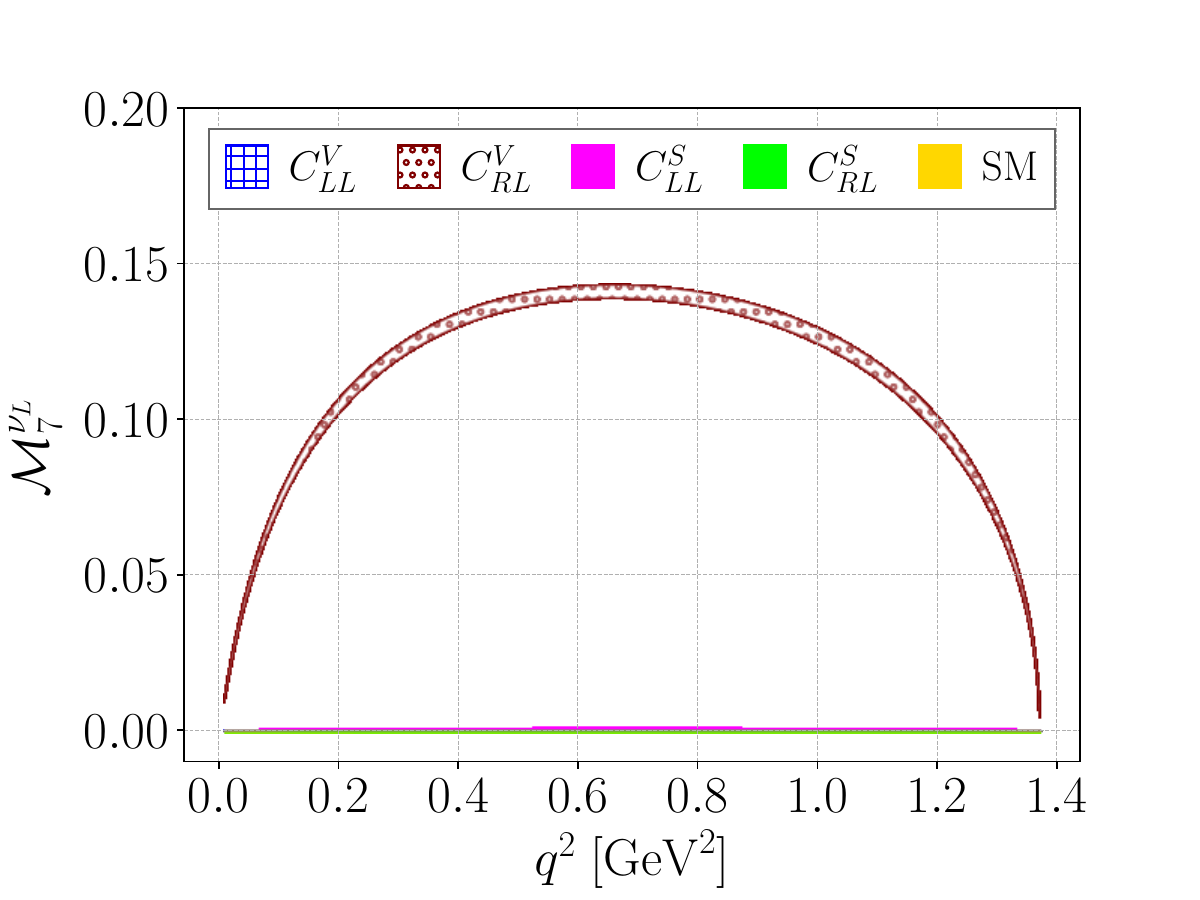}
    \includegraphics[width=0.49\textwidth]{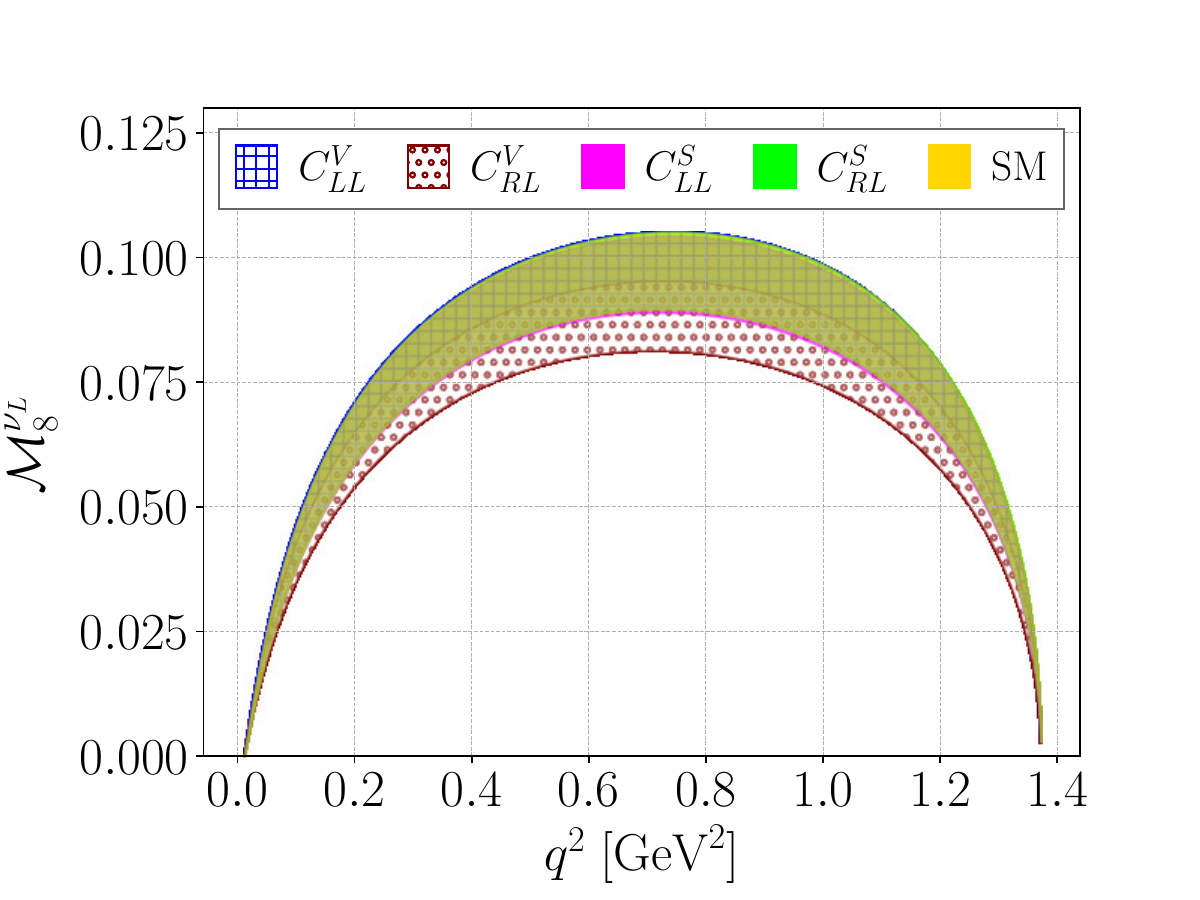}
	\caption{LHN: $q^2$ spectra for the $\mathcal{M}_4^{\nu_L}$ (top left), $\mathcal{M}_5^{\nu_L}$ (top right), $\mathcal{M}_6^{\nu_L}$ (middle left), $\mathcal{M}_7^{\nu_L}$ (middle right) and $\mathcal{M}_8^{\nu_L}$ (bottom) for four-body decay.}
	\label{fig-4body_4to8}
\end{figure}

So far, we have considered three particles in the final state, $\Lambda$, $\mu^-$, and $\bar \nu_\mu$. We now consider $\Lambda \to p \pi$ decay, resulting in 4 particles in the final state: $p$, $\pi$, $\mu$, and $\nu$. As discussed in Sec. \ref{4_body_obs}, this offers more number of angular observables. We present the SM and NP predicted values for the observables ${\cal M}^{\nu_L}_{0}$, ${\cal M}^{\nu_L}_{1}$, ${\cal M}^{\nu_L}_{2}$ and ${\cal M}^{\nu_L}_{3}$ in Fig. \ref{fig-4body_0to3}, and ${\cal M}^{\nu_L}_{4}$, ${\cal M}^{\nu_L}_{5}$, ${\cal M}^{\nu_L}_{6}$, ${\cal M}^{\nu_L}_{7}$ and ${\cal M}^{\nu_L}_{8}$ in Fig. \ref{fig-4body_4to8}. 

Apart from ${\cal M}^{\nu_L}_{1}$, ${\cal M}^{\nu_L}_{6}$, and ${\cal M}^{\nu_L}_{7}$, there are no significant deviations from the SM predictions for any of the NP scenarios. In the following, we try to explain the behavior of each of the observables  ${\cal M}^{\nu_L}_{1}$, ${\cal M}^{\nu_L}_{6}$, and ${\cal M}^{\nu_L}_{7}$.

The observable $\mathcal{M}_1^{\nu_L}$, which corresponds to the forward--backward
asymmetry of the proton in the decay $\Lambda\to p\pi$, is directly related to
the longitudinal polarization fraction of the $\Lambda$ in
$\Lambda_c^-\to\Lambda\,\mu^-\,\bar\nu$. This quantity probes the imbalance between the
$\Lambda$ helicity states are produced in the weak transition and are therefore sensitive to the helicity structure of the underlying quark current.
The numerical expression of ${\cal M}_1^{\nu_L}$ in terms of the WCs for $q^2 =0.1$ GeV$^2$ and $q^2 =0.4$ GeV$^2$ is shown below. 

\begin{equation} \label{M1_0.1}
\begin{aligned}
\mathcal M_1^{\nu_L}(0.1) \;\approx\;&
0.49
\;-\;0.08\,\Re |C_{RL}^V|
\;-\;0.02\,\Re |C_{LL}^S|
\;-\;0.02\,\Re |C_{RL}^S|
\\
&\;-\;1.87\,|C_{RL}^V|^2
\;-\;0.12\,|C_{LL}^S|^2
\;-\;0.01\,|C_{RL}^S|^2
\\
&\;+\;0.08\,\Re[C_{LL}^V\, C_{RL}^{V\ast}]
\;+\;0.02\,\Re[C_{LL}^V\, C_{LL}^{S\ast}]
\;+\;0.02\,\Re[C_{LL}^V\, C_{RL}^{S\ast}]
\\
&\;-\;0.16\,\Re[C_{RL}^V\, C_{LL}^{S\ast}]
\;-\;0.01\,\Re[C_{RL}^V\, C_{RL}^{S\ast}]
\;-\;0.02\,\Re[C_{LL}^S\, C_{RL}^{S\ast}] \;
\end{aligned}
\end{equation}

\begin{equation} \label{M1_0.4}
\begin{aligned}
\mathcal M_1^{\nu_L}(0.4) \;\approx\;&
0.47
\;+\;0.03\,\Re |C_{RL}^V|
\;-\;0.02\,\Re |C_{LL}^S|
\;-\;0.02\,\Re |C_{RL}^S|
\\
&\;-\;1.57\,|C_{RL}^V|^2
\;-\;0.45\,|C_{LL}^S|^2
\;-\;0.07\,|C_{RL}^S|^2
\\
&\;-\;0.03\,\Re[C_{LL}^V\,C_{RL}^{V\ast}]
\;+\;0.02\,\Re[C_{LL}^V\,C_{LL}^{S\ast}]
\;+\;0.02\,\Re[C_{LL}^V\,C_{RL}^{S\ast}]
\\
&\;-\;0.16\,\Re[C_{RL}^V\,C_{LL}^{S\ast}]
\;-\;0.04\,\Re[C_{RL}^V\,C_{RL}^{S\ast}]
\;-\;0.09\,\Re[C_{LL}^S\,C_{RL}^{S\ast}] \;.
\end{aligned}
\end{equation}

Note that $|C^V_{RL}|^2$ has the largest numerical factor, which explains the deviation from SM for the $C^V_{RL}$. This factor becomes smaller at larger $q^2$ values, as can be seen in the following expression of ${\cal M}_1^{\nu_L}$ at $q^2 = 1.2$ GeV$^2$.
\begin{equation} \label{M1_1.2}
\begin{aligned}
\mathcal M_1^{\nu_L}(1.2) \;\approx\;&
0.34
\;+\;0.22\,\Re |C_{RL}^V|
\;-\;0.03\,\Re |C_{LL}^S|
\;-\;0.01\,\Re |C_{RL}^S|
\\
&\;-\;0.17\,|C_{RL}^V|^2
\;-\;0.35\,|C_{LL}^S|^2
\;-\;0.07\,|C_{RL}^S|^2
\\
&\;-\;0.11\,\Re[C_{LL}^V\, C_{RL}^{V\ast}]
\;+\;0.02\,\Re[C_{LL}^V\,C_{LL}^{S\ast}]
\;+\;0.01\,\Re[C_{LL}^V\,C_{RL}^{S\ast}]
\\
&\;-\;0.07\,\Re[C_{RL}^V\, C_{LL}^{S\ast}]
\;-\;0.04\,\Re[C_{RL}^V\, C_{RL}^{S\ast}]
\;-\;0.17\,\Re[C_{LL}^S\,C_{RL}^{S\ast}]
\;
\end{aligned}
\end{equation}
For larger $q^2$, ${\cal M}_1^{\nu_L}$ values for different NP scenarios align with the SM expectation.

For operators involving a left--handed neutrino, the asymmetry $\mathcal{M}_1^{\nu_L}$
retains strong sensitivity to the chirality of the quark current. In the SM,
$\mathcal{M}_1^{\nu_L}$ has a nonzero baseline value arising from the $V\!-\!A$
structure, which favors left-handed $\Lambda$ helicity states. New physics
operators with a right-handed quark current, such as $C_{RL}^V$, interfere
linearly with the SM contribution. This interference modifies the relative
weights of vector and axial-vector form factors and hence the longitudinal
polarization fraction of the $\Lambda$, making $\mathcal{M}_1^{\nu_L}$ particularly
sensitive to $C_{RL}^V$. By contrast, scalar operators contribute only through
helicity-flip amplitudes, which are suppressed by $m_\mu$ and light-quark
masses, and therefore have a much smaller effect.

Now we discuss the observable ${\cal M}_6^{\nu_L}$. We observe that for $C^V_{RL}$, it deviates slightly from the SM at higher $q^2$ values. The numerical expression for ${\cal M}_6^{\nu_L}$ at $q^2 =1.2 $GeV$^2$ is given as
\begin{equation}
\label{ex-M6_lhn}
\begin{aligned}
{\cal M}_6^{\nu_L}(1.2) \;=\;& -0.29
\;+\; 0.32\,\Re\!\big(C^{V}_{RL}\big)
\;+\; 0.07\,\Re\!\big(C^{S}_{LL}\big)
\;+\; 0.07\,\Re\!\big(C^{S}_{RL}\big) \\[4pt]
&\;+\; 0.48\,\big|C^{V}_{RL}\big|^2
\;+\; 0.17\,\big|C^{S}_{LL}\big|^2
\;+\; 0.17\,\big|C^{S}_{RL}\big|^2 \\[4pt]
&\;-\; 0.80\,\Re\!\big(C^{V}_{LL}\,C^{V\,\ast}_{RL}\big)
\;-\; 0.24\,\Re\!\big(C^{V}_{LL}\,C^{S\,\ast}_{LL}\big)
\;-\; 0.24\,\Re\!\big(C^{V}_{LL}\,C^{S\,\ast}_{RL}\big) \\[4pt]
&\;-\; 0.57\,\Re\!\big(C^{V}_{RL}\,C^{S\,\ast}_{LL}\big)
\;-\; 0.59\,\Re\!\big(C^{V}_{RL}\,C^{S\,\ast}_{RL}\big)
\;-\; 0.07\,\Re\!\big(C^{S}_{LL}\,C^{S\,\ast}_{RL}\big)
\;+\;\mathcal{O}(\mathrm{WC}^3)\,
\end{aligned}
\end{equation}
At large values of $q^2$, the observable ${\cal M}^{\nu_L}_6$ becomes particularly sensitive to
the operator $C^{V}_{RL}$. The reason is that the numerator of $\mathcal{M}^{\nu_L}_6$ contains
terms proportional to $ q^2\, H_{\pm\frac{1}{2},0}$, which grow as $\sqrt{q^2}$ and dominate in
the high-$q^2$ region, while the transverse helicity amplitudes are
kinematically suppressed near the endpoint. Since $C^{V}_{RL}$ flips the quark
chirality relative to the SM $V\!-\!A$ structure, it reverses the relative sign
between vector and axial contributions in the longitudinal channel. This
modifies interference with the SM amplitude precisely in the region where
longitudinal contributions dominate, leading to enhanced deviations of ${\cal M}^{\nu_L}_6$ from its SM expectation at high $q^2$.

Next, we come to the observable ${\cal M}^{\nu_L}_7$ and the explanation of why ${\cal M}^{\nu_L}_7$ vanishes in the SM and most NP scenarios, but not for $C^{V}_{RL}$.
By definition,
\begin{equation}
\label{ex-M7_lhn}
    \begin{aligned}
       {\cal M}^{\nu_L}_7 \;\propto\; \Im\!\Big[
H_{+\frac12,+}^\ast \big(q^2 H_{-\frac12,0} - m_\mu (m_\mu H_{-\frac12,t}+\sqrt{q^2}\,H^{\rm S}_{-\frac12})\big) \\
- H_{-\frac12,-}^\ast \big(m_\mu (m_\mu H_{+\frac12,t}+\sqrt{q^2}\,H^{\rm S}_{+\frac12})+q^2 H_{+\frac12,0}\big)
\Big] 
    \end{aligned}
\end{equation}

In the SM, the hadronic form factors are taken to be real, and there are no strong phases, so all helicity amplitudes carry the same weak phase. Consequently the imaginary part vanishes and ${\cal M}_7^{\rm SM}=0$. The same cancellation persists for $C^{V}_{LL}$ in the LHN scenario, because this operator merely rescales the SM $V\!-\!A$ structure without introducing a new relative phase among helicity amplitudes. Scalar operators ($C^{S}_{LL}, \, C^{S}_{RL}$) contribute mainly through timelike helicity pieces weighted by $m_\mu$, so their interference terms in ${\cal M}^{\nu_L}_7$ are helicity–flip and mass–suppressed; with (effectively) real form factors these pieces remain numerically tiny. 

By contrast, the LHN operator $C^{V}_{RL}$ ($V\!+\!A$ on quarks) modifies the relative $V$ vs.\ $A$ weight in the \emph{transverse} helicity amplitudes with a sign opposite to the SM. This reshuffling prevents the $H_{+\frac12,+}$ and $H_{-\frac12,-}$ contributions from canceling in the bracket above. If $C^{V}_{RL}$ carries a weak phase, the SM--$C^{V}_{RL}$ interference produces a non-vanishing imaginary part, 
\(
{\cal M}^{\nu_L}_7 \sim \Im(C^{V}_{RL}) \times \big[\text{real hadronic factor}\big],
\)
so ${\cal M}^{\nu_L}_7$ becomes the unique T-odd, phase-sensitive probe in which $C^{V}_{RL}$ can generate visible deviations, while other operators either lack linear SM interference (RHN), act as an overall rescaling ($C^{V}_{LL}$), or are $m_\mu$-suppressed (scalars).

In the recent BESIII measurement~\cite{BESIII:2023jxv}, the $T$-odd observable ${\cal M}_7$ was reported with a non-zero value $0.068 \,\pm\, 0.055_{\rm stat} \,\pm\, 0.002_{\rm syst}$. Although the uncertainty is still large, the central value does not include the SM expectation within $1\sigma$. This is significant, because in the SM the observable ${\cal M}_7$ is identically zero as all the helicity amplitudes are real, and hence even a small non-zero value of ${\cal M}_7$ can be interpreted as a potential NP hint. Moreover, ${\cal M}_7$ is directly sensitive to the sign of ${\Im}(C_{RL}^V)$. For our benchmark choice of $C_{RL}^V$, we obtain $|{\cal M}_7| = 0.07 \pm 0.01$. Note that the allowed region for $C_{RL}^V$ is symmetric with respect to the real axis, so the sign of ${\Im}(C_{RL}^V)$ fixes the sign of the predicted ${\cal M}_7$. Therefore, the sign and the size of ${\Im}(C_{RL}^V)$ are constrained by ${\cal M}_7$. 
On the other hand, the observed value of $\mathcal{P}_L^{\Lambda}$, being consistent with the SM expectation, will restrict the magnitude of $C_{RL}^V$. These suggest that a small imaginary value is preferred for $C_{RL}^V$, and it will be clearer with the improved measurement of ${\cal M}_7$ and $\mathcal{P}_L^{\Lambda}$ at low-$q^2$.

These results emphasize the importance of further experimental studies of observables such as ${\cal M}_7$, which are particularly sensitive to right-handed quark currents in the $c \to s\,\ell \,\nu$ transition.

%%%%%%%%%%%%%%%%%%%%%%%%%%%%%%%%%%%%%%%%%%%%%%%%%%%%%%%%
\subsection{Predictions on observables for Right-Handed Neutrinos} \label{pre_rhn}
%%%%%%%%%%%%%%%%%%%%%%%%%%%%%%%%%%%%%%%%%%%%%%%%%%%%%%%%

We now repeat the analysis of Sec.~\ref{observables}, turning to the scenario where right-handed neutrinos are present in addition to the left-handed ones. The NP effects are parametrized in terms of the Wilson coefficients associated with the operators $O^V_{LR}$, $O^V_{RR}$, $O^S_{LR}$, and $O^S_{RR}$, with one coefficient varied at a time. The respective benchmark points for RHN WCs are given in Table \ref{tab:BPs}. The corresponding predictions for the three-body and four-body final states are presented in the following subsections.

%%%%%%%%%%%%%%%%%%%%%%%%%%%%%%%%%%%%%%%%%%%%%%%%%%%%%%%%
\subsubsection{$\Lambda_c^- \to \Lambda \mu^- \nu_{\mu}$ Decay} \label{pre_rhn_3b}
\begin{figure}[h!]
	\centering
    \includegraphics[width=0.49\textwidth]{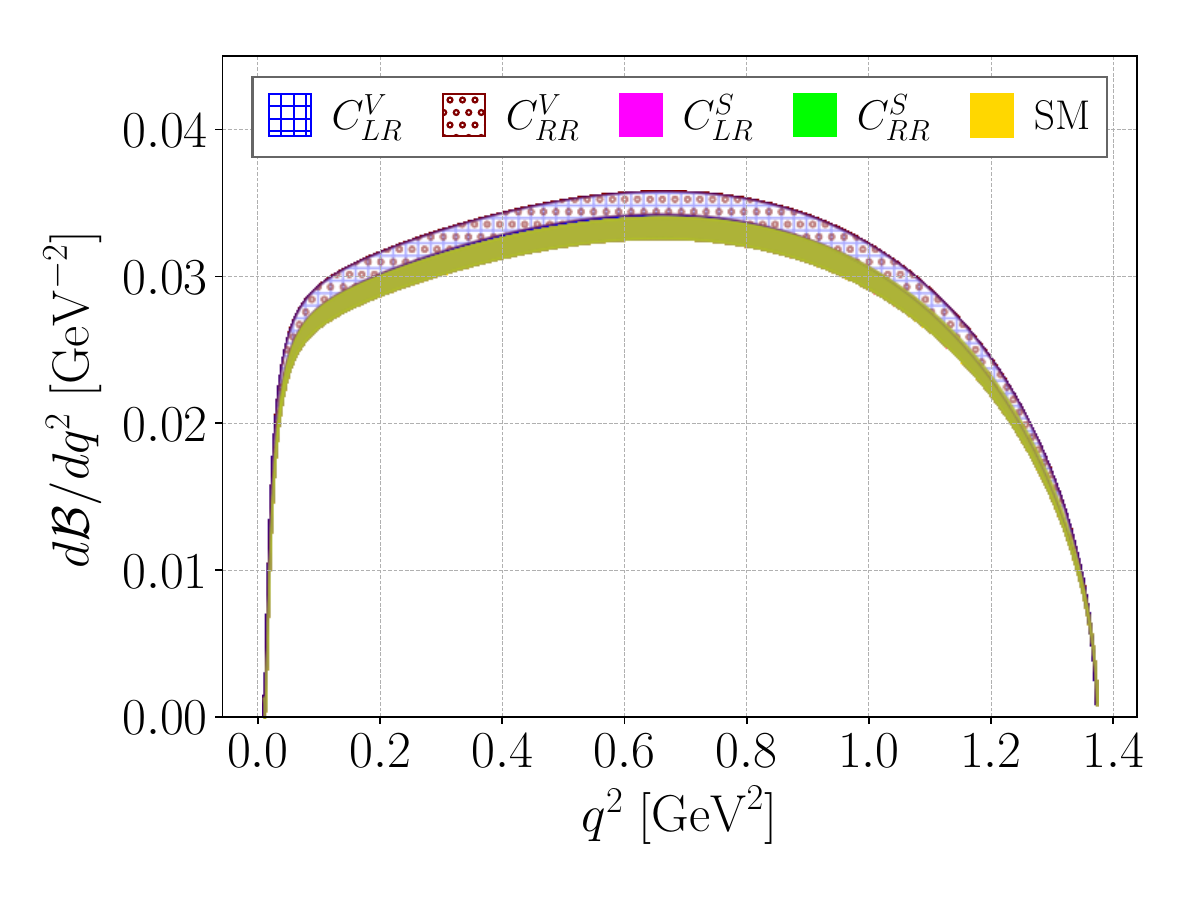}
    \includegraphics[width=0.49\textwidth]{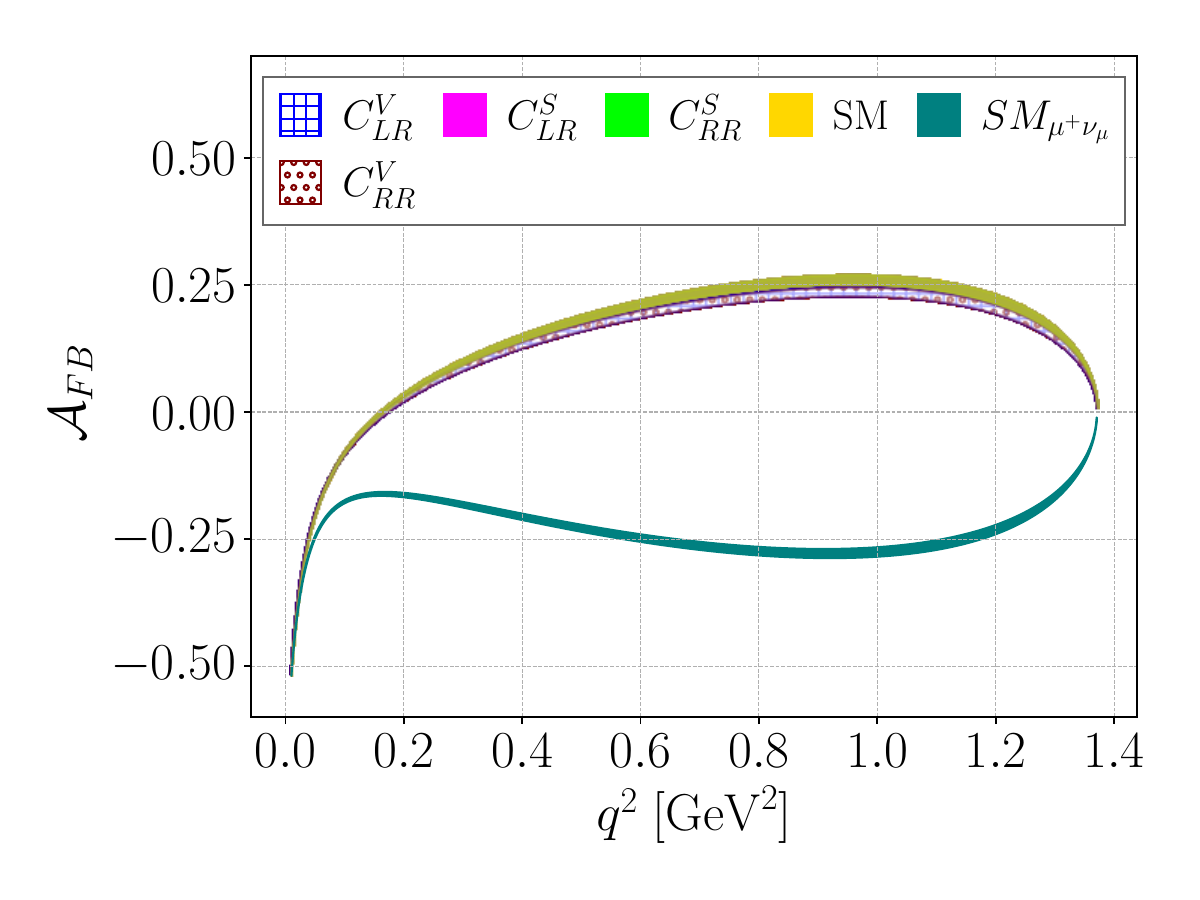}
    \includegraphics[width=0.49\textwidth]{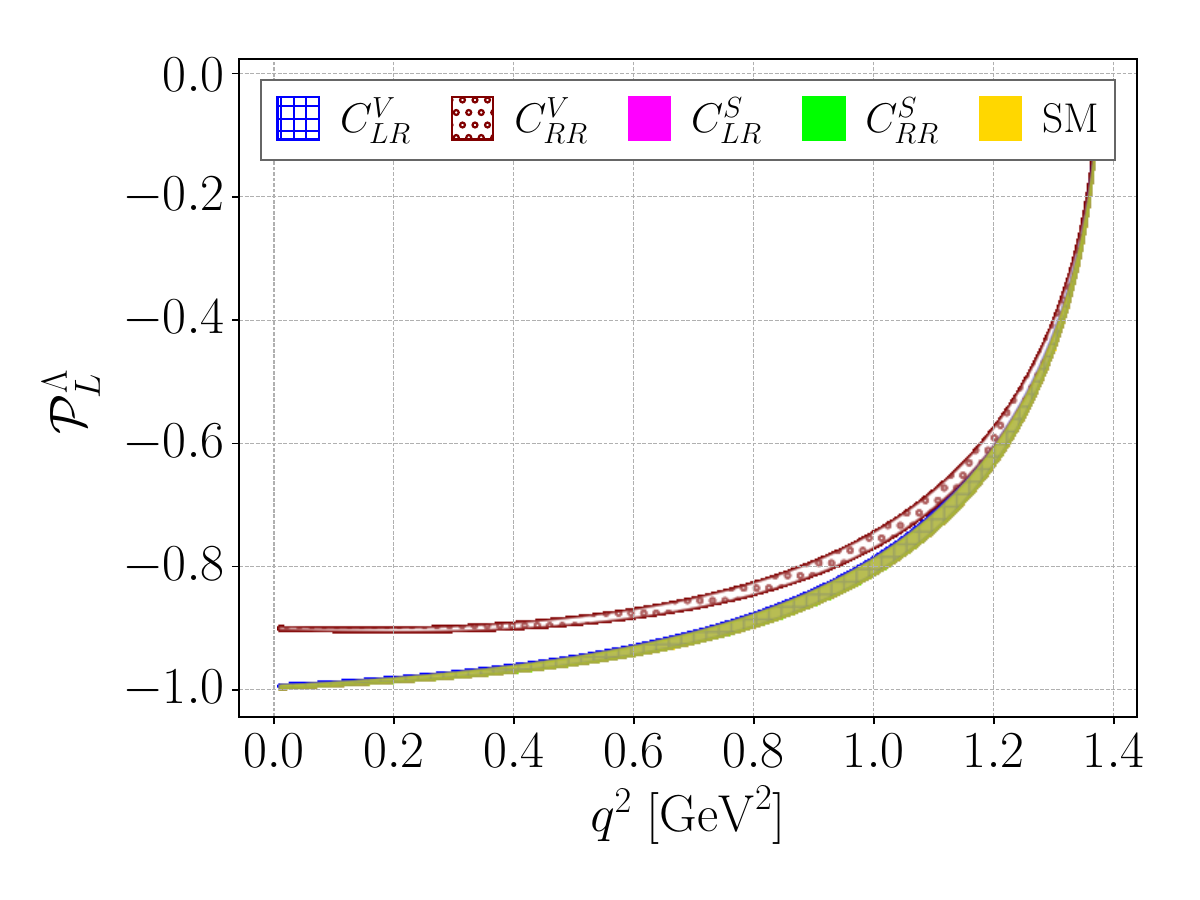}
    \includegraphics[width=0.49\textwidth]{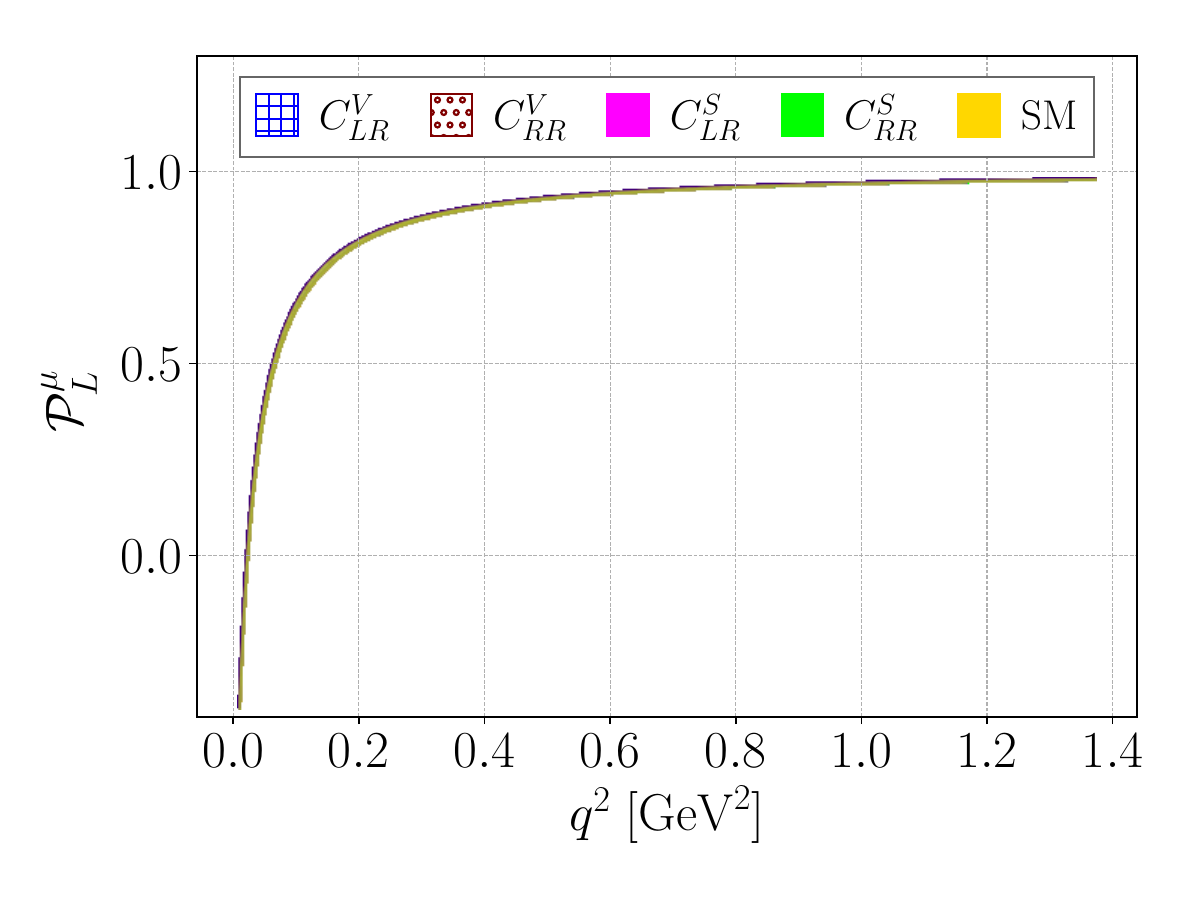}
	\caption{RHN: $q^2$ spectra for the differential branching fraction (top left), forward-backward asymmetry (top right), $\Lambda$ polarization asymmetry (bottom left), and muon polarization asymmetry (bottom right) for three-body decay.}
	\label{fig-3body_rhn}
\end{figure}

%%%%%%%%%%%%%%%%%%%%%%%%%%%%%%%%%%%%%%%%%%%%%%%%%%%%%%%%
In Fig.~\ref{fig-3body_rhn}, we present the observables $d\mathcal{B}/dq^2$, $\mathcal{A}_{FB}$, $\mathcal{P}^\Lambda_L$, and $\mathcal{P}^\mu_L$. As in the LHN case, we find that in the RHN scenario there is no visible deviation from the SM for $d\mathcal{B}/dq^2$, $\mathcal{A}_{FB}$, and $\mathcal{P}^\mu_L$. In contrast, for $\mathcal{P}^\Lambda_L$ we observe a significant deviation from the SM in the NP scenario with $C^V_{RR}$, particularly in the low-$q^2$ region. This behavior can be understood from the analytic expression of $\mathcal{P}^\Lambda_L$ in terms of the NP Wilson coefficients at $q^2 = 0.1~\text{GeV}^2$.
\begin{equation}
\label{ex-P_Lambda_rhn}
\begin{aligned}
\frac{\mathrm{d} \varGamma^{\lambda_{\Lambda}=\frac{1}{2}}} {\mathrm{d} q^2} - \frac{\mathrm{d} \varGamma^{\lambda_{\Lambda} =-\frac{1}{2}}} {\mathrm{d} q^2} =\;&
-838.15\,\bigl(1 + |C^V_{LR}|^2\bigr)
+ 722.03\,|C^V_{RR}|^2
+ 58.71\,|C^S_{LR}|^2
- 58.71\,|C^S_{RR}|^2
\\
&+ 116.11\,\Re\!\bigl[C^{V\ast}_{LR}\,C^V_{RR}\bigr]
+ 11.58\,\Re\!\bigl[C^{V\ast}_{LR}\,C^S_{LR}\bigr]
+ 158.08\,\Re\!\bigl[C^{V\ast}_{LR}\,C^S_{RR}\bigr]
\\&
- 158.08\,\Re\!\bigl[C^{V\ast}_{RR}\,C^S_{LR}\bigr]
- 11.58\,\Re\!\bigl[C^{V\ast}_{RR}\,C^S_{RR}\bigr]
\end{aligned}
\end{equation}

From the above numerical expressions, we see that $C^V_{RR}$ acts for the RHN scenario, the same way $C^V_{RL}$ acted for the LHN scenario -- with a large numerical factor and opposite sign from that of the SM. As a result, we see a significant deviation for $C^V_{RR}$. On the other hand, $C^V_{LR}$, similar to $C^V_{LL}$, has a large numerical factor; however, it only rescales the SM structure, and its contribution to the numerator and the denominator of $\mathcal{P}^\Lambda_L$ cancels out.

%%%%%%%%%%%%%%%%%%%%%%%%%%%%%%%%%%%%%%%%%%%%%%%%
\subsubsection{$\Lambda_c^- \to \Lambda \,(p \pi)\, \mu^- \nu_{\mu}$ Decay} \label{pre_rhn_4b}
%%%%%%%%%%%%%%%%%%%%%%%%%%%%%%%%%%%%%%%%%%%%%%%%

\begin{figure}[h!]
	\centering
    \includegraphics[width=0.49\textwidth]{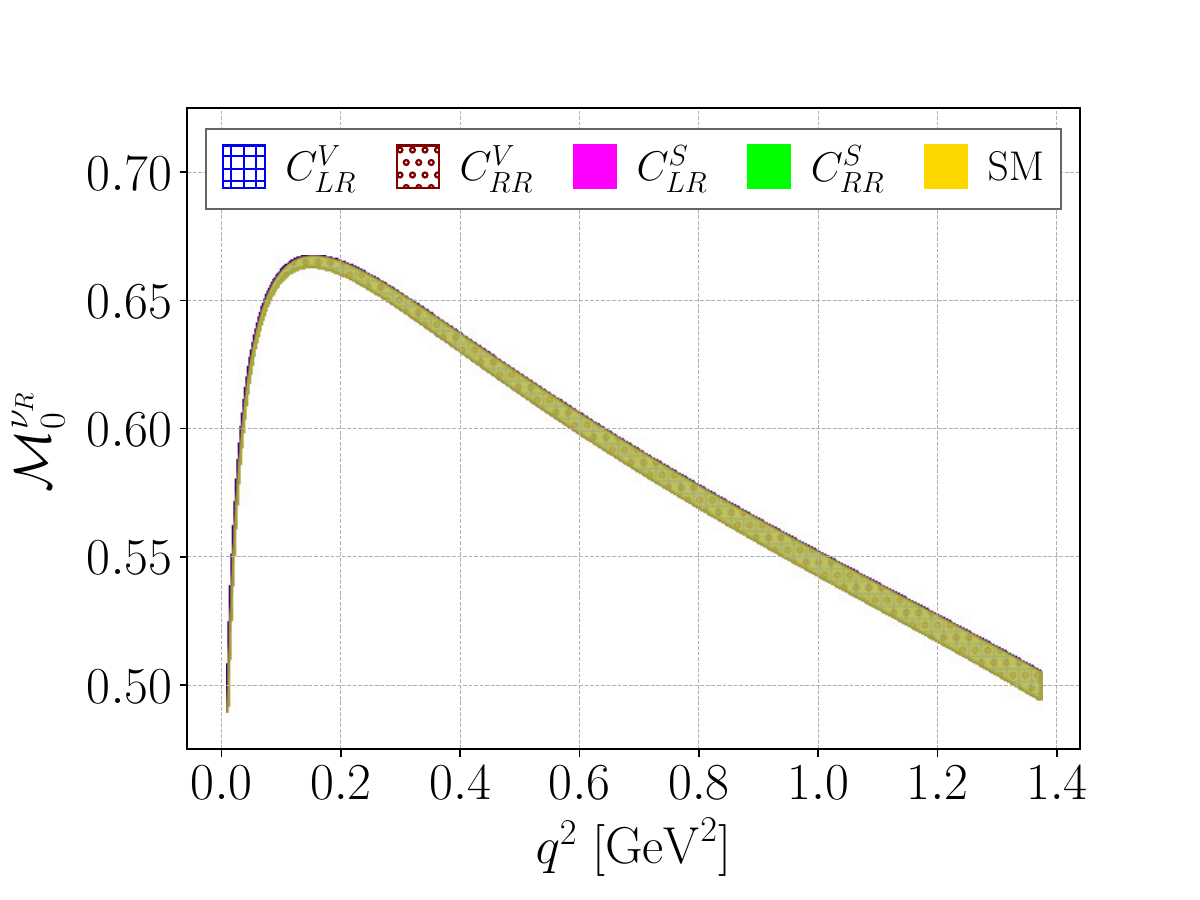}
    \includegraphics[width=0.49\textwidth]{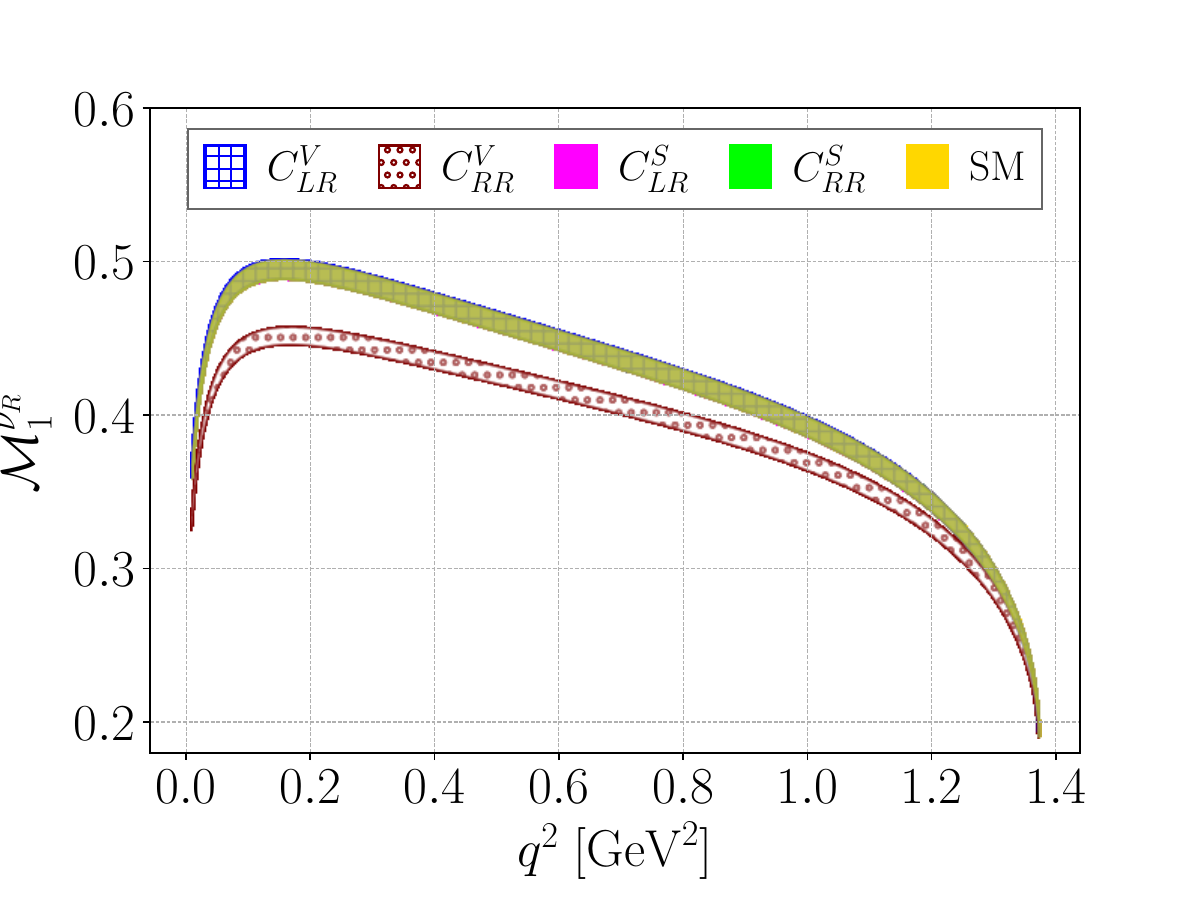}
    \includegraphics[width=0.49\textwidth]{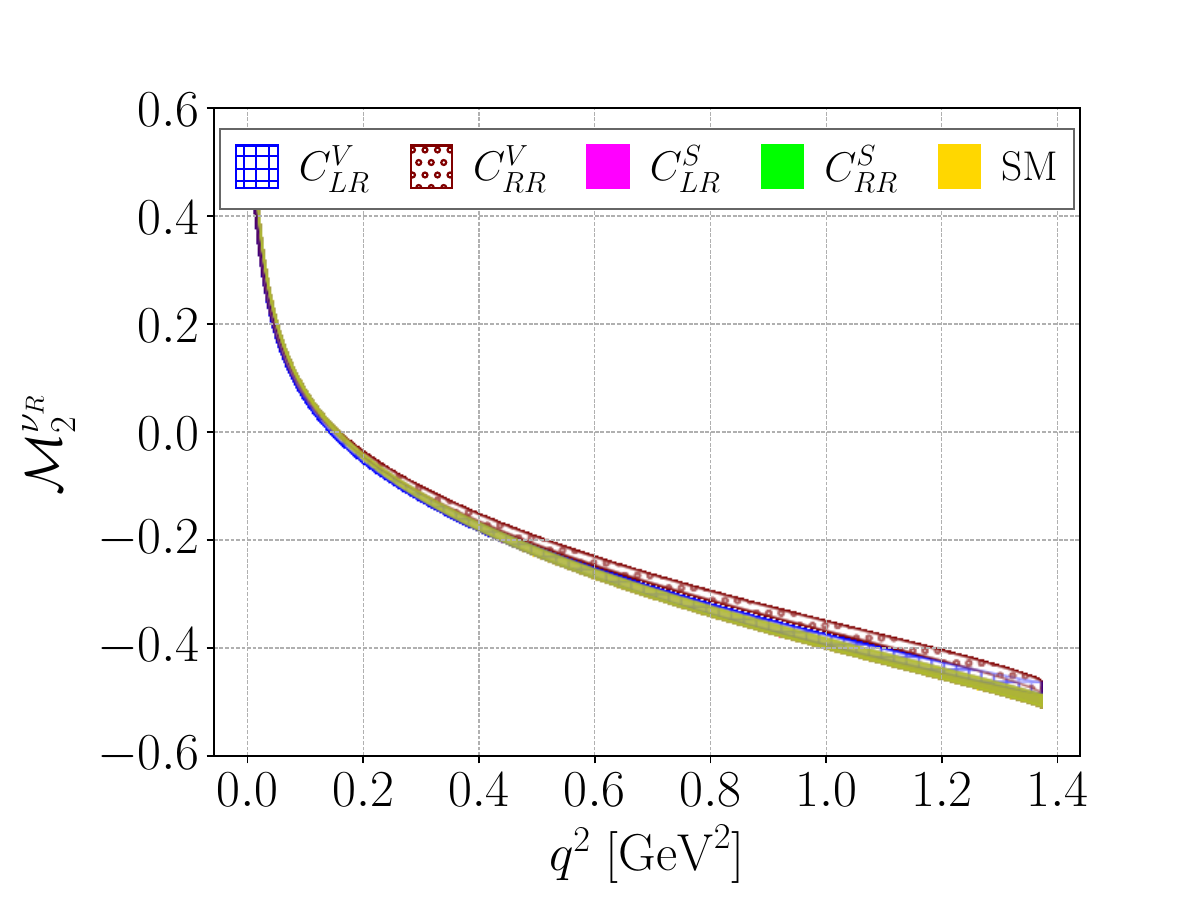}
    \includegraphics[width=0.49\textwidth]{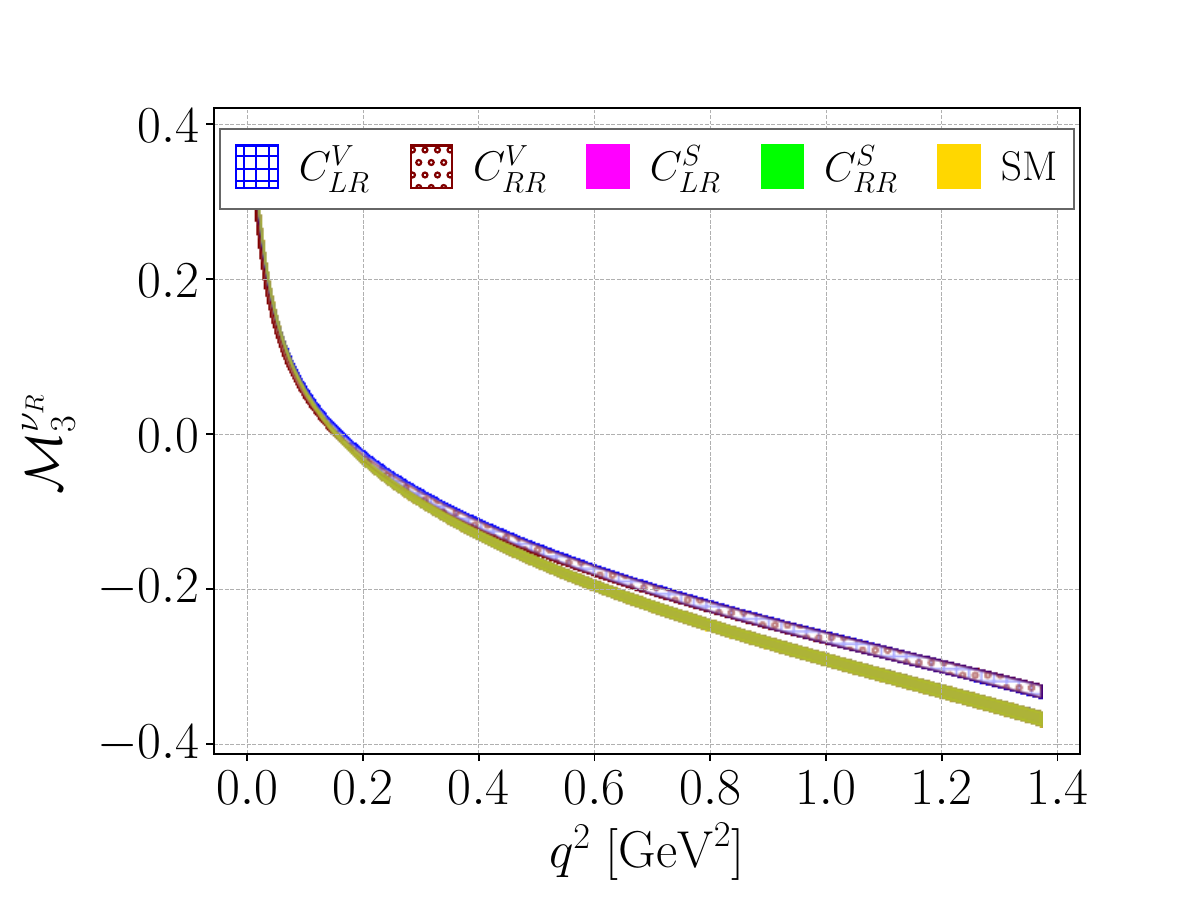}
	\caption{RHN: $q^2$ spectra for the $\mathcal{M}_0^{\nu_R}$ (top left), $\mathcal{M}_1^{\nu_R}$ (top right), $\mathcal{M}_2^{\nu_R}$ (bottom left), and $\mathcal{M}_3^{\nu_R}$ (bottom right) for four-body decay.}
	\label{fig-4body_rhn0to3}
\end{figure}

\begin{figure}[h!]
	\centering
    \includegraphics[width=0.49\textwidth]{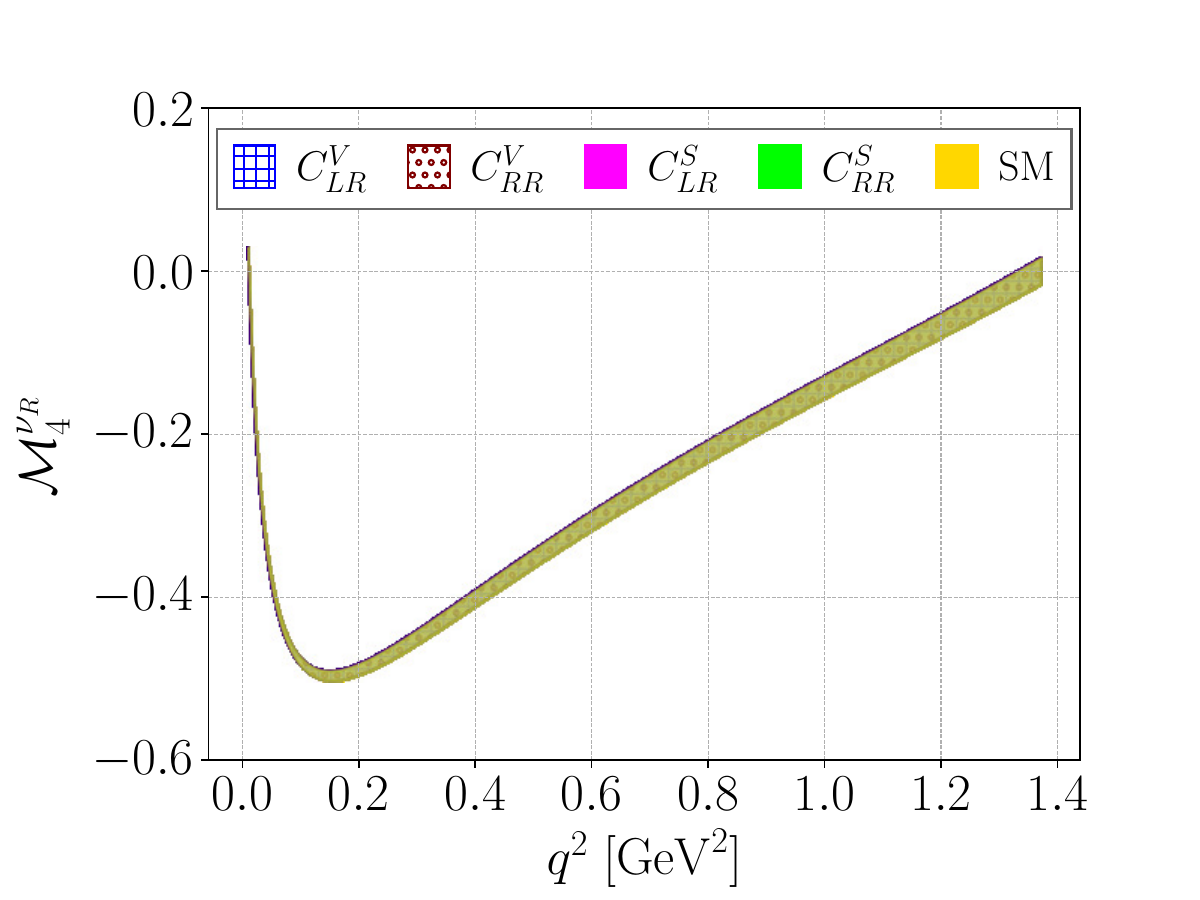}
    \includegraphics[width=0.49\textwidth]{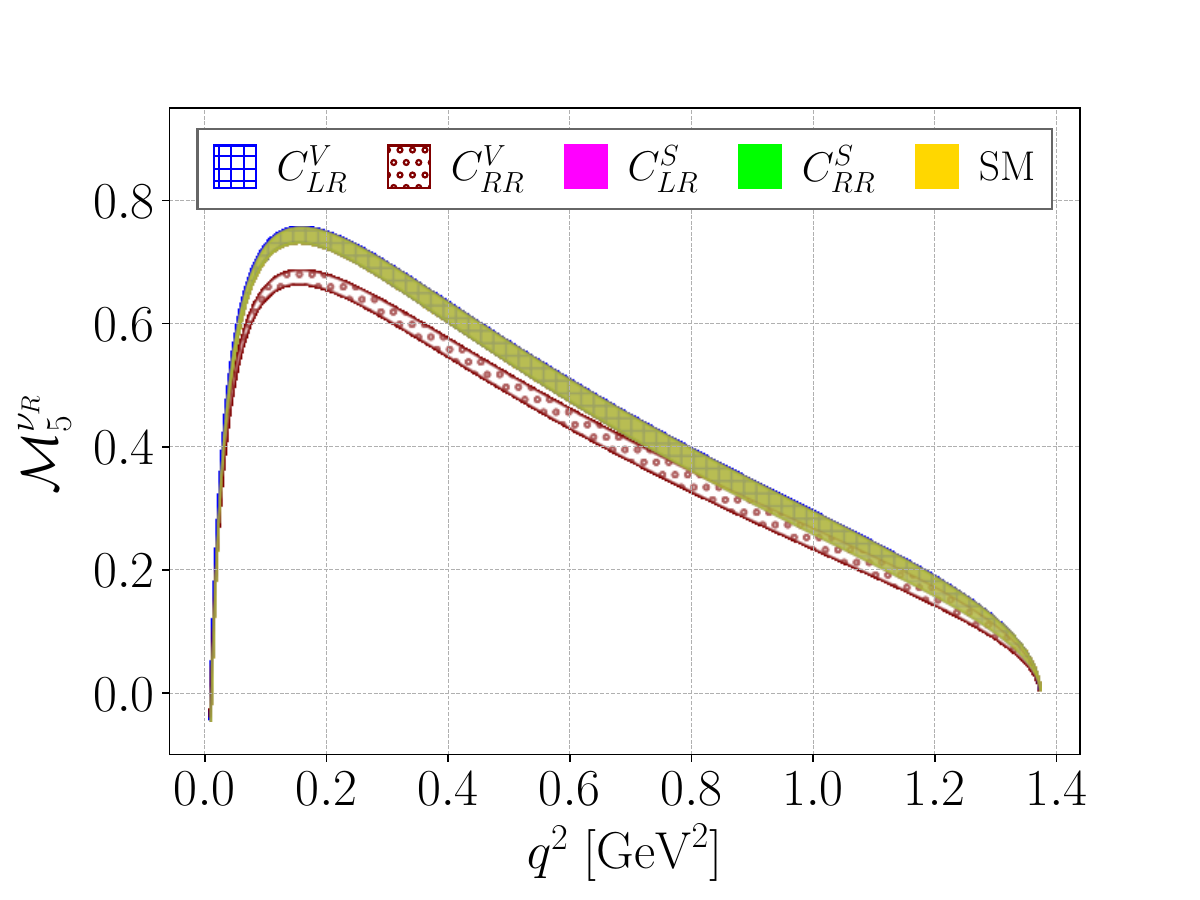}
    \includegraphics[width=0.49\textwidth]{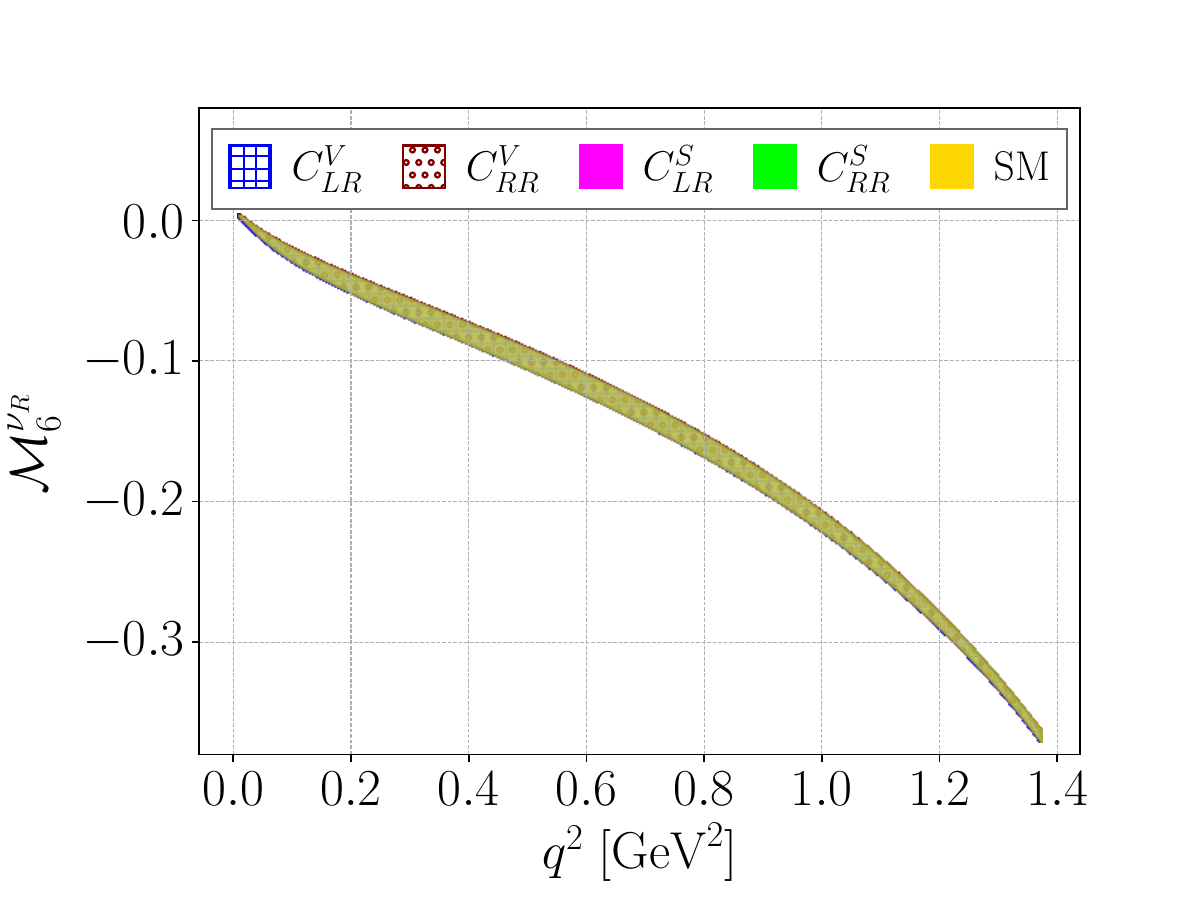}
    \includegraphics[width=0.49\textwidth]{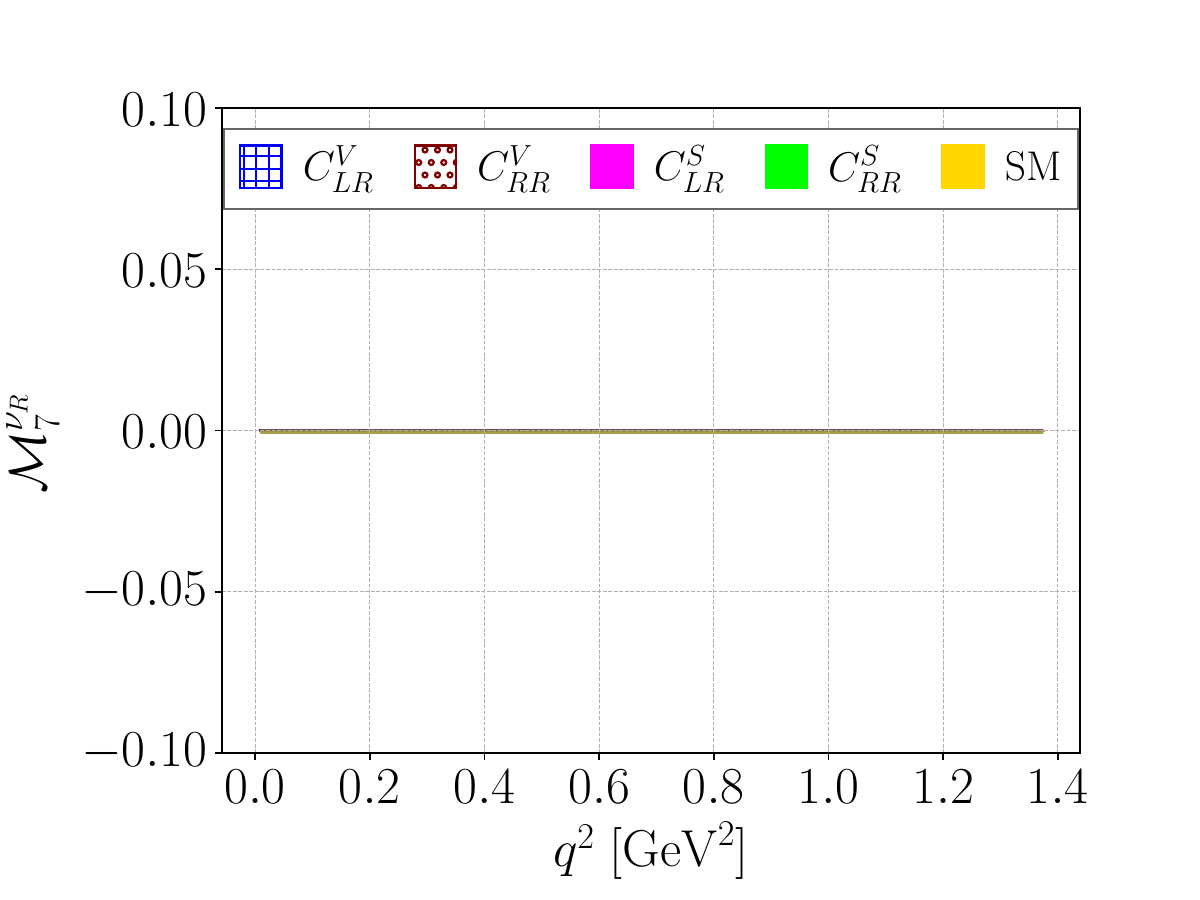}
    \includegraphics[width=0.49\textwidth]{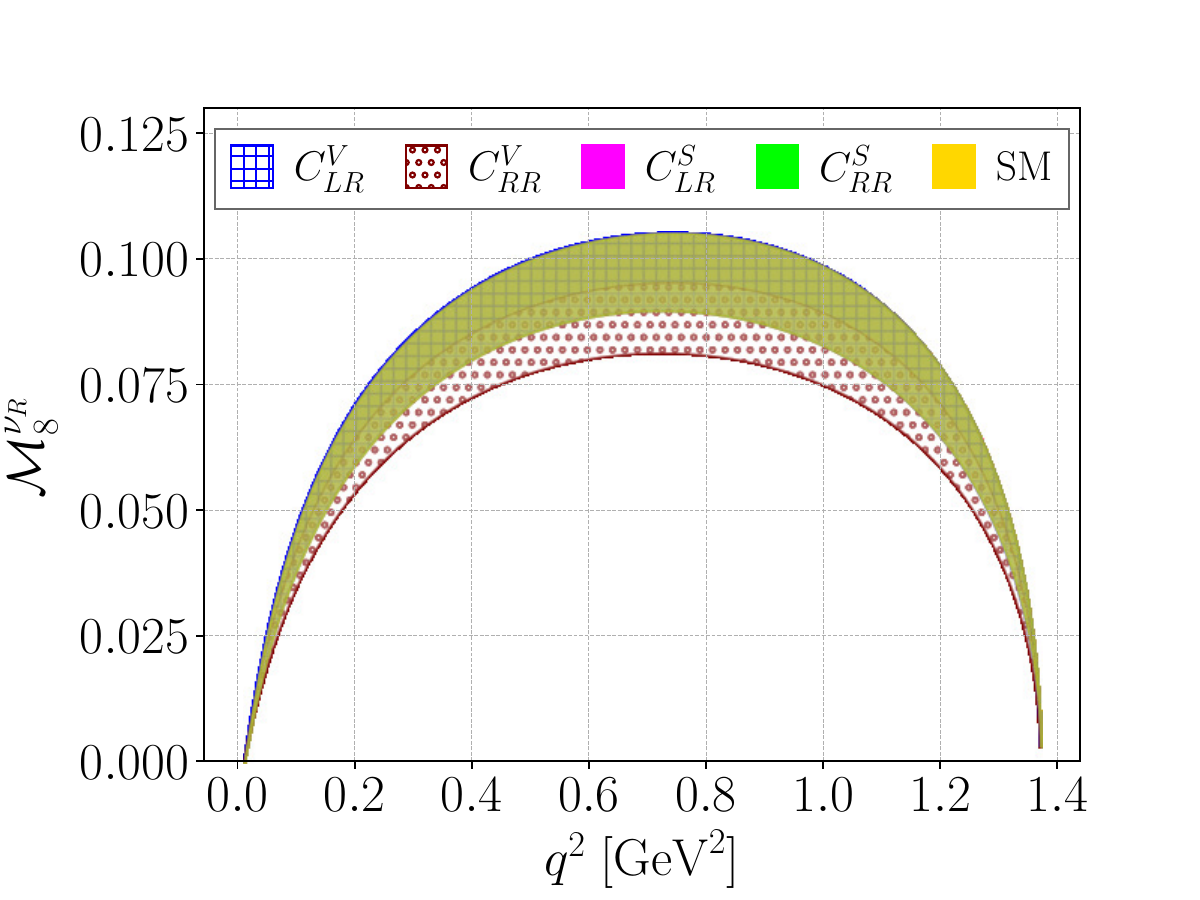}
	\caption{RHN: $q^2$ spectra for the $\mathcal{M}_4^{\nu_R}$ (top left), $\mathcal{M}_5^{\nu_R}$ (top right), $\mathcal{M}_6^{\nu_R}$ (middle left), $\mathcal{M}_7^{\nu_R}$ (middle right) and $\mathcal{M}_8^{\nu_R}$ (bottom) for four-body decay.}
	\label{fig-4body_rhn4to8}
\end{figure}

We present ${\cal M}^{\nu_R}_{0}$, ${\cal M}^{\nu_R}_{1}$, ${\cal M}^{\nu_R}_{2}$ and ${\cal M}^{\nu_R}_{3}$ in Fig.~\ref{fig-4body_rhn0to3}, and ${\cal M}^{\nu_R}_{4}$, ${\cal M}^{\nu_R}_{5}$, ${\cal M}^{\nu_R}_{6}$, ${\cal M}^{\nu_R}_{7}$ and ${\cal M}^{\nu_R}_{8}$ in Fig.~\ref{fig-4body_rhn4to8}. We note that only ${\cal M}^{\nu_R}_{1}$ and ${\cal M}^{\nu_R}_{5}$ shows significant deviations from SM for the NP scenario $C^V_{RR}$.

Note that when the final-state neutrino is right-handed, interference with the SM is forbidden, and all contributions appear only at quadratic order in the corresponding Wilson coefficients. In this setup, operators with a
left-handed quark current, such as $C_{LR}^V$, mimic the SM hadronic helicity
pattern, so their effect largely cancels between the numerator and denominator in
the polarization ratio. As a result, $\mathcal{M}_1^{\nu_R}$ shows no appreciable
dependence on $C_{LR}^V$ as can be seen from eq.\,\ref{M10.4}.

\begin{equation}
\label{M10.4}
\begin{aligned}
\mathcal M_1^{\nu_R}(0.4)
&= 0.47
\;+\; \Big[
\;{\underbrace{0.00}_{\text{cancels}}}\,|C_{LR}^V|^2
\;-\;0.79\,|C_{RR}^V|^2
\;-\;0.23\,|C_{LR}^S|^2
\;-\;0.04\,|C_{RR}^S|^2 \\
&\hspace{3.8em}
+\;0.14\,\Re\!\big(C_{LR}^V\,C_{RR}^{V\ast}\big)
\;-\;0.05\,\Re\!\big(C_{LR}^V\,C_{LR}^{S\ast}\big)
\;-\;0.11\,\Re\!\big(C_{LR}^V\,C_{RR}^{S\ast}\big) \\
&\hspace{3.8em}
\;-\;0.24\,\Re\!\big(C_{RR}^V \, C_{LR}^{S\ast}\big)
\;-\;0.06\,\Re\!\big(C_{RR}^V \, C_{RR}^{S\ast}\big)
\;-\;0.19\,\Re\!\big(C_{LR}^S \, C_{RR}^{S\ast}\big)
\Big]
\;
\end{aligned}
\end{equation}

In contrast, right-handed quark currents
($C_{RR}^V$) alter the relative sign between the vector and axial-vector form
factors, thereby shifting the balance of $\Lambda$ helicity amplitudes and
leading to sizeable modifications of $\mathcal{M}^{\nu_R}_1$. Scalar operators with
right-handed neutrinos again enter only through timelike amplitudes and are
numerically subdominant.

For ${\cal M}^{\nu_R}_5$, there is mild deviation from SM for $C^V_{RR}$ at lower $q^2$. The numerical expression for ${\cal M}^{\nu_R}_5$ is given as

\begin{equation}
\label{ex-M5_0.10}
    \begin{aligned}
       \mathcal{M}^{\nu_R}_5(0.10)\;= & \;0.67
\;-\;0.09\,\Re (C^V_{RL})
\;-\;0.04\,\Re (C^S_{LL})
\;-\;0.18\,\Re (C^S_{RL})
\;-\;1.35\,\bigl|C^V_{RR}\bigr|^2 \\ &
\;-\;0.09\,\Re\!\bigl[C^V_{LR}\,C^{V\ast}_{RR}\bigr] 
    \end{aligned}
\end{equation}

The above relation shows large sensitivity of ${\cal M}^{\nu_R}_5$ for $C^{V}_{RR}$. The deviation of $\mathcal{M}^{\nu_R}_5$ at low $q^2$ comes from the fact that
$C^V_{RR}$ contributes quadratically (with no SM interference) to the dominant
longitudinal helicity amplitudes. These quadratic terms spoil the near
cancellation present in the SM expression for $\mathcal{M}^{\nu_R}_5$, producing a
visible shift. Other Wilson coefficients either cancel out, interfere only
weakly, or are helicity-flip and $m_\mu$-suppressed, so only $C^V_{RR}$
generates a significant deviation in this observable.

Unlike the LHN, ${\cal M}^{\nu_R}_7$ does not show any deviation for any of the NP scenarios in the case of RHN. In the RHN scenarios ($\nu_R$), the new amplitudes do not interfere linearly with the SM, because the SM involves only a left-handed neutrino current. Hence ${\cal M}^{\nu_R}_7$, which is proportional to an imaginary interference term between helicity structures, receives RHN contributions only at quadratic order in the RHN Wilson coefficients (no SM--RHN cross term). With real hadronic form factors and no strong phases, the RHN helicity amplitudes share a common weak phase, so their self-interferences are real and the imaginary parts entering ${\cal M}^{\nu_R}_7$ vanish. Even allowing complex RHN Wilson coefficients, in the absence of relatively strong phases between the relevant helicity structures, the imaginary parts cancel between the $H_{+\frac12,+}$ and $H_{-\frac12,-}$ pieces. Consequently, ${\cal M}_7^{\nu_R}\simeq 0$ across $q^2$, up to tiny effects from muon-mass–suppressed terms or subleading hadronic phases.

%%%%%%%%%%%%%%%%%%%%%%%%%%%%%%%%%%%%%%%%%%%%%%%%%%%%%%%%%%%%%%%
\section{Conclusions} \label{conclusion}
%%%%%%%%%%%%%%%%%%%%%%%%%%%%%%%%%%%%%%%%%%%%%%%%%%%%%%%%%%%%%%%%%
The long-standing deviations observed in semileptonic decays involving third-generation quarks and leptons, together with recent experimental searches in semileptonic processes of the $\Lambda_c$, motivate a comprehensive study of $\Lambda_c^- \to \Lambda (\to p \pi)\,\mu^- \nu_{\mu}$, which involves second-generation leptons and a transition between second-generation quarks. Furthermore, recent advances in calculations within the SMEFT framework, as well as the extension to include right-handed neutrinos in EFTs, call for an updated analysis of the allowed NP parameter space in the $c \to s \mu \nu_{\mu}$ channel.

In this work, we first determine the direct constraints on LEFT operators contributing to $c \to s \mu \nu_{\mu}$ from observables sensitive to this channel. We then demonstrate how these bounds improve once the SMEFT-implied constraints are taken into account, where a much larger set of observables enters through operator mixing when SMEFT coefficients are matched to LEFT.
We find that for all Wilson coefficients, the SMEFT-implied constraints are significantly tighter than those obtained directly from $c \to s \mu \nu_{\mu}$ observables alone. In particular, the allowed parameter space for the scalar coefficient $C^{S}_{RL}$ shrinks substantially once the full set of correlated observables is included. 

Next, we include right-handed neutrinos in the LEFT framework and obtain the allowed region for the corresponding WCs that contribute directly to $c\to s \mu \nu_{\mu}$- mediated observables.

We find that for both the LHN and RHN scenarios, the vector WCs has significant nonzero allowed values. Using these allowed regions, we study the predictions for observables in $\Lambda_c^- \to \Lambda \mu^- \nu$ and $\Lambda_c^- \to \Lambda(\to p \pi)\,\mu^- \nu$. To illustrate the possible size of NP effects, we select benchmark points for each Wilson coefficient within their $1\sigma$ allowed range.

Our analysis shows prominent deviations from the SM appear in the observables $\mathcal{P}^\Lambda_{L}$ and ${\cal M}_1$ for $C^V_{RL}$ and $C^V_{RR}$, especially in the low-$q^2$ region. Both correspond to the longitudinal polarization fraction of the $\Lambda$. We also observe mild deviations in ${\cal M}_5$ and ${\cal M}_6$ for $C^{V}_{RR}$ and $C^{V}_{RL}$, respectively. Moreover, we find that the observable ${\cal M}_7$ shows large deviations throughout the $q^2$ range for imaginary values of $C^V_{RL}$. Since ${\cal M}_7$ vanishes in the SM and remains small in other NP scenarios, this provides a striking probe of NP. These results suggest sensitivity to vector-type NP scenarios involving $C^{V}_{RL}$, $C^{V}_{RR}$ and
NP models that generate such operators, such as vector leptoquarks coupling to $c\to s \mu \nu_{\mu}$, or models with right-handed charged currents from extended gauge sectors, e.g., $W'$ bosons.

A comparison of our predicted values for the observables $\mathcal{P}_L^{\Lambda}$ and ${\cal M}_7$ with their recent BESIII measurements highlights the potential sensitivity of these quantities to new physics effects. In particular, it motivates future high-precision studies of $T$-odd observables such as ${\cal M}_7$, which serve as clean probes of right-handed quark currents.
Such observables could be accessed at BESIII, LHCb, Belle II, and possible future high-luminosity experiments studying charm baryon decays, where polarization-sensitive measurements in $\Lambda_c$ decays are becoming increasingly feasible.
Taken together, our analysis highlights the importance of a complete four-body angular distribution study, which, for the first time, incorporates SMEFT-implied indirect bounds while allowing for right-handed neutrinos in the EFT framework.

%%%%%%%%%%%%%%%%%%%%%%%%%%%%%%%%%%%%%%%%%%%%%%%%%%%%%%%%%%%%
\section*{Acknowledgements}
PB acknowledges to Ministry of Education, Government of India, for providing the Institute Fellowship and the Department of Physics, MNIT Jaipur, for access to its computing facilities.
SK would like to acknowledge the use of the computational facilities of the Department of Theoretical Physics at the Tata Institute of Fundamental Research, Mumbai.
DK acknowledges support from the ANRF, Government of India, under research grant no. SERB/EEQ/2021/000965.
PB and SK also thank the PPC2024 conference, where this project was initiated through fruitful discussions, and the IITGN Flavor Physics Week, during which the work was further developed.
%%%%%%%%%%%%%%%%%%%%%%%%%%%%%%%%%%%%%%%%%%%%%%%%%%%%%%%%%%%%%%%%%%%
\appendix{}
%%%%%%%%%%%%%%%%%%%%%%%%%%%%%%%%%%%%%%%%%%%%%%%%%%%%%%%
%%%%%%%%%%%%%%%%%%%%%%%%%%%%%%%%%%%%%%%%%%%%%%%%%%%%%%
\section{Definition of Helicity Amplitudes and Form Factors} \label{app-A0}
%%%%%%%%%%%%%%%%%%%%%%%%%%%%%%%%%%%%%%%%%%%%%%%

%%%%%%%%%%%%%%%%%%%%%%%%%%%%%%%%%%%%%%%%%%%%%%%%
\subsection{Helicity Amplitudes} \label{decay_amp}
%%%%%%%%%%%%%%%%%%%%%%%%%%%%%%%%%%%%%%%%%%%%%%%%
Here, we collect the information on the hadronic and leptonic helicity amplitudes for the calculation of the decay amplitude. In the SM, the quark level transition of $c \to s \mu \nu_{\mu}$ is processed by the $c \to s W^{*-}$ and $W^{*-} \to \mu^- \bar \nu_{\mu}$ subsequently. For the calculation of the helicity amplitudes of the hadronic and leptonic currents, we choose the $z-$axis along the $W^{*-}$. The off-shell $W^{*-}$ has four helicities $\lambda_W = \pm 1, 0, t$ with polarization vectors as follows:

\begin{align} \label{epsilon}
  \varepsilon^{\mu}(\pm)=\frac{1}{\sqrt{2}}(0,\mp,-i,0), \quad \varepsilon^{\mu}(0)=(0,0,0,1), \quad  \varepsilon^{\mu}(t)=(1,0,0,0) 
\end{align}

The hadronic amplitudes of the $\Lambda_c^- \to \Lambda W^{*-}$ process for the vector, axial-vector, scalar, and pseudo-scalar currents are given as

\begin{eqnarray} 
&&{H}_{\lambda_{\Lambda},\lambda_W}^V=\varepsilon^*_\mu(\lambda_W)\langle \Lambda(p_{\Lambda},\lambda_{\Lambda})|\bar{c}\gamma^{\mu}s |\Lambda_c(p_{\Lambda_c},\lambda_{\Lambda_c})\rangle, \label{hadronic_hel_V}\\
&&{H}_{\lambda_{\Lambda},\lambda_W}^A= \varepsilon^*_\mu(\lambda_W)\langle \Lambda(p_{\Lambda},\lambda_{\Lambda})|\bar{c}\gamma^{\mu}\gamma_5 s |\Lambda_c(p_{\Lambda_c},\lambda_{\Lambda_c})\rangle, \label{hadronic_hel_A}\\
&&{H}^{SPL(R)}_{\lambda_{\Lambda}}= \langle{\Lambda(p_{\Lambda},\lambda_{\Lambda})}|\bar{c}(1\mp\gamma_5)s|\Lambda_c(p_{\Lambda_c},\lambda_{\Lambda_c})\rangle \label{hadronic_hel_SP}
\end{eqnarray}

These hadronic amplitudes depend only on the form factors and are independent of whether the neutrino is left- or right-handed; RHN effects enter exclusively through the leptonic current.

Now for the leptonic part $W^{*-} \to \mu^- \bar \nu_{\mu}$: The helicity amplitudes corresponding to the vector and scalar operators for left-handed neutrinos (helicity = + 1/2) are expressed as

\begin{align}
&L^{L, \nu_L}_{\lambda_\mu,\lambda_W}=\varepsilon^{\mu}(\lambda_W)\bar \mu(p_\mu,\lambda_\mu)\,\gamma^\mu(1-\gamma_5)\, \bar\nu_{\mu_L}(p_{\bar\nu},+\tfrac12), \label{vector_LH}\\
    &L^{SPL, \nu_L}_{\lambda_\mu}=\bar \mu(p_\mu, \lambda_\mu)\,(1-\gamma_5)\, \bar\nu_{\mu_L}(p_{\bar\nu},+\tfrac12) \label{scalar_LH},
\end{align}

where $\lambda_\mu = \pm 1/2$ is the helicity of the muon lepton. Similarly, for the right-handed neutrinos (helicity = - 1/2), the helicity amplitudes corresponding to the vector and scalar currents are

\begin{align}
&L^{R, \nu_R}_{\lambda_\mu,\lambda_W}=\varepsilon^{\mu}(\lambda_W)\bar \mu(p_\mu,\lambda_\mu)\,\gamma^\mu(1+\gamma_5)\, \bar\nu_{\mu_R}(p_{\bar\nu},-\tfrac12), \label{vector_RH}\\
    &L^{SPR, \nu_R}_{\lambda_\mu}=\bar \mu(p_\mu, \lambda_\mu)\,(1+\gamma_5)\, \bar\nu_{\mu_R}(p_{\bar\nu},-\tfrac12) \label{scalar_RH},
\end{align}

Closed-form expressions for all hadronic and leptonic helicity amplitudes used in the three-body
$\Lambda_c^-\to\Lambda\,\mu^- \nu_{\mu}$ and four-body
$\Lambda_c^-\to\Lambda(\to p\pi)\,\mu^- \nu_{\mu}$ analyses are collected in
Appendices~\ref{LHN_hel_amp}, ~\ref{RHN_hel_amp} and \ref{HA_3body} (see also the notation of Ref.~\cite{Datta:2017aue}).

%%%%%%%%%%%%%%%%%%%%%%%%%%%%%%%%%%%%%%%%%%%%%%%%%%%%%%
\subsection{Form Factors} \label{ffs}
%%%%%%%%%%%%%%%%%%%%%%%%%%%%%%%%%%%%%%%%%%%%%%%%%%
 In $\Lambda_c \to \Lambda$ decay, the vector, axial-vector, scalar, and pseudo-scalar currents are expressed in terms of the six form factors $f_0, f_+, f_{\perp}, g_0, g_+, g_{\perp}$ \cite{Feldmann:2011xf, Datta:2017aue} as follows

\begin{eqnarray}
   \langle \Lambda(p_{\Lambda},\lambda_{\Lambda})|\bar{c}\gamma^{\mu}s|\Lambda_c(p_{\Lambda_c},\lambda_{\Lambda_c})\rangle &=&\bar{u}_2(p_{\Lambda},\lambda_{\Lambda})\Bigg[f_0(q^2)(m_{\Lambda_c}-m_\Lambda)\frac{q^\mu}{q^2} \nonumber\\
   &&+f_+(q^2)\frac{m_{\Lambda_c}+m_\Lambda}{Q_+}\bigg(p_{\Lambda_c}^{\mu}
    +p_{\Lambda}^{\mu}-(m_{\Lambda_c}^2-m_\Lambda^2)\frac{q^{\mu}}{q^2}\bigg)\nonumber \\
   &&+f_{\perp}(q^2)\bigg(\gamma^{\mu}-\frac{2m_{\Lambda}}{Q_+}p^{\mu}_{\Lambda}-\frac{2m_{\Lambda_c}}{Q_+}p_{\Lambda}^{\mu}\bigg)\Bigg]u_1(p_{\Lambda_c},\lambda_{\Lambda_c}),\\
  \langle \Lambda(p_{\Lambda},\lambda_{\Lambda})|\bar{c}\gamma^{\mu}\gamma_5s|\Lambda_c(p_{\Lambda_c},\lambda_{\Lambda_c})\rangle
   &=&-\bar{u}_2(p_{\Lambda},\lambda_{\Lambda})\gamma_5\Bigg[g_0(q^2)(m_{\Lambda_c}+m_\Lambda)\frac{q^{\mu}}{q^2}\nonumber\\
 &&+g_+(q^2)\frac{m_{\Lambda_c}-m_\Lambda}{Q_-}\bigg(p_{\Lambda_c}^{\mu}+p_{\Lambda}^{\mu}
    -(m_{\Lambda_c}^2-m_\Lambda^2)\frac{q^{\mu}}{q^2}\bigg)\nonumber\\
  &&+g_{\perp}(q^2)\bigg(\gamma^{\mu}+\frac{2m_\Lambda}{Q_-}p_{\Lambda_c}^{\mu}-\frac{2m_{\Lambda_c}}{Q_-}p_{\Lambda}^{\mu}\bigg)\Bigg]u_1(p_{\Lambda_c},\lambda_{\Lambda_c}), \\
  \langle \Lambda(p_{\Lambda},\lambda_{\Lambda})|\bar{c}s|\Lambda_c(p_{\Lambda_c},\lambda_{\Lambda_c})\rangle&=&f_0(q^2)
   \frac{m_{\Lambda_c}-m_\Lambda}{m_c-m_s}\bar{u}_2(p_{\Lambda},\lambda_{\Lambda})u_1(p_{\Lambda_c},\lambda_{\Lambda_c}),
   \label{eq:scalar ffs}\\
    \langle \Lambda(p_{\Lambda},\lambda_{\Lambda})|\bar{c}\gamma_5 s|\Lambda_c(p_{\Lambda_c},\lambda_{\Lambda_c})\rangle &=& g_0(q^2) \frac{m_{\Lambda_c}+m_\Lambda}{m_c+m_s} \bar{u}_2(p_{\Lambda},\lambda_{\Lambda})\gamma_5u_1(p_{\Lambda_c},\lambda_{\Lambda_c}) \label{eq:pseudo-scalar ffs}
\end{eqnarray}

where $q = p_{\Lambda_c} - p_{\Lambda}$, $\lambda_{\Lambda_c(\Lambda)} = \pm \frac{1}{2}$ denotes the helicities of the $\Lambda_c$ and $\Lambda$ hadron respectively. And other parameters are $Q_+ = (m_{\varLambda_{c}} + m_{\varLambda})^2 - q^2$, $Q_- = (m_{\varLambda_{c}} - m_{\varLambda})^2 - q^2$ and $M_+ = m_{\varLambda_{c}} + m_{\varLambda}$, $M_- = m_{\varLambda_{c}} - m_{\varLambda}$ respectively. 

In this work, we have used the  Lattice QCD results of Ref.~\cite{Meinel:2016dqj}. Here, for completeness, we provide the explicit expression of the form factor in terms of $z$-expansion:
\begin{equation}
	f(q^2) = \frac{1}{1 - \frac{q^2}{{m_{pole}^f}^2}}\sum_{n = 0}^{n_{max}} a_n^f \Big[z(q^2)\Big]^n
\end{equation}
where the function $z(q^{2})=\frac{\sqrt{t_{+}-q^{2}}-\sqrt{{t}_{+} - t_0}}{\sqrt{t_{+}-q^{2}}+\sqrt{{t}_{+} - t_0}}$ with $t_0 = (m_{\varLambda_c} - m_{\varLambda})^2$, $t_+ = (m_{D} + m_{K})^2$. And $m_{pole}^{f_+, f_{\perp}} = 2.112$, $m_{pole}^{g_+, g_{\perp}} = 2.460$ and  $m_{pole}^{f_0} = 2.318$, $m_{pole}^{g_0} = 1.968$ respectively.
We consider statistical uncertainties in the form factors, using the nominal fit (of order two, $n_{max} = 2$) \cite{Meinel:2016dqj}.

%%%%%%%%%%%%%%%%%%%%%%%%%%%%%%%%%%%%%%%%%%%%%%%%%%%%%%%%%%%%%%
\section{Explicit expression of Helicity Amplitudes} \label{app-A}
%%%%%%%%%%%%%%%%%%%%%%%%%%%%%%%%%%%%%%%%%%%%%%%%%%%%%%%%%%
Here we compile all the Leptonic and Hadronic helicity amplitudes for the $\Lambda^- \to \Lambda (p \pi) \mu^- \nu_{\mu}$ decay.
\subsection{Leptonic Helicity Amplitudes}
\label{app-A.1}
The leptonic helicity amplitudes for the left and right-handed neutrinos corresponding to $W^{*-} \to \mu^- \bar \nu_{\mu}$ are given as follows

%%%%%%%%%%%%%%%%%%%%%%%%%%%%%%%%%%%%%%%%%%%%%%%%%%%%%%%%%%%%%%%%%%%%%
\subsubsection{Left-handed neutrinos} \label{LHN_hel_amp}
%%%%%%%%%%%%%%%%%%%%%%%%%%%%%%%%%%%%%%%%%%%%%%%%%%%%%%%%%%%%%%%%%%%%%%%%%%

\begin{equation}
\label{lhn_amp}
\begin{aligned}
L_{-\tfrac12, -}^{L, \nu_L} &= 2\sqrt{2}\,\sqrt{q^2 - m_\mu^2}, &
L_{-\tfrac12, +}^{L, \nu_L} &= 0, \\
L_{-\tfrac12, 0}^{L, \nu_L} &= 0, &
L_{-\tfrac12, t}^{L, \nu_L} &= 0, \\
L_{+\tfrac12, 0}^{L, \nu_L} &= 2\,m_\mu\,\frac{\sqrt{q^2  - m_\mu^2}}{\sqrt{q^2 }}, &
L_{+\tfrac12, t}^{L, \nu_L} &= -\,2\,m_\mu\,\frac{\sqrt{q^2  - m_\mu^2}}{\sqrt{q^2 }}, \\
L_{+\tfrac12, +}^{L, \nu_L} &= 0, &
L_{+\tfrac12, -}^{L, \nu_L} &= 0 \\
L^{SPL, \nu_L}_{-\tfrac12} &= 0, & L^{SPL, \nu_L}_{+\tfrac12} &= 2\,\sqrt{q^2  - m_\mu^2}
\end{aligned}
\end{equation}

%%%%%%%%%%%%%%%%%%%%%%%%%%%%%%%%%%%%%%%%%%%%%%%%%%%%%%%%%%%%
\subsubsection{Right-handed neutrinos} \label{RHN_hel_amp}
%%%%%%%%%%%%%%%%%%%%%%%%%%%%%%%%%%%%%%%%%%%%%%%%%%%%%%%%%%%%

\begin{equation}
\label{lhn_amp}
\begin{aligned}
L_{-\tfrac12, -}^{R, \nu_R} &= 0, &
L_{-\tfrac12, +}^{R, \nu_R} &= 0, \\
L_{-\tfrac12, 0}^{R, \nu_R} &= 2  \, m_{\mu}\,\frac{\sqrt{q^2  - m_{\mu}^2}}{\sqrt{q^2 }}, &
L_{-\tfrac12, t}^{R, \nu_R} &= - \, 2\,m_\mu\,\frac{\sqrt{q^2  - m_\mu^2}}{\sqrt{q^2}}, \\
L_{+\tfrac12, 0}^{R, \nu_R} &= 0, &
L_{+\tfrac12, t}^{R, \nu_R} &= 0, \\
L_{+\tfrac12, +}^{R, \nu_R} &= 2\,\sqrt{2} \,\sqrt{q^2  - m_\mu^2}, &
L_{+\tfrac12, -}^{R, \nu_R} &= 0 \\
L^{SPR, \nu_R}_{-\tfrac12} &= 2\,\sqrt{q^2  - m_\mu^2}, & L^{SPR, \nu_R}_{+\tfrac12} &= 0
\end{aligned}
\end{equation}

%%%%%%%%%%%%%%%%%%%%%%%%%%%%%%%%%%%%%%%%%%%%%%%%%%%%%%%%%%%%%%
\subsection{Hadronic Helcity Amplitude} \label{app-A.2}
%%%%%%%%%%%%%%%%%%%%%%%%%%%%%%%%%%%%%%%%%%%%%%%%%%%%%%%%%%%%%%%%%%%%%%%
\subsubsection{$\Lambda_c^- \to \Lambda \mu^- \bar \nu_{\mu}$ Decay}
\label{HA_3body}
%%%%%%%%%%%%%%%%%%%%%%%%%%%%%%%%%%%%%%%%%%%%%%%%%%%%%%%%%%%%%%%%%%%%%%%%%%%%%
The hadronic helicity amplitudes for three-body decay are given in the following expressions:

\begin{widetext}
\begin{align}
    H_{\frac12, t}^L = & f_0(q^2) M_{-} \sqrt{\frac{Q_{+}}{q^2}} - g_0(q^2) M_{+} \sqrt{\frac{Q_{-}}{q^2}}, ~\label{HL1} \\
    H_{\frac12, +}^L = & - f_{\perp}(q^2) \sqrt{2 Q_{-}} + g_{\perp}(q^2) \sqrt{2 Q_{+}}, ~ \label{HL2} \\
   H_{\frac12, 0}^L = & f_{+}(q^2) M_{+} \sqrt{\frac{Q_{-}}{q^2}} - g_{+}(q^2) M_{-} \sqrt{\frac{Q_{+}}{q^2}}, ~  \label{HL3} \\
   H_{-\frac12, t}^L = & f_0(q^2) M_{-} \sqrt{\frac{Q_{+}}{q^2}} + g_0(q^2) M_{+} \sqrt{\frac{Q_{-}}{q^2}} , ~  \label{HL4} \\
   H_{-\frac12, -}^L = & - f_{\perp}(q^2) \sqrt{2 Q_{-}} - g_{\perp}(q^2) \sqrt{2 Q_{+}} , ~ \label{HL5} \\
   H_{-\frac12, 0}^L = & f_{+}(q^2) M_{+} \sqrt{\frac{Q_{-}}{q^2}} + g_{+}(q^2) M_{-} \sqrt{\frac{Q_{+}}{q^2}} , ~ \label{HL6}
\end{align}
\end{widetext}

\begin{widetext}
\begin{align}
 H_{\frac12, t}^R = & f_0(q^2) M_{-} \sqrt{\frac{Q_{+}}{q^2}} + g_0(q^2) M_{+} \sqrt{\frac{Q_{-}}{q^2}}  , ~ \label{HR1}   \\
 H_{\frac12, +}^R = & f_{\perp}(q^2) \sqrt{2 Q_{-}} - g_{\perp}(q^2) \sqrt{2 Q_{+}} , ~  \label{HR2} \\
 H_{\frac12, 0}^R = & f_{+}(q^2) M_{+} \sqrt{\frac{Q_{-}}{q^2}} + g_{+}(q^2) M_{-} \sqrt{\frac{Q_{+}}{q^2}} , ~   \label{HR3} \\
 H_{-\frac12, t}^R = & f_0(q^2) M_{-} \sqrt{\frac{Q_{+}}{q^2}} - g_0(q^2) M_{+} \sqrt{\frac{Q_{-}}{q^2}} , ~  \label{HR4} \\
 H_{-\frac12, -}^R = & f_{\perp}(q^2) \sqrt{2 Q_{-}} + g_{\perp}(q^2) \sqrt{2 Q_{+}} , ~ \label{HR5} \\
 H_{-\frac12, 0}^R = & f_{+}(q^2) M_{+} \sqrt{\frac{Q_{-}}{q^2}} - g_{+}(q^2) M_{-} \sqrt{\frac{Q_{+}}{q^2}} , ~ \label{HR6}
\end{align}
\end{widetext}

\begin{widetext}
\begin{align}
H_{\frac12}^{SPL} = & f_0(q^2) \frac{M_{-}}{m_c - m_s} \sqrt{Q_{+}} 
+ g_0(q^2) \frac{M_{+}}{m_c + m_s} \sqrt{Q_{-}}  , ~ \label{HSP1} \\
 H_{-\frac12}^{SPL} = & f_0(q^2) \frac{M_{-}}{m_c - m_s} \sqrt{Q_{+}} 
- g_0(q^2) \frac{M_{+}}{m_c + m_s} \sqrt{Q_{-}} , ~  \label{HSP2}   \\
H_{\frac12}^{SPR} = & f_0(q^2) \frac{M_{-}}{m_c - m_s} \sqrt{Q_{+}} 
- g_0(q^2) \frac{M_{+}}{m_c + m_s} \sqrt{Q_{-}} , ~ \label{HSP3} \\
H_{-\frac12}^{SPR} = & f_0(q^2) \frac{M_{-}}{m_c - m_s} \sqrt{Q_{+}} 
+ g_0(q^2) \frac{M_{+}}{m_c + m_s} \sqrt{Q_{-}}  ~ \label{HSP4}
\end{align}    
\end{widetext}

%%%%%%%%%%%%%%%%%%%%%%%%%%%%%%%%%%%%%%%%%%%%%%%%%
\subsubsection{$\Lambda_c^- \to \Lambda (p \pi) \mu^- \bar \nu_{\mu}$ Decay} \label{HA_4body}
%%%%%%%%%%%%%%%%%%%%%%%%%%%%%%%%%%%%%%%%%%%%%%%
The Hadronic helicity amplitudes for the four-body decay of a left-handed neutrino ($\nu_L$) and right-handed neutrino ($\nu_R$) are given as follows:

\begin{align}
   H_{\frac12, t}^{VA, \nu_L} = & (1 + C^V_{LL} + C^V_{RL}) f_0(q^2) M_{-} \sqrt{\frac{Q_{+}}{q^2}} - (1 + C^V_{LL} - C^V_{RL}) g_0(q^2) M_{+} \sqrt{\frac{Q_{-}}{q^2}}  , ~ \label{HVA1nuL} \\
   H_{\frac12, 1}^{VA, \nu_L} = & - (1 + C^V_{LL} + C^V_{RL}) f_{\perp}(q^2) \sqrt{2 Q_{-}} + (1 + C^V_{LL} - C^V_{RL}) g_{\perp}(q^2) \sqrt{2 Q_{+}} , ~  \label{HVA2nuL} \\
   H_{\frac12, 0}^{VA, \nu_L} = & (1 + C^V_{LL} + C^V_{RL}) f_{+}(q^2) M_{+} \sqrt{\frac{Q_{-}}{q^2}} - (1 + C^V_{LL} - C^V_{RL}) g_{+}(q^2) M_{-} \sqrt{\frac{Q_{+}}{q^2}}  , ~ \label{HVA3nuL} \\
   H_{-\frac12, t}^{VA, \nu_L} = & (1 + C^V_{LL} + C^V_{RL}) f_0(q^2) M_{-} \sqrt{\frac{Q_{+}}{q^2}} + (1 + C^V_{LL} - C^V_{RL}) g_0(q^2) M_{+} \sqrt{\frac{Q_{-}}{q^2}} , ~   \label{HVA4nuL} \\
   H_{-\frac12, -1}^{VA, \nu_L} = & - (1 + C^V_{LL} + C^V_{RL}) f_{\perp}(q^2) \sqrt{2 Q_{-}} - (1 + C^V_{LL} - C^V_{RL}) g_{\perp}(q^2) \sqrt{2 Q_{+}} , ~ \label{HVA5nuL} \\
   H_{-\frac12, 0}^{VA, \nu_L} = & (1 + C^V_{LL} + C^V_{RL}) f_{+}(q^2) M_{+} \sqrt{\frac{Q_{-}}{q^2}} + (1 + C^V_{LL} - C^V_{RL}) g_{+}(q^2) M_{-} \sqrt{\frac{Q_{+}}{q^2}}   , ~  \label{HVA6nuL}  \\
   H_{\frac12, t}^{SP, \nu_L} = & (C^S_{LL} + C^S_{RL}) f_0(q^2) \frac{M_{-}}{m_c - m_s} \sqrt{Q_{+}} 
- (C^S_{LL} - C^S_{RL}) g_0(q^2) \frac{M_{+}}{m_c + m_s} \sqrt{Q_{-}}   , ~  \label{HSPanuL} \\
 H_{-\frac12, t}^{SP, \nu_L} = & (C^S_{LL} + C^S_{RL}) f_0(q^2) \frac{M_{-}}{m_c - m_s} \sqrt{Q_{+}} 
+ (C^S_{LL} - C^S_{RL}) g_0(q^2) \frac{M_{+}}{m_c + m_s} \sqrt{Q_{-}} , ~ \label{HSPbnuL}
\end{align}

%%%%%%%%%%%%%%%%

\begin{align}
   H_{\frac12, t}^{VA, \nu_R} = & (C^V_{LR} + C^V_{RR}) f_0(q^2) M_{-} \sqrt{\frac{Q_{+}}{q^2}} - (C^V_{LR} - C^V_{RR}) g_0(q^2) M_{+} \sqrt{\frac{Q_{-}}{q^2}}  , ~ \label{HVA1nuR} \\
   H_{\frac12, 1}^{VA, \nu_R} = & - (C^V_{LR} + C^V_{RR}) f_{\perp}(q^2) \sqrt{2 Q_{-}} + (C^V_{LR} - C^V_{RR}) g_{\perp}(q^2) \sqrt{2 Q_{+}} , ~  \label{HVA2nuR} \\
   H_{\frac12, 0}^{VA, \nu_R} = & (C^V_{LR} + C^V_{RR}) f_{+}(q^2) M_{+} \sqrt{\frac{Q_{-}}{q^2}} - (C^V_{LR} - C^V_{RR}) g_{+}(q^2) M_{-} \sqrt{\frac{Q_{+}}{q^2}} , ~  \label{HVA3nuR} \\
   H_{-\frac12, t}^{VA, \nu_R} = & (C^V_{LR} + C^V_{RR}) f_0(q^2) M_{-} \sqrt{\frac{Q_{+}}{q^2}} + (C^V_{LR} - C^V_{RR}) g_0(q^2) M_{+} \sqrt{\frac{Q_{-}}{q^2}} , ~    \label{HVA4nuR} \\
   H_{-\frac12, -1}^{VA, \nu_R} = & - (C^V_{LR} + C^V_{RR}) f_{\perp}(q^2) \sqrt{2 Q_{-}} - (C^V_{LR} - C^V_{RR}) g_{\perp}(q^2) \sqrt{2 Q_{+}} ,~ \label{HVA5nuR} \\
   H_{-\frac12, 0}^{VA, \nu_R} = & (C^V_{LR} + C^V_{RR}) f_{+}(q^2) M_{+} \sqrt{\frac{Q_{-}}{q^2}} + (C^V_{LR} - C^V_{RR}) g_{+}(q^2) M_{-} \sqrt{\frac{Q_{+}}{q^2}} , ~ \label{HVA6nuR} \\
   H_{\frac12, t}^{SP, \nu_R} = & (C^S_{LR} + C^S_{RR}) f_0(q^2) \frac{M_{-}}{m_c - m_s} \sqrt{Q_{+}} 
   - (C^S_{LR} - C^S_{RR}) g_0(q^2) \frac{M_{+}}{m_c + m_s} \sqrt{Q_{-}} , ~  \label{HSPanuR} \\
  H_{-\frac12, t}^{SP, \nu_R} = & (C^S_{LR} + C^S_{RR}) f_0(q^2) \frac{M_{-}}{m_c - m_s} \sqrt{Q_{+}} 
  + (C^S_{LR} - C^S_{RR}) g_0(q^2) \frac{M_{+}}{m_c + m_s} \sqrt{Q_{-}}  ~ \label{HSPbnuR}
\end{align}

%%%%%%%%%%%%%%%%%%%%%%%%%%%%%%%%%
\section{Functions of total amplitude for $\Lambda_c^- \to \Lambda \mu^- \nu_{\mu}$ Decay} \label{app-B}
%%%%%%%%%%%%%%%%%%%%%%%%%%%%%%%%%%%%%%%%
The functions of total amplitude for left-handed ($\nu_L$) and right-handed neutrinos ($\nu_R$) are given as follows:
%%%%%%%%%%%%%%%%%%%%%%%%%%%%%%%%%%%%%%%%%%%%%%%%%%%%%
\subsection{Left-handed Neutrino Functions} \label{lhn_fun}
%%%%%%%%%%%%%%%%%%%%%%%%%%%%%%%%%%%%%%%%%%%%%%%%%

\begin{align}
\mathcal{A}_{V L}^{\nu_L}= & 2 \sin ^2 \theta\left(\left|H_{\frac{1}{2}, 0}^L\right|^2+\left|H_{-\frac{1}{2}, 0}^L\right|^2\right)+(1-\cos \theta)^2\left|H_{\frac{1}{2},+}^L\right|^2+(1+\cos \theta)^2\left|H_{-\frac{1}{2},-}^L\right|^2  \nonumber \\ & +\frac{m_{\mu}^2}{q^2}\left[2 \cos ^2 \theta\left(\left|H_{\frac{1}{2}, 0}^L\right|^2+\left|H_{-\frac{1}{2}, 0}^L\right|^2\right)+\sin ^2 \theta\left(\left|H_{\frac{1}{2},+}^L\right|^2+\left|H_{-\frac{1}{2},-}^L\right|^2\right)\right. \nonumber \\ & 
    \left.+2\left(\left|H_{\frac{1}{2}, t}^L\right|^2+\left|H_{-\frac{1}{2}, t}^L\right|^2\right)-4 \cos \theta \Re \left(H_{\frac{1}{2}, t}^{L*} H_{\frac{1}{2}, 0}^L+H_{-\frac{1}{2}, t}^{L*} H_{-\frac{1}{2}, 0}^L\right)\right], ~ \label{MVLnuL} \\
    \mathcal{A}_{V R}^{\nu_L}= & 2 \sin ^2 \theta\left(\left|H_{\frac{1}{2}, 0}^R\right|^2+\left|H_{-\frac{1}{2}, 0}^R\right|^2\right)+(1-\cos \theta)^2\left|H_{\frac{1}{2},+}^R\right|^2+(1+\cos \theta)^2\left|H_{-\frac{1}{2},-}^R\right|^2 \nonumber \\ & +\frac{m_{\mu}^2}{q^2}\left[2 \cos ^2 \theta\left(\left|H_{\frac{1}{2}, 0}^R\right|^2+\left|H_{-\frac{1}{2}, 0}^R\right|^2\right)+\sin ^2 \theta\left(\left|H_{\frac{1}{2},+}^R\right|^2+\left|H_{-\frac{1}{2},-}^R\right|^2\right)\right. \nonumber \\ & 
    \left.+2\left(\left|H_{\frac{1}{2}, t}^R\right|^2+\left|H_{-\frac{1}{2}, t}^R\right|^2\right)-4 \cos \theta \Re \left(H_{\frac{1}{2}, t}^{R*} H_{\frac{1}{2}, 0}^R+H_{-\frac{1}{2}, t}^{R*} H_{-\frac{1}{2}, 0}^R\right)\right], ~ \label{MVRnuL} \\
     \mathcal{A}_{S L}^{\nu_L}= & 2\left(\left|H_{-\frac{1}{2}}^{S P L}\right|^2+\left|H_{\frac{1}{2}}^{S P L}\right|^2\right) \quad \quad
  \mathcal{A}_{S R}^{\nu_L} =  2\left(\left|H_{-\frac{1}{2}}^{S P R}\right|^2+\left|H_{\frac{1}{2}}^{S P R}\right|^2\right), ~ \label{MSLRnuL} \\
  \mathcal{A}_{VL,VR}^{\nu_L, \mathrm{int}} =\; 
& 2 \sin^2\theta \left(H_{-\frac{1}{2},0}^L H_{-\frac{1}{2},0}^R + H_{\frac{1}{2},0}^L H_{\frac{1}{2},0}^R \right) 
+ (1 - \cos\theta)^2 H_{\frac{1}{2},+}^L H_{\frac{1}{2},+}^R  \nonumber \\&
+ (1 + \cos\theta)^2 H_{-\frac{1}{2},-}^L H_{-\frac{1}{2},-}^R 
 + \frac{m_\mu^2}{q^2} \Big[ 
    2 \left( H_{-\frac{1}{2},t}^L H_{-\frac{1}{2},t}^R + H_{\frac{1}{2},t}^L H_{\frac{1}{2},t}^R \right) \nonumber \\& + 2 \sin^2\theta \left( H_{\frac{1}{2},+}^L H_{\frac{1}{2},+}^R + H_{-\frac{1}{2},-}^L H_{-\frac{1}{2},-}^R \right)
  + 2 \cos^2\theta \left( H_{-\frac{1}{2},0}^L H_{-\frac{1}{2},0}^R + H_{\frac{1}{2},0}^L H_{\frac{1}{2},0}^R \right) \nonumber \nonumber \\ &
  - 2 \cos\theta \left( H_{-\frac{1}{2},0}^L H_{-\frac{1}{2},t}^R + H_{\frac{1}{2},0}^L H_{\frac{1}{2},t}^R \right. \left. 
  + H_{-\frac{1}{2},t}^L H_{-\frac{1}{2},0}^R + H_{\frac{1}{2},t}^L H_{\frac{1}{2},0}^R \right) \Big], ~ \label{MVLVRintNUl} \\
   \mathcal{A}_{V L, S L}^{\nu_L, \mathrm{int}} = & \left(-\frac{2 m_{\mu}}{\sqrt{q^2}}\right)\left[H_{-\frac{1}{2}, t}^L H_{-\frac{1}{2}}^{S P L}+H_{\frac{1}{2}, t}^L H_{\frac{1}{2}}^{S P L}-\cos \theta\left(H_{-\frac{1}{2}, 0}^L H_{-\frac{1}{2}}^{S P L}+H_{\frac{1}{2}, 0}^L H_{\frac{1}{2}}^{S P L}\right)\right], ~ \label{MVLSLnuLint} \\
   \mathcal{A}_{V L, S R}^{\nu_L, \mathrm{int}} = & \left(-\frac{2 m_{\mu}}{\sqrt{q^2}}\right)\left[H_{-\frac{1}{2}, t}^L H_{-\frac{1}{2}}^{S P R}+H_{\frac{1}{2}, t}^L H_{\frac{1}{2}}^{S P R}-\cos \theta\left(H_{-\frac{1}{2}, 0}^L H_{-\frac{1}{2}}^{S P R}+H_{\frac{1}{2}, 0}^L H_{\frac{1}{2}}^{S P R}\right)\right] , ~ \label{MVLSRnuLint} \\
    \mathcal{A}_{V R, S L}^{\nu_L, \mathrm{int}}= & \left(-\frac{2 m_{\mu}}{\sqrt{q^2}}\right)\left[H_{-\frac{1}{2}, t}^R H_{-\frac{1}{2}}^{S P L}+H_{\frac{1}{2}, t}^R H_{\frac{1}{2}}^{S P L} - \cos \theta\left(H_{-\frac{1}{2}, 0}^R H_{-\frac{1}{2}}^{S P L}+H_{\frac{1}{2}, 0}^R H_{\frac{1}{2}}^{S P L}\right)\right] , ~ \label{MVRSLnuL}    \\
     \mathcal{A}_{V R, S R}^{\nu_L, \mathrm{int}}= & \left(-\frac{2 m_{\mu}}{\sqrt{q^2}}\right)\left[H_{-\frac{1}{2}, t}^R H_{-\frac{1}{2}}^{S P R}+H_{\frac{1}{2}, t}^R H_{\frac{1}{2}}^{S P R} - \cos \theta\left(H_{-\frac{1}{2}, 0}^R H_{-\frac{1}{2}}^{S P R}+H_{\frac{1}{2}, 0}^R H_{\frac{1}{2}}^{S P R}\right)\right] , ~ \label{MVRSRnuL} \\
     \mathcal{A}_{S L, S R}^{\nu_L, \mathrm{int}}= & 2\left(H_{-\frac{1}{2}}^{S P L} H_{-\frac{1}{2}}^{S P R}+H_{\frac{1}{2}}^{S P L} H_{\frac{1}{2}}^{S P R}\right) ~ \label{MSLSRnuLint}   
\end{align}

where hadronic helicity amplitudes are given in the Appendix in \ref{app-A.2}.
%%%%%%%%%%%%%%%%%%%%%%%%%%%%%%%%%%%%%%%%%%%%
\subsection{Right-handed Neutrino Functions} \label{rhn_fun}
%%%%%%%%%%%%%%%%%%%%%%%%%%%%%%%%%%%%%%%%%%%%%%%%%%
\begin{align}
   \mathcal{A}_{V L}^{\nu_R}= & 2 \sin ^2 \theta\left(\left|H_{\frac{1}{2}, 0}^L\right|^2+\left|H_{-\frac{1}{2}, 0}^L\right|^2\right)+(1+\cos \theta)^2\left|H_{\frac{1}{2},+}^L\right|^2+(1-\cos \theta)^2\left|H_{-\frac{1}{2},-}^L\right|^2 \nonumber \\ & +\frac{m_{\mu}^2}{q^2}\left[2 \cos ^2 \theta\left(\left|H_{\frac{1}{2}, 0}^L\right|^2+\left|H_{-\frac{1}{2}, 0}^L\right|^2\right)+\sin ^2 \theta\left(\left|H_{\frac{1}{2},+}^L\right|^2+\left|H_{-\frac{1}{2},-}^L\right|^2\right)\right. \nonumber \\ & 
    \left.+2\left(\left|H_{\frac{1}{2}, t}^L\right|^2+\left|H_{-\frac{1}{2}, t}^L\right|^2\right)-4 \cos \theta \Re \left(H_{\frac{1}{2}, t}^{L*} H_{\frac{1}{2}, 0}^L+H_{-\frac{1}{2}, t}^{L*} H_{-\frac{1}{2}, 0}^L\right)\right], ~ \label{MVLnuR} \\
    \mathcal{A}_{V R}^{\nu_R} = & 2 \sin ^2 \theta\left(\left|H_{\frac{1}{2}, 0}^R\right|^2+\left|H_{-\frac{1}{2}, 0}^R\right|^2\right)+(1+\cos \theta)^2\left|H_{\frac{1}{2},+}^R\right|^2+(1-\cos \theta)^2\left|H_{-\frac{1}{2},-}^R\right|^2 \nonumber \\ & +\frac{m_{\mu}^2}{q^2}\left[2 \cos ^2 \theta\left(\left|H_{\frac{1}{2}, 0}^R\right|^2+\left|H_{-\frac{1}{2}, 0}^R\right|^2\right)+\sin ^2 \theta\left(\left|H_{\frac{1}{2},+}^R\right|^2+\left|H_{-\frac{1}{2},-}^R\right|^2\right)\right. \nonumber \\ & \left.+ 2\left(\left|H_{\frac{1}{2}, t}^R\right|^2+\left|H_{-\frac{1}{2}, t}^R\right|^2\right)-4 \cos \theta \Re \left(H_{\frac{1}{2}, t}^{R*} H_{\frac{1}{2}, 0}^R+H_{-\frac{1}{2}, t}^{R*} H_{-\frac{1}{2}, 0}^R\right)\right]  , ~ \label{MVRnuR} \\
    \mathcal{A}_{S L}^{\nu_R}= & 2\left(\left|H_{-\frac{1}{2}}^{S P L}\right|^2+\left|H_{\frac{1}{2}}^{S P L}\right|^2\right) \quad \quad
  \mathcal{A}_{S R}^{\nu_R}=  2\left(\left|H_{-\frac{1}{2}}^{S P R}\right|^2+\left|H_{\frac{1}{2}}^{S P R}\right|^2\right)  , ~ \label{MSLRnuR} \\
  \mathcal{A}_{VL,VR}^{\nu_R, \mathrm{int}} =\; 
& 2 \sin^2\theta \left(H_{-\frac{1}{2},0}^L H_{-\frac{1}{2},0}^R + H_{\frac{1}{2},0}^L H_{\frac{1}{2},0}^R \right) 
+ (1 + \cos\theta)^2 H_{\frac{1}{2},+}^L H_{\frac{1}{2},+}^R \nonumber \\ &
+ (1 - \cos\theta)^2 H_{-\frac{1}{2},-}^L H_{-\frac{1}{2},-}^R + \frac{m_\mu^2}{q^2} \Big[ 
    2 \left( H_{-\frac{1}{2},t}^L H_{-\frac{1}{2},t}^R + H_{\frac{1}{2},t}^L H_{\frac{1}{2},t}^R \right) \nonumber 
\\ &  + 2 \sin^2\theta \left( H_{\frac{1}{2},+}^L H_{\frac{1}{2},+}^R + H_{-\frac{1}{2},-}^L H_{-\frac{1}{2},-}^R \right) 
  + 2 \cos^2\theta \left( H_{-\frac{1}{2},0}^L H_{-\frac{1}{2},0}^R + H_{\frac{1}{2},0}^L H_{\frac{1}{2},0}^R \right) \nonumber
\\ &  - 2 \cos\theta \left( H_{-\frac{1}{2},0}^L H_{-\frac{1}{2},t}^R + H_{\frac{1}{2},0}^L H_{\frac{1}{2},t}^R \right. \left. 
  + H_{-\frac{1}{2},t}^L H_{-\frac{1}{2},0}^R + H_{\frac{1}{2},t}^L H_{\frac{1}{2},0}^R 
  \right) 
\Big] , ~ \label{MVLVRintnuR} \\
  \mathcal{A}_{V L, S L}^{\nu_R, \mathrm{int}} = & \left(-\frac{2 m_{\mu}}{\sqrt{q^2}}\right)\left[H_{-\frac{1}{2}, t}^L H_{-\frac{1}{2}}^{S P L}+H_{\frac{1}{2}, t}^L H_{\frac{1}{2}}^{S P L}-\cos \theta\left(H_{-\frac{1}{2}, 0}^L H_{-\frac{1}{2}}^{S P L}+H_{\frac{1}{2}, 0}^L H_{\frac{1}{2}}^{S P L}\right)\right] , ~ \label{MVLSLintnuR} \\
    \mathcal{A}_{V L, S R}^{\nu_R, \mathrm{int}}= & \left(-\frac{2 m_{\mu}}{\sqrt{q^2}}\right)\left[H_{-\frac{1}{2}, t}^L H_{-\frac{1}{2}}^{S P R}+H_{\frac{1}{2}, t}^L H_{\frac{1}{2}}^{S P R}-\cos \theta\left(H_{-\frac{1}{2}, 0}^L H_{-\frac{1}{2}}^{S P R}+H_{\frac{1}{2}, 0}^L H_{\frac{1}{2}}^{S P R}\right)\right] , ~ \label{MVLSRintnuR}   \\
         \mathcal{A}_{V R, S R}^{\nu_R, \mathrm{int}}= & \left(-\frac{2 m_{\mu}}{\sqrt{q^2}}\right)\left[H_{-\frac{1}{2}, t}^R H_{-\frac{1}{2}}^{S P R}+H_{\frac{1}{2}, t}^R H_{\frac{1}{2}}^{S P R}-\cos \theta\left(H_{-\frac{1}{2}, 0}^R H_{-\frac{1}{2}}^{S P R}+H_{\frac{1}{2}, 0}^R H_{\frac{1}{2}}^{S P R}\right)\right], ~ \label{MVRSRnuR} \\
         \mathcal{A}_{V R, S L}^{\nu_R, \mathrm{int}}= & \left(-\frac{2 m_{\mu}}{\sqrt{q^2}}\right)\left[H_{-\frac{1}{2}, t}^R H_{-\frac{1}{2}}^{S P L}+H_{\frac{1}{2}, t}^R H_{\frac{1}{2}}^{S P L}-\cos \theta\left(H_{-\frac{1}{2}, 0}^R H_{-\frac{1}{2}}^{S P L}+H_{\frac{1}{2}, 0}^R H_{\frac{1}{2}}^{S P L}\right)\right], ~ \label{MVRSLnuR} \\
          \mathcal{A}_{S L, S R}^{\nu_R, \mathrm{int}}= & 2\left(H_{-\frac{1}{2}}^{S P L} H_{-\frac{1}{2}}^{S P R}+H_{\frac{1}{2}}^{S P L} H_{\frac{1}{2}}^{S P R}\right) ~ \label{MSLSRintnuR}
\end{align}

Similar to LHN, the hadronic helicity amplitudes used in RHN are given in the Appendix \ref{app-A.2}.

%%%%%%%%%%%%%%%%%%%%%%%%%%%%%%%%%%%%%%%%%%%%%%%%%%%%%%%%%%%%%%%
\section{Kinematics of Decay Process} \label{app-kin}
%%%%%%%%%%%%%%%%%%%%%%%%%%%%%%%%%%%%%%%%%%%%%%%%%%%%%%%%%%%%%%%
The kinematics of the decay is given as follows:

\begin{equation*}
\begin{tikzpicture}[baseline=(current bounding box.center)]
    % 1. The Main Reaction Line
    \node (start) at (0,0) {$\Lambda_c^- (p_{\Lambda_c})$};
    \node[right=0.2cm of start] (arrow) {$\longrightarrow$};
    \node[right=0.2cm of arrow] (lambda) {$\Lambda (p_{\Lambda})$};
    \node[right=0.5cm of lambda] (plus) {$+$};
    \node[right=0.7cm of plus] (W) {$W^*(q)$};

    % 2. The Secondary Decay (Lambda -> p pi)
    % "draw" command creates the L-shape:
    % (start point) -- (go down) -- (go right) node {text}
    \draw[->, >=stealth] (lambda.south) -- ++(0,-0.8) -- ++(0.5,0) 
        node[right] {$p(p_p) \; \pi(p_{\pi})$};

    % 3. The Secondary Decay (W -> mu nu)
    \draw[->, >=stealth] (W.south) -- ++(0,-0.8) -- ++(0.5,0) 
        node[right] {$\mu^-(p_\mu) \nu_\mu(p_\nu)$};
        
\end{tikzpicture}
\end{equation*}

In the rest frame of the $\Lambda_c$ \\

\begin{align}
p_{\Lambda_c} = (m_{\Lambda_c}, 0, 0, 0),  \quad
p_\Lambda = (E_\Lambda, 0, 0, |\vec{p_\Lambda}|) , \quad
q = (q_0, 0, 0, - |\vec{p_\Lambda}|) 
\end{align}

where

\begin{align}
E_\Lambda  =\frac{1}{2 m_{\Lambda_c}}\left(m_{\Lambda_c}^2 + m_{\Lambda}^2 - q^2\right), \quad
q_0  = \frac{1}{2 m_{\Lambda_c}}\left(m_{\Lambda_c}^2-m_{\Lambda}^2+q^2\right) ,\quad
|q|  =\left|\vec{p_{\Lambda}}\right|=\frac{1}{2 m_{\Lambda_c}} \sqrt{Q_{+} Q_{-}}
\end{align}

and
$$
Q_{ \pm}=\left(m_{\Lambda_c} \pm m_{\Lambda}\right)^2-q^2 .
$$

In the rest frame of the $\Lambda$ momentum of the p and $\pi$ are 

\begin{align}
p_p = (E_p, \, |\vec{p}_p| sin \theta_2 cos \phi_2, \, |\vec{p}_p| sin \theta_2 sin \phi_2, \, |\vec{p}_p| cos \theta_2) \\
p_\pi = (E_\pi, \, - |\vec{p}_p| sin \theta_2 cos \phi_2, \, - |\vec{p}_p| sin \theta_2 sin \phi_2, \, - |\vec{p}_p| cos \theta_2)
\end{align}

where 

\begin{align}
E_p =\frac{1}{2 m_{\Lambda}}\left(m_p^2 + m_{\pi}^2 - q^2\right), \quad
E_\pi =\frac{1}{2 m_{\Lambda}}\left(m_p^2 - m_{\pi}^2 + q^2\right), \quad
|\vec{p_p}| =\frac{1}{2 m_{\Lambda}} \sqrt{Q_{+}^\prime Q_{-}^\prime}
\end{align}

and
$$
Q_{\pm}^\prime =\left(m_{\Lambda} \pm m_{p}\right)^2-p_\pi^2 
$$

Lepton and anti-neutrino momenta in the $W^*$ rest frame:

\begin{align}
p_\mu &= (E_\mu, |\vec{p_\mu}| \, sin \, \theta_3 \, cos \phi_3, \, |\vec{p_\mu}| \, sin \, \theta_3 \, sin \phi_3, \, |\vec{p_\mu}| \, cos \, \theta_3), \\
p_{\bar\nu} &= (|\vec{p_\mu}|, \, - |\vec{p_\mu}| \, sin \, \theta_3 \, cos \phi_3, \, - |\vec{p_\mu}| \, sin \, \theta_3 \, sin \phi_3, \, - |\vec{p_\mu}| \, cos \, \theta_3)
\end{align}

where
\begin{align}
E_\mu &= \frac{q^2 + m_\mu^2}{2\sqrt{q^2}}, &
|\vec{p_\mu}| &= \frac{q^2 - m_\mu^2}{2\sqrt{q^2}}
\end{align}

%%%%%%%%%%%%%%%%%%%%%%%%%%%%%%%%%
\section{Angular Observables of $\Lambda_c^- \to \Lambda (p\pi)\,\mu^- \nu_{\mu}$ Decay} \label{app-C}
%%%%%%%%%%%%%%%%%%%%%%%%%%%%%%%%%%%%%%%%
The angular observables of four-body decay, considering left-handed ($\nu_L$) \cite{Karmakar:2023rdt} and right-handed neutrinos ($\nu_R$), are as follows 

%%%%%%%%%%%%%%%%%%%%%%%%%%%%%%%%%%%%%%%%%
\subsection{Angular observables with Left-Handed Neutrino} \label{4body_ang_obs_lhn}
%%%%%%%%%%%%%%%%%%%%%%%%%%%%%%%%%%%%%%%%%%%%

\begin{align}
\mathcal{M}_0^{\nu_L}  = &  \frac{1}{\Gamma_{0}^{\nu_L}}\left(\frac{1}{2}\left(m_\mu^2+q^2\right)\left|H_{\frac{1}{2}, 1}^{V A, \nu_L}\right|^2+\frac{1}{2}\left(m_\mu^2+q^2\right)\left|H_{-\frac{1}{2},-1}^{V A, \nu_L}\right|^2+ \left|m_\mu H_{\frac{1}{2}, t}^{V A, \nu_L}+\sqrt{q^2} H_{\frac{1}{2}, t}^{S P, \nu_L}\right|^2\right. \nonumber\\
& \left.+\left|m_\mu H_{-\frac{1}{2}, t}^{V A, \nu_L}+\sqrt{q^2} H_{-\frac{1}{2}, t}^{S P, \nu_L}\right|^2+q^2\left|H_{\frac{1}{2}, 0}^{V A, \nu_L}\right|^2+q^2\left|H_{-\frac{1}{2}, 0}^{V A, \nu_L}\right|^2\right),~\label{M0nuL} \\
 \mathcal{M}_1^{\nu_L}  = & \frac{\alpha_P}{2 \Gamma_{0}^{\nu_L}}\left(\left(m_\mu^2+q^2\right)\left|H_{\frac{1}{2}, 1}^{V A, \nu_L}\right|^2+\left(m_\mu^2+q^2\right)\left|H_{-\frac{1}{2},-1}^{V A, \nu_L}\right|^2-2\left|m_\mu H_{\frac{1}{2}, t}^{V A, \nu_L}+\sqrt{q^2} H_{\frac{1}{2}, t}^{S P, \nu_L}\right|^2\right. \nonumber \\
& \left.+2\left|m_\mu H_{-\frac{1}{2}, t}^{V A, \nu_L}+\sqrt{q^2} H_{-\frac{1}{2}, t}^{S P, \nu_L}\right|^2-2 q^2\left|H_{\frac{1}{2}, 0}^{V A, \nu_L}\right|^2+2 q^2\left|H_{-\frac{1}{2}, 0}^{V A, \nu_L}\right|^2\right),~\label{M1nuL} \\
 \mathcal{M}_2^{\nu_L}
= & -\frac{1}{\Gamma_{0}^{\nu_L}}\Bigg(
   q^2\left|H_{\tfrac{1}{2},1}^{V A,\nu_L}\right|^2
 + q^2\left|H_{-\tfrac{1}{2},-1}^{V A,\nu_L}\right|^2
 - 2 \operatorname{\Re}\Big\{\,
    \big(m_\mu H_{\tfrac{1}{2},0}^{V A,\nu_L}\big)
    \big(m_\mu H_{\tfrac{1}{2},t}^{V A,\nu_L}+\sqrt{q^2}\,H_{\tfrac{1}{2},t}^{S P,\nu_L}\big)^{\!*} \nonumber \\ &
   + \big(m_\mu H_{-\tfrac{1}{2},0}^{V A,\nu_L}\big)
     \big(m_\mu H_{-\tfrac{1}{2},t}^{V A,\nu_L}+\sqrt{q^2}\,H_{-\tfrac{1}{2},t}^{S P,\nu_L}\big)^{\!*}
   \,\Big\}
\Bigg)\, \label{M2nuL} \\
\mathcal{M}_3^{\nu_L}
= & -\frac{\alpha_P}{\Gamma_{0}^{\nu_L}}\Bigg(
   q^2\left|H_{\tfrac{1}{2},1}^{V A,\nu_L}\right|^2
 + q^2\left|H_{-\tfrac{1}{2},-1}^{V A,\nu_L}\right|^2
 + 2 m_\mu \operatorname{\Re}\Big\{\,
     H_{\tfrac{1}{2},0}^{V A,\nu_L}
       \big(m_\mu H_{\tfrac{1}{2},t}^{V A,\nu_L}
        + \sqrt{q^2}\,H_{\tfrac{1}{2},t}^{S P,\nu_L}\big)^{\!*} \nonumber \\ &
 - H_{-\tfrac{1}{2},0}^{V A,\nu_L}
       \big(m_\mu H_{-\tfrac{1}{2},t}^{V A,\nu_L}
        + \sqrt{q^2}\,H_{-\tfrac{1}{2},t}^{S P,\nu_L}\big)^{\!*}
   \,\Big\}
\Bigg)\, \label{M3nuL} \\
\mathcal{M}_4^{\nu_L}  = & -\frac{1}{2 \Gamma_{0}^{\nu_L}}\left(m_\mu^2-q^2\right)\left(\left|H_{\frac{1}{2}, 1}^{V A, \nu_L}\right|^2+\left|H_{-\frac{1}{2},-1}^{V A, \nu_L}\right|^2-2\left(\left|H_{\frac{1}{2}, 0}^{V A, \nu_L}\right|^2+\left|H_{-\frac{1}{2}, 0}^{V A, \nu_L}\right|^2\right)\right), ~\label{M4nuL}  \\
 \mathcal{M}_5^{\nu_L}  = & \frac{\alpha_P}{\Gamma_{0}^{\nu_L}}\left(m_\mu^2-q^2\right)\left(2\left|H_{\frac{1}{2}, 0}^{V A, \nu_L}\right|^2-\left|H_{\frac{1}{2}, 1}^{V A, \nu_L}\right|^2+\left|H_{-\frac{1}{2},-1}^{V A, \nu_L}\right|^2-2\left|H_{-\frac{1}{2}, 0}^{V A, \nu_L}\right|^2\right), ~\label{M5nuL} \\
 \mathcal{M}_6^{\nu_L} = & -\frac{\alpha_P}{\sqrt{2} \Gamma_{0}^{\nu_L}} 2 \operatorname{\Re}\left\{\left(H_{-\frac{1}{2},-1}^{V A, \nu_L}\right)^* \left(m_\mu\left(m_\mu H_{\frac{1}{2}, t}^{V A, \nu_L}+\sqrt{q^2} H_{\frac{1}{2}, t}^{S P}\right)+q^2 H_{\frac{1}{2}, 0}^{V A, \nu_L}\right)\right. \nonumber \\ &
 \left.+\left(H_{\frac{1}{2}, 1}^{V A, \nu_L}\right)^* \left(q^2 H_{-\frac{1}{2}, 0}^{V A, \nu_L}-m_\mu\left(m_\mu H_{-\frac{1}{2}, t}^{V A, \nu_L}+\sqrt{q^2} H_{-\frac{1}{2}, t}^{S P, \nu_L}\right)\right)\right\}, ~\label{M6nuL}   \\
   \mathcal{M}_7^{\nu_L} = & -\frac{\alpha_P}{\sqrt{2} \Gamma_{0}^{\nu_L}} 2 \operatorname{\Im}\left\{\left(H_{\frac{1}{2}, 1}^{V A, \nu_L}\right)^* \left(q^2 H_{-\frac{1}{2}, 0}^{V A, \nu_L}-m_\mu\left(m_\mu H_{-\frac{1}{2}, t}^{V A, \nu_L}+\sqrt{q^2} H_{-\frac{1}{2}, t}^{S P, \nu_L}\right)\right)\right. \nonumber \\ &
 \left.-\left(H_{-\frac{1}{2},-1}^{V A, \nu_L}\right)^* \left(m_\mu\left(m_\mu H_{\frac{1}{2}, t}^{V A, \nu_L}+\sqrt{q^2} H_{\frac{1}{2}, t}^{S P, \nu_L}\right)+q^2 H_{\frac{1}{2}, 0}^{V A, \nu_L}\right)\right\}, ~\label{M7nuL} \\
  \mathcal{M}_8^{\nu_L} =  & \frac{\alpha_P}{\sqrt{2} \Gamma_{0}^{\nu_L}}\left(m_\mu^2-q^2\right) 2 \operatorname{\Re}\left\{\left(H_{-\frac{1}{2},-1}^{V A, \nu_L}\left(H_{\frac{1}{2}, 0}^{V A, \nu_L}\right)^* -H_{-\frac{1}{2}, 0}^{V A, \nu_L}\left(H_{\frac{1}{2}, 1}^{V A, \nu_L}\right)^* \right)\right\}, ~\label{M8nuL} \\
  \mathcal{M}_9^{\nu_L} = & \frac{\alpha_P}{\sqrt{2} \Gamma_{0}^{\nu_L}}\left(m_\mu^2-q^2\right) 2 \operatorname{\Im}\left(H_{-\frac{1}{2},-1}^{V A, \nu_L}\left(H_{\frac{1}{2}, 0}^{V A, \nu_L}\right)^* +H_{\frac{1}{2}, 1}^{V A, \nu_L}\left(H_{-\frac{1}{2}, 0}^{V A, \nu_L}\right)^* \right) ~\label{M9nuL}  
\end{align}

where $\Gamma^{\nu_L}_{total}$ is the total decay width written as in eq. \ref{total_decay_width}.

%%%%%%%%%%%%%%%%%%%%%%%%%%%%%%%%%%%%%%%%%
\subsection{Angular observables with Right-Handed Neutrino} \label{4body_ang_obs_rhn}
%%%%%%%%%%%%%%%%%%%%%%%%%%%%%%%%%%%%%%%%%%%%

\begin{align}
     \mathcal{M}_0^{\nu_R} & =\frac{1}{\Gamma_{0}^{\nu_R}}\left( \frac{1}{2}\left(m_{\mu}^2+q^2\right) \Big| H^{VA,\nu_R}_{\frac{1}{2},1} \Big| ^{2}+\frac{1}{2} \left(m_{\mu}^2+q^2\right) \Big| H^{VA,\nu_R}_{-\frac{1}{2},-1} \Big| ^{2}+ \Big| m_{\mu}H^{VA,\nu_R}_{\frac{1}{2},t} + \sqrt{q^2}\,H^{SP,\nu_R}_{\frac{1}{2},t} \Big| ^{2}\right. \nonumber \\
		&\left.+ \Big|m_{\mu} H^{VA,\nu_R}_{-\frac{1}{2},t} + \sqrt{q^2}\,H^{SP,\nu_R}_{-\frac{1}{2},t} \Big| ^{2}+q^2 \Big| H^{VA,\nu_R}_{\frac{1}{2},0} \Big| ^{2}+q^2 \Big| H^{VA,\nu_R}_{-\frac{1}{2},0}\Big| ^{2}\right), ~\label{M0_nuR}    \\  
\mathcal{M}_1^{\nu_R} &=\frac{\alpha_P}{2\Gamma_{0}^{\nu_R}}\left(\left(m_{\mu}^2+q^2\right) \Big| H^{VA,\nu_R}_{\frac{1}{2},1} \Big| ^{2}+\left(m_{\mu}^2+q^2\right) \Big| H^{VA,\nu_R}_{-\frac{1}{2},-1} \Big| ^{2}-2 \Big|m_{\mu} H^{VA,\nu_R}_{\frac{1}{2},t} + \sqrt{q^2}\,H^{SP}_{\frac{1}{2},t} \Big| ^{2}\right. \nonumber \\
		&\left.+2 \Big| m_{\mu}H^{VA,\nu_R}_{-\frac{1}{2},t} + \sqrt{q^2}\,H^{SP,\nu_R}_{-\frac{1}{2},t} \Big| ^{2}-2 q^2 \Big| H^{VA,\nu_R}_{\frac{1}{2},0} \Big| ^{2}+2 q^2 \Big| H^{VA,\nu_R}_{-\frac{1}{2},0} \Big| ^{2}\right), ~ \label{M1_nuR}  \\
   \mathcal{M}_2^{\nu_R}
&= -\frac{1}{\Gamma_{0}^{\nu_R}}\Bigg(
   -q^2 \Big|H^{VA,\nu_R}_{\tfrac{1}{2},1}\Big|^2
   + q^2 \Big|H^{VA,\nu_R}_{-\tfrac{1}{2},-1}\Big|^2 
   + 2\,\Re\Big\{
      m_{\mu}H^{VA,\nu_R}_{\tfrac{1}{2},0}
      \Big(m_{\mu}H^{VA,\nu_R}_{\tfrac{1}{2},t}
      + \sqrt{q^2}\,H^{SP,\nu_R}_{\tfrac{1}{2},t}\Big)^{\!*} \nonumber \\
&      + m_{\mu}H^{VA,\nu_R}_{-\tfrac{1}{2},0}
      \Big(m_{\mu}H^{VA,\nu_R}_{-\tfrac{1}{2},t}
      + \sqrt{q^2}\,H^{SP,\nu_R}_{-\tfrac{1}{2},t}\Big)^{\!*}
   \Big\} \Bigg)\ , \label{M2_nuR}   \\ 
   \mathcal{M}_3^{\nu_R}
&= \frac{\alpha_P}{\Gamma_{0}^{\nu_R}}\Bigg(
   q^2 \Big|H^{VA,\nu_R}_{\tfrac{1}{2},1}\Big|^2
 + q^2 \Big|H^{VA,\nu_R}_{-\tfrac{1}{2},-1}\Big|^2
 - 2 m_{\mu}\,\Re\Big\{
     H^{VA,\nu_R}_{\tfrac{1}{2},0}
     \Big(m_{\mu}H^{VA,\nu_R}_{\tfrac{1}{2},t}
       + \sqrt{q^2}\,H^{SP,\nu_R}_{\tfrac{1}{2},t}\Big)^{\!*} \nonumber \\ &
   - H^{VA,\nu_R}_{-\tfrac{1}{2},0}
     \Big(m_{\mu}H^{VA,\nu_R}_{-\tfrac{1}{2},t}
       + \sqrt{q^2}\,H^{SP,\nu_R}_{-\tfrac{1}{2},t}\Big)^{\!*}
   \Big\} \Bigg)\ , \label{M3_nuR} \\
    \mathcal{M}_4^{\nu_R}  = & -\frac{1}{2\Gamma_{0}^{\nu_R}} \left(m_{\mu}^2-q^2\right) \left( \Big| H^{VA,\nu_R}_{\frac{1}{2},1} \Big|^{2} + \Big| H^{VA,\nu_R}_{-\frac{1}{2},-1} \Big|^{2} -2 \left( \Big| H^{VA,\nu_R}_{\frac{1}{2},0} \Big| {}^2+ \Big| H^{VA,\nu_R}_{-\frac{1}{2},0} \Big| {}^2\right)\right), ~\label{M4_nuR} \\
     \mathcal{M}_5^{\nu_R} = & \frac{\alpha_P}{\Gamma_{0}^{\nu_R}}(m_{\mu}^2-q^2)\left(2 \Big| H^{VA,\nu_R}_{\frac{1}{2},0} \Big|^2 - \Big| H^{VA,\nu_R}_{\frac{1}{2},1} \Big|^2 + \Big| H^{VA,\nu_R}_{-\frac{1}{2},-1} \Big|^2 -2 \Big| H^{VA,\nu_R}_{-\frac{1}{2},0} \Big|^2\right), ~\label{M5_nuR}  \\
     \mathcal{M}_6^{\nu_R} &=-\frac{\alpha_P}{\sqrt{2}\Gamma_{0}^{\nu_R}} 2 \Re\left\{\left(H^{VA,\nu_R}_{-\frac{1}{2},-1}\right){}^* \left( - m_{\mu}(m_{\mu}H^{VA,\nu_R}_{\frac{1}{2},t}+\sqrt{q^2}\,H^{SP,\nu_R}_{\frac{1}{2},t})+q^2 H^{VA,\nu_R}_{\frac{1}{2},0}\right)\right. \nonumber \\
		&\left. ~~+\left(H^{VA,\nu_R}_{\frac{1}{2},1}\right){}^* \left(q^2 H^{VA,\nu_R}_{-\frac{1}{2},0}  + m_{\mu}(m_{\mu}H^{VA,\nu_R}_{-\frac{1}{2},t}+\sqrt{q^2}\,H^{SP,\nu_R}_{-\frac{1}{2},t})\right)\right\}, ~\label{M6_nuR} \\
        \mathcal{M}_7^{\nu_R} & = -\frac{\alpha_P}{\sqrt{2}\Gamma_{0}^{\nu_R}}2 \Im\left\{\left(H^{VA,\nu_R}_{\frac{1}{2},1}\right){}^* \left(q^2 H^{VA,\nu_R}_{-\frac{1}{2},0} + m_{\mu}(m_{\mu}H^{VA,\nu_R}_{-\frac{1}{2},t}+\sqrt{q^2}\,H^{SP,\nu_R}_{-\frac{1}{2},t})\right)\right. \nonumber \\
		&\left.~~-\left(H^{VA,\nu_R}_{-\frac{1}{2},-1}\right){}^* \left( - m_{\mu}(m_{\mu}H^{VA,\nu_R}_{\frac{1}{2},t}+\sqrt{q^2}\,H^{SP,\nu_R}_{\frac{1}{2},t})+q^2 H^{VA,\nu_R}_{\frac{1}{2},0}\right)\right\}, ~\label{M7_nuR}  \\
        \mathcal{M}_8^{\nu_R} = & \frac{\alpha_P}{\sqrt{2}\Gamma_{0}^{\nu_R}}(m_{\mu}^2-q^2)2 \Re{\left(H^{VA,\nu_R}_{-\frac{1}{2},-1} \left(H^{VA,\nu_R}_{\frac{1}{2},0}\right){}^*-H^{VA,\nu_R}_{-\frac{1}{2},0} \left(H^{VA,\nu_R}_{\frac{1}{2},1}\right){}^*\right)}, ~\label{M8_nuR} \\
        \mathcal{M}_9^{\nu_R} = & \frac{\alpha_P}{\sqrt{2}\Gamma_{0}^{\nu_R}}(m_{\mu}^2-q^2)2 \Im \left(H^{VA,\nu_R}_{-\frac{1}{2},-1} \left(H^{VA,\nu_R}_{\frac{1}{2},0}\right){}^*+H^{VA,\nu_R}_{\frac{1}{2},1} \left(H^{VA,\nu_R}_{-\frac{1}{2},0}\right){}^*\right)  ~\label{M9_nuR}
\end{align}

where the expression of total decay width $\Gamma_{0}^{\nu_L(\nu_R)}$ for the left-handed and right-handed neutrinos is written as

\begin{equation}
\label{total_decay_width}
\begin{aligned}
\Gamma_{0}^{\nu_L(\nu_R)} & \equiv \frac{2}{3}\left\{( m _ { \mu } ^ { 2 } + 2 q ^ { 2 } ) \left(\left|H_{-\frac{1}{2},-}^{V A, \nu_L(\nu_R)}\right|^2+\left|H_{-\frac{1}{2}, 0}^{V A, \nu_L(\nu_R)}\right|^2+\left|H_{\frac{1}{2}, 0}^{V A, \nu_L(\nu_R)}\right|^2\right.\right. \left.+\left|H_{\frac{1}{2}, +}^{V A, \nu_L(\nu_R)}\right|^2\right) \\ & + 3\left(\left|m_\mu H_{-\frac{1}{2}, t}^{V A, \nu_L(\nu_R)}+\sqrt{q^2} H_{-\frac{1}{2}, t}^{S P, \nu_L(\nu_R)}\right|^2\right. \left.\left.+\left|m_\mu H_{\frac{1}{2}, t}^{V A, \nu_L(\nu_R)}+\sqrt{q^2} H_{\frac{1}{2}, t}^{S P, \nu_L(\nu_R)}\right|^2\right)\right\} 
\end{aligned}
\end{equation}

The hadronic helicity amplitude for the left-handed and right-handed neutrinos is given in the Appendix \ref{app-A.2}.

%%%%%%%%%%%%%%%%%%%%%%%%%%%%%%%%%%%%%%%%%%%%%%%%%%%%%%%%%%%%%%%%%%%%%%%%%%%%%%%

\bibliographystyle{JHEPsid}
\bibliography{reference}

\end{document}